\documentclass{aa}  

\usepackage{graphicx}
\usepackage{booktabs}
\usepackage{supertabular}
\usepackage[version=4]{mhchem}
\usepackage{ulem}

\usepackage[varg]{txfonts}

\begin{document}

   \title{Complex organic molecules in low-mass protostars on Solar System scales}
    \subtitle{II. Nitrogen-bearing species}
   \author{P. Nazari
          \inst{1},
          M. L. van Gelder
          \inst{1},
          E. F. van Dishoeck
          \inst{1,2}
          \and
          B. Tabone
          \inst{1}
          \and
          M. L. R. van ’t Hoff
          \inst{3}
          \and
          N. F. W. Ligterink
          \inst{4}
          \and
          H. Beuther
          \inst{5}
          \and
          A. C. A. Boogert
          \inst{6}
          \and
          A. Caratti o Garatti
          \inst{7}
          \and
          P. D. Klaassen
          \inst{8}
          \and
          H. Linnartz
          \inst{9}
          \and
          V. Taquet
          \inst{10}
          \and
          {\L}. Tychoniec \inst{1,11}
          }

   \institute{Leiden Observatory, Leiden University, P.O. Box 9513, 2300 RA Leiden, the Netherlands\\ 
        \email{nazari@strw.leidenuniv.nl}
    \and
        Max Planck Institut f\"{u}r Extraterrestrische Physik (MPE), Giessenbachstrasse 1, 85748 Garching, Germany
        \and
        Department of Astronomy, University of Michigan, 1085 S. University Ave., Ann Arbor, MI 48109, USA
        \and
        Space Research and Planetary Sciences, Physics Institute, University of Bern, Sidlerstrasse 5, 3012 Bern, Switzerland
        \and
        Max Planck Institute for Astronomy, K\"{o}nigstuhl 17, 69117 Heidelberg, Germany
        \and
        Institute for Astronomy, University of Hawaii at Manoa, 2680 Woodlawn Drive, Honolulu, HI 96822, USA
        \and
        Dublin Institute for Advanced Studies, School of Cosmic Physics, Astronomy and Astrophysics Section, 31 Fitzwilliam Place, D04C932 Dublin 2, Ireland
        \and
        UK Astronomy Technology Centre, Royal Observatory Edinburgh, Blackford Hill, Edinburgh EH9 3HJ, UK
        \and 
        Laboratory for Astrophysics, Leiden Observatory, Leiden University, P.O. Box 9513, 2300 RA Leiden, the Netherlands
        \and
        INAF, Osservatorio Astrofisico di Arcetri, Largo E. Fermi 5, 50125 Firenze, Italy
        \and
        ESO/European Southern Observatory, Karl-Schwarzschild-Strasse 2, D-85748 Garching bei M\"{u}nchen, Germany}

   \date{Received XXX; accepted YYY}

 
  \abstract
   {The chemical inventory of planets is determined by the physical and chemical processes that govern the early phases of star formation. Nitrogen-bearing species are of interest as many provide crucial precursors in the formation of life-related matter.}
   {The aim is to investigate nitrogen-bearing complex organic molecules towards two deeply embedded Class 0 low-mass protostars (Perseus B1-c and Serpens S68N) at millimetre wavelengths with the Atacama Large Millimeter/submillimeter Array (ALMA). Next, the results of the detected nitrogen-bearing species are compared with those of oxygen-bearing species for the same and other sources. The similarities and differences are used as further input to investigate the underlying formation pathways.}
   {ALMA observations of B1-c and S68N in Band 6 (${\sim} 1\rm \, mm$) and Band 5 (${\sim} 2\,\rm mm$) are studied at ${\sim} 0.5\arcsec$ resolution, complemented by Band 3 (${\sim} 3$ mm) data in a  ${\sim}2.5\arcsec$ beam. The spectra are analysed for nitrogen-bearing species using the CASSIS spectral analysis tool, and the column densities and excitation temperatures are determined. A toy model is developed to investigate the effect of source structure on the molecular emission.}
   {Formamide ($\rm NH_{2}CHO$), ethyl cyanide ($\rm C_{2}H_{5}CN$), isocyanic acid (HNCO, $\rm HN^{13}CO$, DNCO), and methyl cyanide (CH$_{3}$CN, $\rm CH_{2}DCN$, and $\rm CHD_{2}CN$) are identified towards the investigated sources. Their abundances relative to CH$_3$OH and HNCO are similar for the two sources, with column densities that are typically an order of magnitude lower than those of oxygen-bearing species. The largest variations, of an order of magnitude, are seen for $\rm NH_{2}CHO$ abundance ratios with respect to HNCO and $\rm CH_{3}OH$ and do not correlate with the protostellar luminosity. In addition, within uncertainties, the nitrogen-bearing species have similar excitation temperatures to those of oxygen-bearing species (${\sim} 100-300\,\rm K$).  The measured excitation temperatures are larger than the sublimation temperatures for the respective species.}
   {The similarity of most abundances with respect to HNCO for the investigated sources, including those of $\rm CH_{2}DCN$ and $\rm CHD_{2}CN$, hints at a shared chemical history, especially the high D-to-H ratio in cold regions prior to star formation. However, some of the variations in abundances may reflect the sensitivity of the chemistry to local conditions such as temperature (e.g. $\rm NH_{2}CHO$), while others may arise from differences in the emitting areas of the molecules linked to their different binding energies in the ice. The excitation temperatures likely reflect the mass-weighted kinetic temperature of a gas that follows a power law structure. The two sources discussed in this work add to the small number of sources that have been subjected to such a detailed chemical analysis on Solar System scales. Future data from the \textit{James Webb Space Telescope} will allow a direct comparison between the ice and gas abundances of both smaller and larger nitrogen-bearing species.}


\titlerunning{N-bearing Complex organic molecules in low-mass protostars}
\authorrunning{P. Nazari et al.}
   \maketitle

%

\section{Introduction}

Interstellar molecules with six or more atoms containing carbon atoms as well as hydrogen and oxygen or nitrogen are known as O- or N-bearing complex organic molecules (COMs). Other COMs comprise both oxygen and nitrogen atoms and/or other elements such as sulphur and phosphorus. The Class 0 protostellar stage is the warmest stage during star formation and thus the richest in gas-phase COMs due to the thermal sublimation of ices in hot corinos (\citealt{vantHoff2020c}). Class 0 sources are thus prime targets to unveil their associated chemistry through the observation of (sub-)millimetre lines. Over the past decades, single dish and interferometric studies on cloud scales have indeed revealed that both low-mass (e.g. \citealt{Ewine1995}; \citealt{Cazaux2003}; \citealt{Bottinelli2004}; \citealt{Bottinelli2004IRAS}; \citealt{Bisschop2008}; \citealt{Jorgensen2016}; \citealt{Ceccarelli2017}; \citealt{Bergner2017}; \citealt{Bianchi2019}) and high-mass protostars (\citealt{Blake1987}; \citealt{Gibb2000}; \citealt{Nummelin2000}; \citealt{fontani2007}; \citealt{Belloche2013}; \citealt{Ilee2016}; \citealt{Bogelund2019}; \citealt{Taniguchi2020}) are rich in COMs. It has also recently been found that planet formation likely starts at earlier stages of low-mass star formation, during the Class 0$/$I stage (\citealt{Harsono2018}; \citealt{Manara2018}; \citealt{Tychoniec2018}; \citealt{Tychoniec2020}; \citealt{Tobin2020}). Therefore, to understand the chemical enrichment of forming planets in complex compounds, one needs to study the formation of such species at these early phases on Solar System scales (${\sim} 50\,\rm au$). 

Complex organic molecules start forming at the dawn of star formation, in molecular clouds of gas and dust with temperatures of ${\sim} 10-20\, \rm K$ and initial densities of ${\sim} 10^{3}-10^{4}\, \rm cm^{-3}$. Under these conditions, first simple molecules, such as water ($\rm H_{2}O$), carbon dioxide ($\rm CO_{2}$), and ammonia ($\rm NH_{3}$), form on icy grains (\citealt{Boogert2015}; \citealt{Linnartz2015}). Once CO freezes out, O-bearing COMs start forming through the hydrogenation of CO, resulting in the formation of formaldehyde ($\rm H_{2}CO$) and methanol ($\rm CH_{3}OH$; \citealt{Watanabe2002}; \citealt{Fuchs2009}). Species as complex as glycolaldehyde ($\rm HC(O)CH_{2}OH$) and ethylene glycol ($\rm H_{2}C(OH)CH_{2}OH$) can form through the recombination of HCO radicals at low temperatures (\citealt{Fedoseev2015}). It is expected that other radical recombinations result in the formation of even larger O-containing COMs, such as glycerol ($\rm HOCH_{2}CH(OH)CH_{2}OH$; \citealt{Fedoseev2017}). As the temperature increases, more COMs, such as ethanol ($\rm CH_{3}CH_{2}OH$; \citealt{Oberg2009}) and propanal ($\rm CH_{3}CH_{2}CHO$; \citealt{Qasim2019}), can form in the ice, perhaps with the assistance of some UV radiation.

Much less is known about the formation of N-bearing species in ices.
Nitrogen-bearing molecules are important as nitrogen is a crucial element in developing biomolecules such as amino acids and nucleobases, and they are thus essential for the emergence of life. N-bearing COMs have been detected in high-mass (e.g. \citealt{Isokoski2013}; \citealt{Belloche2016}; \citealt{Bogelund2019ngc}; \citealt{Csengeri2019}; \citealt{Ligterink2020}) and low-mass protostars (e.g. \citealt{Bottinelli2008}; \citealt{Ligterink2018}; \citealt{Marcelino2018}; \citealt{calcutt2018}; \citealt{Lee2019}; \citealt{Belloche2020}). Some are thought to form through the recombination of radicals in the solid state, such as methylamine ($\rm CH_{3}NH_{2}$) from $\rm NH_{2}$ and $\rm CH_{3}$ produced by the photodissociation of NH$_3$ and CH$_4$ ice, although these radicals can also result from hydrogen addition reactions to N and C atoms (\citealt{Garrod2008}). Other species can form via isocyanic acid (HNCO) in the solid state \citep[e.g. methyl isocyanate,
$\rm CH_{3}NCO$;][]{Cernicharo2016, Ligterink2017}. Formamide ($\rm NH_{2}CHO$) is thought to have gas-phase formation routes (\citealt{Barone2015}; \citealt{Song2016}; \citealt{Codella2017}; \citealt{Skouteris2017}) as well as two main ice pathways, namely the hydrogenation of HNCO (e.g. \citealt{Raunier2004}; \citealt{Haupa2019}) and the radical-radical addition of $\rm NH_{2}$ and CHO (\citealt{Jones2011}, \citealt{Dulieu2019}; \citealt{Martin2020}). HNCO itself can also form on the surface of interstellar ices via the reaction of CO and NH radicals (\citealt{Fedoseev2015}). Another molecule put forward as a parent molecule for many N-bearing COMs is methyl cyanide \citep[CH$_{3}$CN;][]{Bulak2021}, which may be produced via reactions of $\rm CH_{3}$ and CN radicals in ices. One route for the formation of ethyl cyanide ($\rm C_{2}H_{5}CN$) is thought to be a piecewise addition of its functional groups on dust grains (\citealt{Belloche2009}). In order to elucidate the formation pathways to N-bearing COMs, more observational constraints on their abundances are needed.

Another clue regarding formation routes comes from observed deuteration fractions.  The low temperature in the dense core stage causes the $\rm D/H$ ratio in molecules to increase to higher values than the elemental ratio in the interstellar medium (ISM) of ${\sim} 2 \times 10^{-5}$
(\citealt{Prodanovic2010}). The chemical reaction $\ce{H_{3}^{+} + HD <=> H_{2}D^{+} + H_{2} + \Delta E}$ increases the amount of $\rm H_{2}D^{+}$ in the gas phase at low temperatures (${\sim} 10-20\, \rm K$) since this reaction has a small activation barrier for its reverse reaction (\citealt{Watson1976}; \citealt{Aikawa1999}; \citealt{Tielens2013}; \citealt{Ceccarelli2014}; \citealt{Sipila2015}). Moreover, this process can be enhanced by CO freeze-out as CO is one of the main molecules to destroy H$_{3}^{+}$ and H$_{2}$D$^{+}$ in the gas phase (\citealt{Brown1989}; \citealt{Roberts2003}). Dissociative recombination of $\rm H_{2}D^{+}$ will enhance the D/H ratio, and, subsequently, D atoms can be transferred onto grains, enriching the ice. Therefore, D/H values provide clues on the temperature at which N-bearing COMs are formed, and thus on their formation history \citep{Taquet2012, Taquet2014, Furuya2016}.

To assess whether many of the N- and O-bearing COMs are indeed produced in ices, a smoking gun would be to detect their ice features directly at mid-infrared wavelengths. The total number of unambiguously identified ice species  is, however, relatively small and so far only comprises one securely identified COM \citep[$\rm CH_{3}OH$;][]{Grim1991, Taban2003}. This is largely a consequence of observational limitations (see the review by \citealt{Boogert2015}); even with more laboratory ice data for COMs currently available (see e.g. \citealt{Terwisscha2018}), only the more abundant molecules in ices can be detected. There are upper limit measurements of some N-bearing COMs in ices towards massive protostars, for example aminomethanol $\rm (NH_{2}CH_{2}OH$; \citealt{Bossa2009}) and $\rm NH_{2}CHO$ (\citealt{Schutte1999}). However, $\rm OCN^{-}$, a direct derivative of HNCO (\citealt{Broekhuizen2004}; \citealt{Fedoseev2016}), has been detected in the ISM in ices (\citealt{Grim1987}; \citealt{vanBroekhuizen2005}; \citealt{Oberg2011}). Observations of most molecules in the solid state (at near- and mid-infrared) are not possible from the ground because a large part of the wavelength range is blocked by the Earth's atmosphere. Moreover, moderate spectral resolution is needed in the critical $3-10\,\mu\rm m$ wavelength range for observation of most molecules in ices, which the \textit{Spitzer Space Telescope} did not have. The \textit{James Webb Space Telescope} (JWST), with its unique sensitivity and appropriate spectral resolution, will transform the study of COMs in the solid state.

The present work focuses on gas-phase sub-millimetre identifications of N-bearing species and has only become possible because of the superb performance of the Atacama Large Millimeter/submillimeter Array \citep[ALMA;][]{Jorgensen2020review}. This is because ALMA has a much higher sensitivity and spatial resolution than the pre-existing telescopes, which enables the study of low-mass protostars on Solar System scales. Moreover, the high sensitivity of ALMA allows the observation of molecule isotopologues. This is especially important for highly abundant molecules that show optically thick emission as their abundances can be measured more accurately using their optically thin isotopologues. ALMA observations provide information on the gas-phase chemical inventory in the hot core, where all ices have sublimated, and these data can then eventually be compared with JWST observations of the ice composition to directly link gas and ice chemistry.

One of the most well-studied low-mass protostars in complex chemistry is IRAS 16293-2422 (hereafter IRAS 16293),  which was investigated as part of the ALMA Protostellar Interferometric Line Survey (PILS) programme (\citealt{Jorgensen2016}). The PILS gives the most complete inventory of nitrogen- and oxygen-bearing COMs in low-mass protostars to date (\citealt{Coutens2016}; \citealt{Ligterink2017}; \citealt{Ligterink2018}; \citealt{calcutt2018}; \citealt{Manigand2020}). \cite{Jorgensen2018} and \cite{Manigand2020} suggest that there are two categories for O-bearing and N-bearing species: Some molecules desorb at temperatures of ${\sim} 100\,\rm K$ and others at ${\sim} 300\, \rm K$, closer to the central protostar. Apart from IRAS 16293, N-bearing COMs have been observed with ALMA towards a handful of low-mass sources: HH 212 (\citealt{Lee2019}), NGC 1333 IRAS 4A2 (\citealt{Lopez-Sepulcre2017}), B1b-S (\citealt{Marcelino2018}), B335 (\citealt{Imai2016}), and L483 (\citealt{Oya2017}). This list has also been supplemented using the Plateau de Bure Interferometer (PdBI) and its upgraded version, the Northern Extended Millimeter Array \citep[NOEMA;][]{Taquet2015,Belloche2020}.

In this paper, ALMA observations are used to study two Class 0 objects: B1-c in the Perseus Barnard 1 cloud and S68N in the Serpens Main star-forming region. These sources are targeted in ALMA Band 6, Band 5, and Band 3. The luminosity of B1-c at its distance of 321\,pc is $\rm 6.0\, L_{\odot}$ (\citealt{Karska2018}; \citealt{oritz2018}). S68N has a luminosity of $\rm 5.4\, L_{\odot}$ (\citealt{Enoch2011}) at its distance of 436\,pc (\citealt{Ortiz2017}). Both S68N and B1-c have been observed and studied with ALMA. Very recently, \cite{vangelder2020} presented observational data for O-bearing COMs towards both sources. In this paper we focus on N-bearing molecules to investigate how similarities and differences between N-bearing and O-bearing species reveal information on the involved chemical processes. Both sources are also targets of the guaranteed time observation (GTO) programme (project ID 1290) of JWST/MIRI (\citealt{Wright2015}). Therefore, in the near future, it will be possible to directly compare the ice observations of these sources obtained by JWST with what has been done in this work for their gas-phase counterparts in the hot corino, where these ices have sublimated.

The layout of this paper is as follows. Section \ref{sec:observations} describes the observations. Section \ref{sec:results} presents the methods and the results. In Sect. \ref{sec:Discussion} we discuss our results and put the sources studied here in the context of what has been done so far in the literature. We also compare the measured excitation temperature with the sublimation temperature of each molecule. Moreover, a simple toy model is constructed to understand how source structure may affect abundance ratios. Finally, a summary is given in Sect.  \ref{sec:conclusion}.     

\section{Observations and methods}
\subsection{The data}
\label{sec:observations}
Two COM-rich protostars (B1-c and S68N) were observed by ALMA (project code: 2017.1.01174.S; principal investigator: E.F. van Dishoeck). The data reduction and first results from these observations are explained in \cite{vangelder2020}. Here we only give a brief overview of the data. B1-c (RA$_{\rm J2000}$: 03:33:17.88, Dec$_{\rm J2000}$: 31:09:31.8) and S68N (RA$_{\rm J2000}$: 18:29:48.08, Dec$_{\rm J2000}$: 01:16:43.3) were observed during ALMA Cycle 5 at $3\,\rm mm$ (Band 3) and $1\,\rm mm$ (Band 6) using the 12m array. Two bands were used here to be sensitive to both more extended (Band 3, maximum baseline of ${\sim} 400\,\rm m$) and hence colder COM emission and the more compact (Band 6, maximum baseline of ${\sim} 800\, \rm m$) and thus warmer COM emission. The targeted N-bearing molecules were originally HNCO and $\rm NH_{2}CHO$, species with a likely solid-state formation origin \citep{Fedoseev2015, Fedoseev2016}. The other N-bearing (complex organic) molecules discussed in this work were serendipitously observed, and hence this work does not aim at a complete inventory of N-bearing molecules. The observational parameters and the lines covered in the data are presented in Appendix \ref{app:lines_tabs}.

The Band 3 data were taken using ALMA configurations C43-2 (S68N) and C43-3 (B1-c) with an angular resolution of ${\sim} 1.5-2.5\arcsec$. In Band 6 the C43-4 configuration was used with an angular resolution of ${\sim} 0.45\arcsec$, corresponding to radii of 72\,au and 98\,au for B1-c and S68N, respectively. The spectral resolution for most spectral windows is ${\sim} 0.2\, \rm km\, s^{-1}$. A few spectral windows in Band 3 have a spectral resolution of ${\sim} 0.3-0.4\, \rm km\, s^{-1}$. The maximum recoverable scales for Band 3 and Band 6 are ${\sim} 20\arcsec$ and ${\sim} 6\arcsec$, respectively. The line rms is ${\sim} 0.15$\,K in the Band 6 data. The absolute flux calibration uncertainty is $\leq 15\,\%$. 

Additionally, Band 5 data (project code: 2017.1.01371.S; principal investigator: M.L.R; van 't Hoff) are included in the analysis for B1-c to confirm identifications and get more accurate measurements of the excitation temperatures. The data were reduced using the ALMA pipeline (CASA version 5.1.1), after which line-free regions were carefully selected for continuum subtraction. The data were then imaged using a robust weighting of 0.5. This dataset has a similar angular resolution (${\sim} 0.45\arcsec$) to our Band 6 dataset, with a maximum baseline of ${\sim} 1.3\, \rm km$. The spectral resolution of this dataset is mostly ${\sim} 0.1 \rm \, km \, s^{-1}$, but for some of the data cubes it is ${\sim} 1.6 \rm \, km \, s^{-1}$. The line rms ranges from ${\sim} 0.1$\,K for spectral windows with ${\sim} 1.6 \rm \, km \, s^{-1}$ spectral resolution to ${\sim} 0.4$\,K for spectral windows with ${\sim} 0.1 \rm \, km \, s^{-1}$ spectral resolution in the Band 5 data. The covered frequency ranges for Bands 5 and 6 are given in Table \ref{tab:freq}.  

\subsection{Spectral modelling}
\label{sec:spec_modelling}

A molecule is considered to be detected when at least three lines are identified at a 3$\sigma$ level without over-predicting any line emission. It is called `tentatively detected' when it has one or two lines at a 3$\sigma$ level in the spectrum. An upper limit is reported when no lines are identified. 

We followed the approach in \cite{vangelder2020} to determine the column densities ($N$) and excitation temperatures ($T_{\rm ex}$) of molecules identified in the spectra. First, a grid of $N$ and $T_{\rm ex}$ for each molecule was set. Although the full width half maximums (FWHMs) are fixed for the final fits, they were varied first; however, it was found that either they do not vary significantly for unblended lines or they are non-constrained due to line blending. Therefore, they were fixed to the best-fit value found for clean, single lines. Assuming a single component origin, the simplest assumption is that a single excitation temperature describes the level populations of a molecule. This condition is referred to as `local thermodynamic equilibrium' (LTE) when the densities are high enough that the excitation temperature approaches the kinetic temperature and one can use the Boltzmann distribution to describe the population of all levels at a single temperature. \cite{Jorgensen2016} found that the assumption of LTE conditions is reasonable on scales of 100\,au for low-mass protostars such as IRAS 16293, where densities are ${\sim} 10^{8} - 10^{9}\,\rm cm^{-3}$ or higher. Assuming LTE conditions, the corresponding spectrum for each grid point is calculated using the CASSIS\footnote{\url{http://cassis.irap.omp.eu/}} (\citealt{Vastel2015}) spectral analysis tool. For each molecule, we used its corresponding line list from the Jet Propulsion Laboratory (JPL) database (\citealt{JPL}) and the Cologne Database for Molecular Spectroscopy (CDMS; \citealt{CDMS2001}; \citealt{muller2005}). Subsequently, the resulting modelled spectrum of each grid point was overlaid onto the observed spectrum of each source and its $\chi^{2}$ was computed. In the computation of the best-fit model, we did not include blended or optically thick lines. In fitting all the molecules in the Band 5 and 6 data, we used 20\% for the flux uncertainty. The uncertainty used here is a conservative estimate to take into account potential errors caused by continuum contamination due to the line richness of the sources.

We investigated a grid with a large range of column densities, from $10^{12}\,\rm cm^{-2}$ to $10^{16}\,\rm cm^{-2}$, with large spacings (for some molecules, a larger initial range was used for the grid). This was to ensure that the full parameter space was covered. Once the range of the final column density was found, a finer grid, with 0.05 spacings in logarithmic scale for the column density, was made. The excitation temperature in our grids mostly ranges from $10\,\rm K$ to $600\,\rm K$ with 10\,K spacings on a linear scale. For $\rm NH_{2}CHO$, $\rm C_{2}H_{5}CN$, and $\rm CHD_{2}CN$ towards B1-c, the maximum of the temperature grid was larger to guarantee that the resulting excitation temperature is not biased. We only fitted the temperature when there were several lines that covered a range of upper energy levels. Otherwise, we fixed the temperature to 200\,K for the Band 5 and 6 data. This is because \cite{vangelder2020} found 200\,K to be a typical excitation temperature for O-bearing species in Band 6. Where it is possible to fit for the temperature, the results from the fits were inspected more closely via the $\chi^{2}$ plots and by manually changing the temperature for a small range of column densities to make sure that the most accurate excitation temperature is found (see Appendix \ref{app:spec_fit}). Typically, the FWHM was fixed to $3\,\rm km\,  s^{-1}$ for B1-c and S68N unless broader lines were observed for a certain molecule. In that case, the FWHM was fixed to $4.5\,\rm km\, s^{-1}$ ($\rm HN^{13}CO$, $\rm C_{2}H_{5}CN$, and $\rm CH_{2}DCN$ in S68N).

In this procedure we fixed the source velocities ($V_{\rm lsr}$) to $6.0\,\rm km\, s^{-1}$ and $8.5\,\rm km\, s^{-1}$ for B1-c and S68N, respectively (\citealt{vangelder2020}). Furthermore, we followed the same method as in \cite{vangelder2020} and assumed that the source size for both sources is the same as the beam size for the Band 6 data ($0.45\arcsec$) since the compact emission from the inner envelope is unresolved (see Sect. \ref{sec:extent}). The beam dilution factor is given by $(\theta_{\rm b}^{2} + \theta_{\rm s}^{2})/\theta_{\rm s}^{2}$, where $\theta_{\rm b}$ is the beam size and $\theta_{\rm s}$ is the source size. Hence, assuming a source size of $0.45\arcsec$, the beam dilution factors are ${\sim}20$ and ${\sim}2$ in the Band 3 data and the Band 5 and 6 data, respectively. The source size affects the measured column densities but not their ratios, unless the lines become optically thick. This is discussed further in Sect. \ref{sec:sources_size}.  

We derived the uncertainties on the column densities using the $\chi^{2}$ error calculation of the grid ($2\sigma$) or the variation in the column densities when manually fitting for temperature. The $2\sigma$ uncertainty on the temperatures using the $\chi^{2}$ error calculation of the grid is reported when the temperatures derived from both the $\chi^{2}$ calculations and the inspection by eye are consistent. However, in cases where the $\chi^{2}$ method is not constraining (e.g. because there are not enough lines covering a large range in upper energy levels), the temperature is reported based on the fit-by-eye method, where we typically find a ${\sim}\rm 50\,K$ or ${\sim}\rm 100\,K$ error (see Appendix \ref{app:spec_fit} for a description of the method).

\section{Results}
\label{sec:results}

\subsection{Line identification and spatial extent}
\label{sec:extent}

\begin{table*}
    \caption{Column densities and excitation temperatures for B1-c and S68N in a $0.45\arcsec$ beam.}
    \label{tab:B1c_NT}
    \resizebox{\textwidth}{!}{\begin{tabular}{@{\extracolsep{1mm}}*{10}{c}}
          \toprule
          \toprule    
          & & \multicolumn{4}{c}{B1-c}&\multicolumn{4}{c}{S68N}\\
          \cmidrule{3-6} \cmidrule{7-10}
                Species & Catalogue & $T_{\rm ex} (\rm K)$ & $N (\rm cm^{-2})$ & $N/\rm{CH_{3}OH} (\%)$ & $N/\rm{HNCO} (\%)$ & $T_{\rm ex} (\rm K)$ & $N (\rm cm^{-2})$ & $N/\rm{CH_{3}OH} (\%)$ & $N/\rm{HNCO} (\%)$\\
                \midrule     

HNCO & CDMS &-- & $(1.6 \pm 0.7)\times 10^{16}$ & $0.9 \pm 0.5$ & $\equiv 100.0$ & -- & ${\sim} 1.2\times 10^{16}$ & $0.9 \pm 0.4$ & $\equiv 100.0$ \\
$\rm HN^{13}CO$ & JPL &$200 \pm 100^{*}$ & $(2.4 \pm 1.1)\times 10^{14}$ & $0.013 \pm 0.007$ & $\equiv 1.47$ &$[200]^{*}$ & ${\sim} 1.8\times 10^{14}$ & $0.013 \pm 0.005$ & $\equiv 1.47$ \\
DNCO & JPL &$[200]^{*}$ & $(3.0 \pm 0.2)\times 10^{14}$ & $0.016 \pm 0.005$ & $1.8 \pm 0.8$ &$[200]$ & $<6.8\times 10^{13}$ & $<0.005$ & $<0.6$ \\
$\rm NH_{2}CHO$ & JPL &$>70$ & $(5.1 \pm 2.4)\times 10^{14}$ & $0.03 \pm 0.02$ & $3.1 \pm 2.1$ &$[200]^{*}$ & $(2.7 \pm 0.3)\times 10^{14}$ & $0.019 \pm 0.009$ & $2.2 \pm 0.3$ \\
$\rm C_{2}H_{5}CN$ & CDMS &$275 \pm 125$ & $(1.1 \pm 0.3)\times 10^{15}$ & $0.06 \pm 0.02$ & $6.5 \pm 3.5$ &$170^{+100}_{-110}$ & $(1.1 \pm 0.6)\times 10^{15}$ & $0.08 \pm 0.05$ & $8.9 \pm 4.7$ \\
$\rm CH_{3}CN$ & CDMS &-- & $(7.4 \pm 2.2)\times 10^{15}$ & $0.4 \pm 0.2$ & $45.5 \pm 24.9$ &-- & $(2.9 \pm 1.0)\times 10^{15}$ & $0.2 \pm 0.1$ & $23.6 \pm 8.3$ \\
$\rm CH_{2}DCN$ & CDMS &$210^{+50}_{-40}$ & $(2.6 \pm 0.2)\times 10^{14}$ & $0.014 \pm 0.004$ & $1.6 \pm 0.7$ &$220^{+60}_{-110}$ & $(1.0 \pm 0.2)\times 10^{14}$ & $0.007 \pm 0.003$ & $0.8 \pm 0.2$ \\
$\rm CHD_{2}CN$ & CDMS &$>200$ & $(1.4 \pm 0.2)\times 10^{14}$ & $0.007 \pm 0.003$ & $0.9 \pm 0.4$ &$[200]$ & $<3.8\times 10^{13}$ & $<0.003$ & $<0.3$ \\
$\rm NH_{2}CN$ & JPL &$[200]$ & $<7.1\times 10^{13}$ & $<0.004$ & $<0.4$ &$[200]$ & $<1.0\times 10^{13}$ & $<0.0007$ & $<0.1$ \\
$\rm CH_{3}NCO$ & CDMS &$[200]$ & $<8.8\times 10^{14}$ & $<0.05$ & $<5.4$ &$[200]$ & $<3.8\times 10^{14}$ & $<0.03$ & $<3.1$ \\
$\rm CH_{3}NH_{2}$ & JPL &$[200]$ & $<5.1\times 10^{15}$ & $<0.3$ & $<31.5$ &$[200]$ & $<1.9\times 10^{15}$ & $<0.1$ & $<15.6$ \\
$\rm HOCH_{2}CN$ & CDMS &$[200]$ & $<2.0\times 10^{15}$ & $<0.1$ & $<12.2$ &$[200]$ & $<6.4\times 10^{14}$ & $<0.05$ & $<5.3$ \\

\bottomrule
        \end{tabular}}
        \tablefoot{Band 5 and 6 data are used to fit B1-c, and Band 6 data are used to fit S68N parameters. The column densities for $\rm CH_{3}OH$ are taken from \cite{vangelder2020}, who use the optically thin isotopologue $\rm CH_{3}^{18}OH$ to determine the column density of $\rm CH_{3}OH$.} The square brackets around the $T_{\rm ex}$ values indicate the molecules for which the temperature was fixed. The column densities for HNCO and $\rm CH_{3}CN$ are found from $\rm HN^{13}CO$ and $\rm CH_{2}DCN$ as their main isotopologues are optically thick. Species with a tentative detection are indicated by a star next to their excitation temperatures.
\end{table*}

\begin{figure*}
    \centering
    \includegraphics[width=17cm]{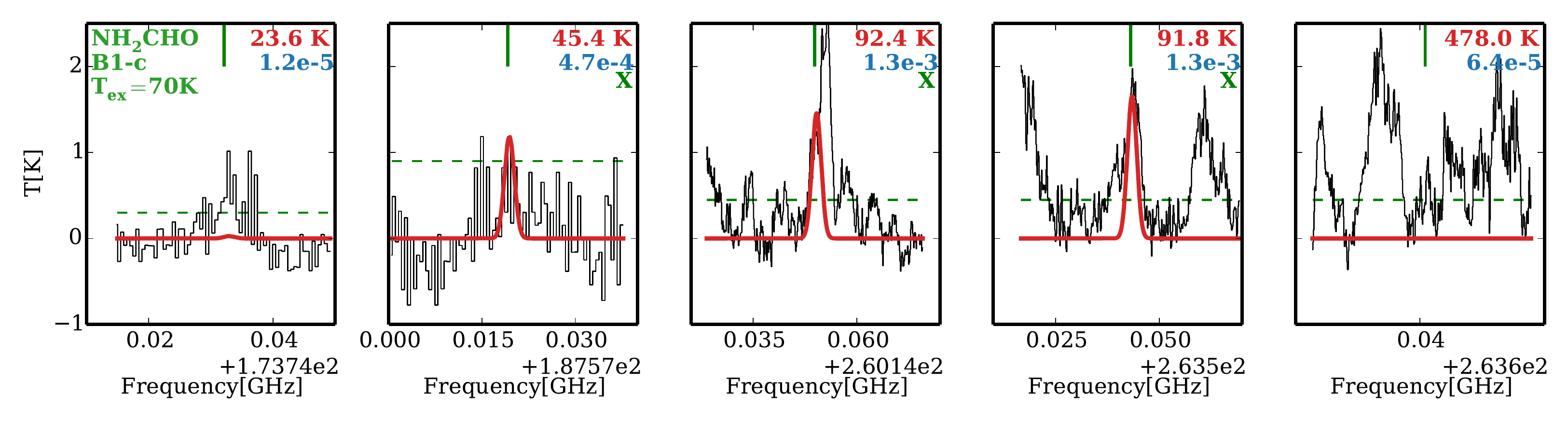}
    \caption{Best fitted model to combined Band 5 and 6 $\rm NH_{2}CHO$ data  for B1-c in red and data in black. Each graph shows one line of $\rm NH_{2}CHO$ with its upper state energy level and the $A_{ij}$ coefficient at the top right in red and blue, respectively. The dashed green line indicates the 3$\sigma$ level. The lines above the 3$\sigma$ level that were used in the fitting are indicated by a green X. The lines with upper energy levels above 1000\,K and/or $A_{ij}$ below $10^{-5}$ are not plotted.}
    \label{fig:NH2CHO_fit_inside}
\end{figure*}

\begin{figure*}
    \centering
    \includegraphics[width=17cm]{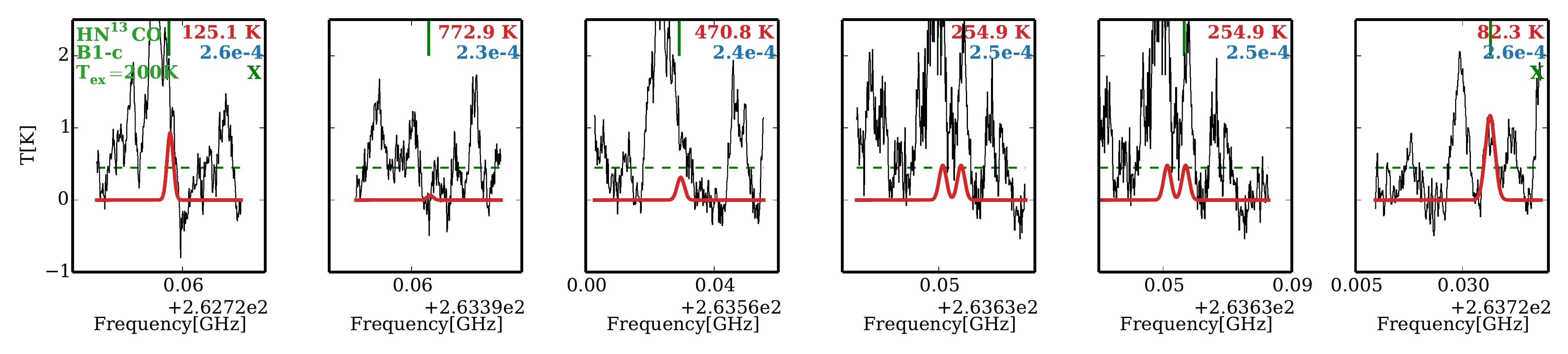}
    \caption{Same as Fig. \ref{fig:NH2CHO_fit_inside} but for $\rm HN^{13}CO$.} 
    \label{fig:HN13CO_fit_inside}
\end{figure*}

We find four N-bearing molecules and some of their isotopologues towards both sources. Towards B1-c, we detect HNCO, $\rm NH_{2}CHO$, $\rm C_{2}H_{5}CN$, $\rm CH_{2}DCN,$ and $\rm CHD_{2}CN$ and tentatively detect $\rm HN^{13}CO$ and DNCO in the Band 6 and 5 data. Moreover, $\rm CH_{3}CN$ is clearly detected in the Band 3 data. Towards S68N, we detect HNCO, $\rm C_{2}H_{5}CN$, and $\rm CH_{2}DCN$ and tentatively detect $\rm NH_{2}CHO$ and $\rm HN^{13}CO$ with upper limits for DNCO and $\rm CHD_{2}CN$ in the Band 6 data. In addition, $\rm CH_{3}CN$ is identified in the Band 3 data. Among the molecules searched for, we also find upper limits on $\rm NH_{2}CN$, $\rm CH_{3}NH_{2}$, $\rm CH_{3}NCO$, and $\rm HOCH_{2}CN$ for both sources in the Band 5 and 6 data. Given that in the Band 3 data only $\rm CH_{3}CN$ is securely detected, with a tentative detection of HNCO, we focused on the Band~6 data for S68N and the combined Band~5 and 6 data for B1-c for most of the analyses.

\begin{figure*}
    \centering
    \includegraphics[width=17cm]{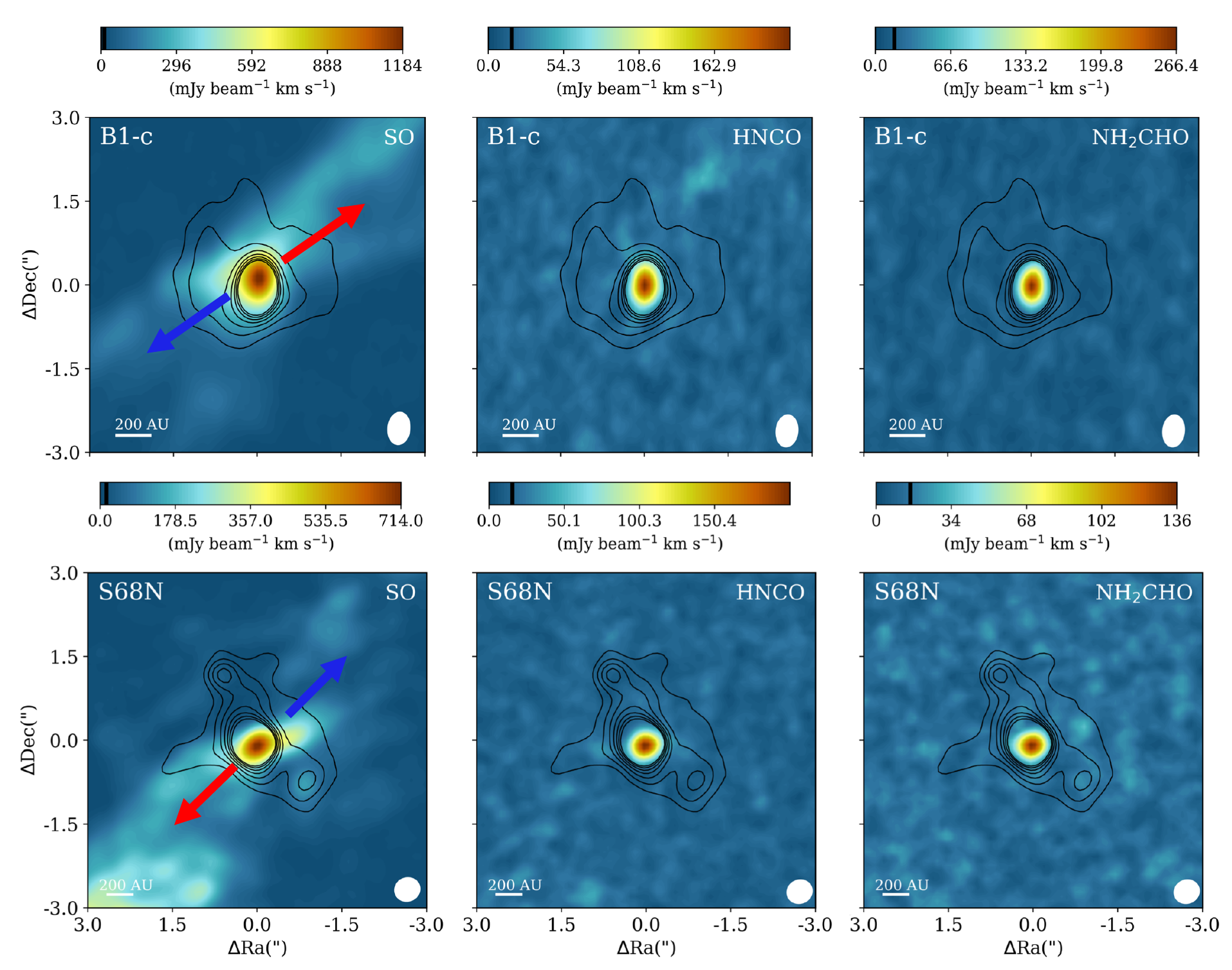}
    \caption{Moment zero maps of the lines of SO $6_{7} - 5_{6}$ ($E_{\rm up}=47.6\,\rm K$), HNCO $12_{0,12} - 11_{0,11}$ ($E_{\rm up}=82.3\,\rm K$), and $\rm NH_{2}CHO$ $13_{1,13} - 12_{1,12}$ ($E_{\rm up}=91.8\,\rm K$) (from left to right) for B1-c (top row) and S68N (bottom row) in the Band 6 data. The images are made by integrating over $[-10,10]\,\rm{km}\,\rm{s}^{-1}$ with respect to $V_{\rm lsr}$. The black contours show the continuum in levels of [15, 30, 45, 60, 75, 90, 105]$\,\sigma_{\rm cont}$ with a $\sigma_{\rm cont}$ of 0.2\,mJy\,$\rm beam^{-1}$ for B1-c and [30, 45, 60, 75, 90, 105, 120]$\,\sigma_{\rm cont}$ with a $\sigma_{\rm cont}$ of 0.09\,mJy\,$\rm beam^{-1}$ for S68N. The beam size is shown at the right-hand side of each panel. SO traces the extended outflow, while the other two molecules show compact emissions. The approximate directions of the blueshifted and redshifted emission of the outflow are shown with blue and red arrows.} 
    \label{fig:moment0}
\end{figure*}

 A summary of the identified molecules along with the fitted parameters of the models are presented in Table \ref{tab:B1c_NT} for B1-c and S68N. The temperature was fitted for $50\%$ of the molecules in B1-c and $20\%$ of the species in S68N. The FWHMs found for the N-bearing species are similar to those found by \cite{vangelder2020} for O-bearing species. The fits to the spectra for $\rm NH_{2}CHO$ and $\rm HN^{13}CO$ towards B1-c are presented in Figs. \ref{fig:NH2CHO_fit_inside} and \ref{fig:HN13CO_fit_inside}, respectively, as examples. The fits to the rest of the data are shown in Appendix \ref{app:spec_fit}.

Figure \ref{fig:moment0} shows moment zero maps of the SO $6_{7} - 5_{6}$ ($E_{\rm up}=47.6\,\rm K$), HNCO $12_{0,12} - 11_{0,11}$ ($E_{\rm up}=82.3\,\rm K$), and $\rm NH_{2}CHO$ $13_{1,13} - 12_{1,12}$ ($E_{\rm up}=91.8\,\rm K$) lines for B1-c and S68N in Band 6. Both B1-c and S68N are known to have outflows (\citealt{Jorgensen2006}; \citealt{Tychoniec2019}). This is seen from the SO moment zero map, where the spatially extended emission traces the outflow (see e.g. \citealt{Podio2015}). Some more complex molecules, such as $\rm NH_{2}CHO$ and HNCO, can also be detected in the outflow in their low upper energy level lines in the Band 3 data (Tychoniec et al. in prep.), but the analysis in this paper focuses on the compact emission from the inner envelope, and mostly on the Band 5 and 6 data. Figure \ref{fig:moment0} shows that the compact emission from $\rm NH_{2}CHO$ and HNCO is not spatially resolved and that these molecules show emission that does not extend beyond the continuum. These must be located within ${\sim} 200\,\rm au$ of the central protostar. The width of the lines of N-bearing species (see Appendix \ref{app:spec_fit} for the spectra), ${\sim} 3-5 \, \rm km\, s^{-1}$, is in line with a rotating structure or turbulence in the inner envelope; therefore, the molecules studied in this work likely trace the inner envelope and potential warm disk around the protostar.

\subsection{Column densities}
\label{sec:col_exT}

A summary of the derived column densities for B1-c and S68N is presented in Table \ref{tab:B1c_NT}. The most abundant N-bearing molecule detected towards B1-c and S68N in the Band 5 and 6 data is HNCO. The emission of this molecule is optically thick in both sources. This is also seen from a comparison of the $\rm HN^{13}CO$ column density with the fitted HNCO column density, where the latter is under-estimated. The column densities of HNCO measured by directly fitting the spectra for HNCO are ${\sim} 1.8 \times 10^{15} \rm cm^{-2}$ and ${\sim} 10^{15} \rm cm^{-2}$ towards B1-c and S68N, respectively. Therefore, the interstellar ratio of $\rm ^{12}C/^{13}C \sim 68$ (\citealt{Milam2005}) was used to find HNCO column densities of $1.6\pm0.7 \times 10^{16} \rm cm^{-2}$ and ${\sim} 1.2 \times 10^{16} \rm cm^{-2}$ for B1-c and S68N from the $\rm HN^{13}CO$ column densities, assuming the $\rm HN^{13}CO$ lines are optically thin. This assumption can be investigated by searching for $\rm HNC^{18}O$ lines. The Band 5 data cover $\rm HNC^{18}O$ lines in B1-c. However, this molecule is not detected, and a ${\sim} 3\sigma$ upper limit column density of ${\sim} 4 \times 10^{14} \rm cm^{-2}$ at 200\, K is found towards B1-c. This upper limit corresponds to a lower limit for the $\rm HNCO/HNC^{18}O$ ratio of $\gtrsim 40$. This lower limit is smaller than the isotope ratio of $\rm ^{16}O/^{18}O \sim 560$ (\citealt{Wilson1994}), implying that $\rm HN^{13}CO$ emission towards B1-c could potentially be marginally optically thick. Using $\rm ^{16}O/^{18}O \sim 560$ and the upper limit value for HNC$^{18}$O, an upper limit for HNCO of $2.2 \times 10^{17}$\,cm$^{-2}$ can be found. Therefore, the value for the HNCO column density towards B1-c could potentially  differ from the current value by up to an order of magnitude and hence should be taken with care. In the Band 3 data, HNCO is tentatively detected towards both sources. However, as derived above from our Band~5 and~6 data, it is optically thick, and none of its isotopologues are detected in Band 3. Therefore, we refrained from deriving the HNCO column density from the main isotopologue lines.

Unfortunately, our Band 5 and 6 frequency range covers neither $\rm CH_{3}CN$ nor its $\rm ^{15}N$ or $\rm ^{13}C$ isotopologues. The frequency range covers three rotational $\rm CH_{3}CN$ lines originating from the $\rm v_{8}=1$ excited vibrational level, but they are not very constraining as they give very high upper limits on the column density at a temperature of 200\,K ($\lesssim 7 \times 10^{18}\,\rm cm^{-2}$ and $\lesssim 6 \times 10^{18}\,\rm cm^{-2}$ for B1-c and S68N, respectively). However, $\rm CH_{2}DCN$ is detected towards both sources. Given that our sources are hot corinos, in each source we used the ratio of $\rm CH_{2}DCN/CH_{3}CN = 0.035$ from the results of the PILS for IRAS 16293B (\citealt{calcutt2018}) to estimate the column density of $\rm CH_{3}CN$ from our value for $\rm CH_{2}DCN$. This is, however, only an estimate. Interestingly, $\rm CHD_{2}CN$ is also detected towards B1-c, whereas we find an upper limit for this molecule towards S68N. In Band 3, $\rm CH_{3}CN$ is clearly detected towards both sources (see Appendix \ref{app:spec_fit}), but, given that the lines are optically thick and none of its isotopologues are detected, it is only possible to report lower limits to the actual column densities. These are ${\sim} 10^{15}\,\rm cm^{-2}$ and ${\sim} 3 \times 10^{14}\,\rm cm^{-2}$ at a fixed temperature of 100\,K (as \citealt{vangelder2020} found that Band 3 traces colder temperatures) for B1-c and S68N, respectively. These lower limits are consistent with the column densities derived for $\rm CH_{3}CN$ using the Band 6 CH$_2$DCN data towards both sources.

HNCO has the highest column density in both sources. $\rm CH_{2}DCN$ and $\rm CHD_{2}CN$ have the lowest column densities for S68N and B1-c, respectively. The column densities for ${\sim} 85\%$ of the detected and tentatively detected N-bearing species studied here are of the order of ${\sim} 10^{14-15}\rm \, cm^{-2}$ in a $0.45\arcsec$ beam for both sources. These values are on average an order of magnitude lower than those of the O-bearing molecules studied by \cite{vangelder2020}, which is an interesting observation. Similar to what \cite{vangelder2020} found, weaker ($T_{\rm b} \lesssim 1\,\rm K $) and fewer lines are found towards the more distant S68N source compared to B1-c. On average, the column densities of the species studied in this work are ${\sim} 1.5-3$ times lower in S68N than in B1-c. 

\subsection{Excitation temperatures}

A summary of the derived excitation temperatures is presented in Table \ref{tab:B1c_NT} for B1-c and S68N. The excitation temperatures found for the N-bearing species span a range between ${\sim} 100\,\rm K$ and ${\sim} 300\,\rm K$. There is no significant difference, within the uncertainties, between the excitation temperatures for the N-bearing COMs and the O-bearing COMs found by \cite{vangelder2020} (see Sect. \ref{sec:T_ex} for a discussion).

For most molecules, a reasonable range of upper state energy level lines is covered in our data, implying that the excitation temperatures fitted here are not biased. The only exception is $\rm NH_{2}CHO,$ where the lines included in the fit only cover upper energy levels below ${\sim} \rm 200\,K$, and thus the excitation temperature derived here is biased towards lower temperatures. Moreover, to derive the excitation temperature of HN$^{13}$CO towards B1-c, its Band 3 data are used (in addition to the Band 6 data) as an additional constraint, eliminating very low excitation temperatures.

\subsection{Column density ratios}

Table \ref{tab:B1c_NT} presents the column density ratios of the studied species with respect to $\rm CH_{3}OH$ and HNCO. The methanol column density is taken from \cite{vangelder2020}. The column density ratios of the species are not calculated with respect to $\rm H_{2}$ because in low-mass protostars it is difficult to derive accurate values for the warm $\rm H_{2}$ gas column density from the dust continuum or the CO column density. This is due to the fact that the dust continuum becomes optically thick at $\lesssim 100\rm\, au$ scales and may have contributions from a forming, colder disk (\citealt{Yildiz2013}; \citealt{Persson2016}; \citealt{DeSimone2020}). Moreover, in low-mass sources not all CO emission comes from the region with warm gas ($>$100\,K) (\citealt{Yildiz2013}). Therefore, the abundances of the species considered in this work are presented with respect to $\rm CH_{3}OH$ and HNCO instead of $\rm H_{2}$. Column density ratios with respect to methanol are given in order to compare our results with those of \cite{vangelder2020} and other studies. The column density ratios with respect to HNCO are calculated to show whether the formation of species discussed here is linked to HNCO or routes resulting in HNCO formation (see Sect. \ref{sec:compare}). 

The column density ratios for all the detected and tentatively detected molecules with respect to methanol are very similar (within a factor of two) between the two sources. This agrees with what \cite{vangelder2020} found for most O-bearing COMs. Moreover, the column density ratios for $\rm NH_{2}CHO$, $\rm C_{2}H_{5}CN$, HNCO, and $\rm NH_{2}CN$ with respect to methanol agree, within the uncertainties, with the results of \cite{Belloche2020} for S68N observed with the PdBI. It should be noted, though, that our estimated value for $\rm CH_{3}CN/CH_{3}OH$ is ${\sim} 13.5$ times smaller than that derived by \cite{Belloche2020}. This discrepancy could originate from the CH$_{3}$OH lines used in \cite{Belloche2020} being (marginally) optically thick, while the CH$_{3}$OH column density reported by \cite{vangelder2020} is derived from its optically thin $^{18}$O isotopologue. It should also be noted that we find the $\rm CH_{3}CN$ column density from its deuterated isotopologues, which is another source of uncertainty. Finally, the assumed $^{16}$O/$^{18}$O ratio used to derive the CH$_3$OH column density may differ slightly from the value assumed in \cite{vangelder2020}.  

\subsection{Source size and optical depth}
\label{sec:sources_size}

The emission from our data is not spatially resolved.\ Therefore, it is not obvious whether the emission is optically thin because the estimated optical depth depends on the assumed size of the emitting region. In this paper it is assumed that the source size is the same as the ALMA Band 6 beam size of $0.45 \arcsec$. A smaller source size will result in larger column densities, and hence the emission can become optically thick. This can be seen by taking the dilution factor into account. The equation $N_{1}\theta_{\rm s_{1}}^{2}/(\theta_{\rm b}^{2} + \theta_{\rm s_{1}}^{2}) = N_{2}\theta_{\rm s_{2}}^{2}/(\theta_{\rm b}^{2} + \theta_{\rm s_{2}}^{2})$ shows how the measured column density can be scaled for different source sizes (where $\theta_{\rm b}$ is the beam size, $\theta_{\rm s}$ is the source size, and the numbers refer to the two assumed source sizes). Although the column densities would become larger for smaller source sizes that depend on the emitting region of a molecule, column density ratios will stay the same as long as the emission remains optically thin and comes from the same region.

N-bearing species identified in this work show a large range of excitation temperatures (Table \ref{tab:B1c_NT}). It is not unlikely that source sizes are different for molecules with different excitation temperatures. Here, two representative values for the excitation temperatures are considered: one for molecules that are excited at temperatures of ${\sim} 100\,\rm K$ and one for molecules that are excited at temperatures of ${\sim} 200\,\rm K$. Therefore, one can assume two source sizes equal to two radii where temperatures are 100\,K and 200\,K. 

Using the same method as \cite{vangelder2020}, the radius at which the
temperature reaches $\rm 100 \, K$ for a spherically symmetric hot
core region can be estimated from
$R_{T=100\rm K}\simeq 15.4 \sqrt{L/L_{\odot}}\, \rm au$
(\citealt{Bisschop2007}), where $L$ is the source luminosity. The
luminosities of B1-c and S68N are $\rm 6.0\, L_{\odot}$
(\citealt{Karska2018}) and $\rm 5.4\, L_{\odot}$
(\citealt{Enoch2011}), respectively. Therefore, the radii at which the
temperature reaches $\rm 100\, K$ for B1-c and S68N are about 37.7\,au
($0.23\arcsec$) and 35.8\,au ($0.16\arcsec$), respectively. When these radii are used for the line analysis, $\rm NH_{2}CHO$ remains optically thin in the Band 5 and 6 data towards B1-c even for this smaller source size. Moreover, $\rm C_{2}H_{5}CN$ stays optically thin for S68N at this smaller radius.

To calculate the radius at which the temperature is 200\,K, we used a toy model for a spherically symmetric infalling envelope with a power law in temperature and density structure (see Appendix \ref{sec:toy_model}). For B1-c and S68N, $R_{T=200\rm K}$ is about 6.7\,au (0.042\arcsec) and 6.3\,au (0.029\arcsec), respectively. When these radii are used for the line analysis, all the molecules with high $T_{\rm ex}$ (HN$^{13}$CO, $\rm C_{2}H_{5}CN$, and CHD$_{2}$CN towards B1-c and CH$_{2}$DCN towards both sources) become optically thick in Band 5 and 6 data. Moreover, $\rm CH_{3}CN$ becomes optically thick in Band 3 for both sources, assuming its excitation temperature is similar to that of its deuterated versions (${\sim} 200\,\rm K$). It is therefore not possible to derive accurate column densities or column density ratios for such small source sizes.

It is worth noting that some deviation from spherical symmetry due to the presence of a disk can occur at $< 100\rm \,au$ scales, and thus our simple toy model runs into limitations. \cite{Persson2016} find that the amount of gas at temperatures above 100\,K in low-luminosity sources can vary by more than an order of magnitude depending on the disk size and structure, affecting optical depth. Moreover, the extremely small source sizes predicted by the simple toy model are likely unrealistic. Adopting $R_{T=100\rm K}$ as the source size for both sources, all molecules in Band 5 and 6 remain optically thin and CH$_3$CN remains marginally optically thick ($\tau > 0.1$) in Band 3 for both sources as expected (see Sect. \ref{sec:col_exT}). For this reason, assuming a source size of $0.45 \arcsec$ in the rest of this paper does not change our results.

\section{Discussion}
\label{sec:Discussion}

\subsection{Excitation versus desorption temperatures}
\label{sec:T_ex}

\begin{figure*}
    \centering
    \includegraphics[width=17cm]{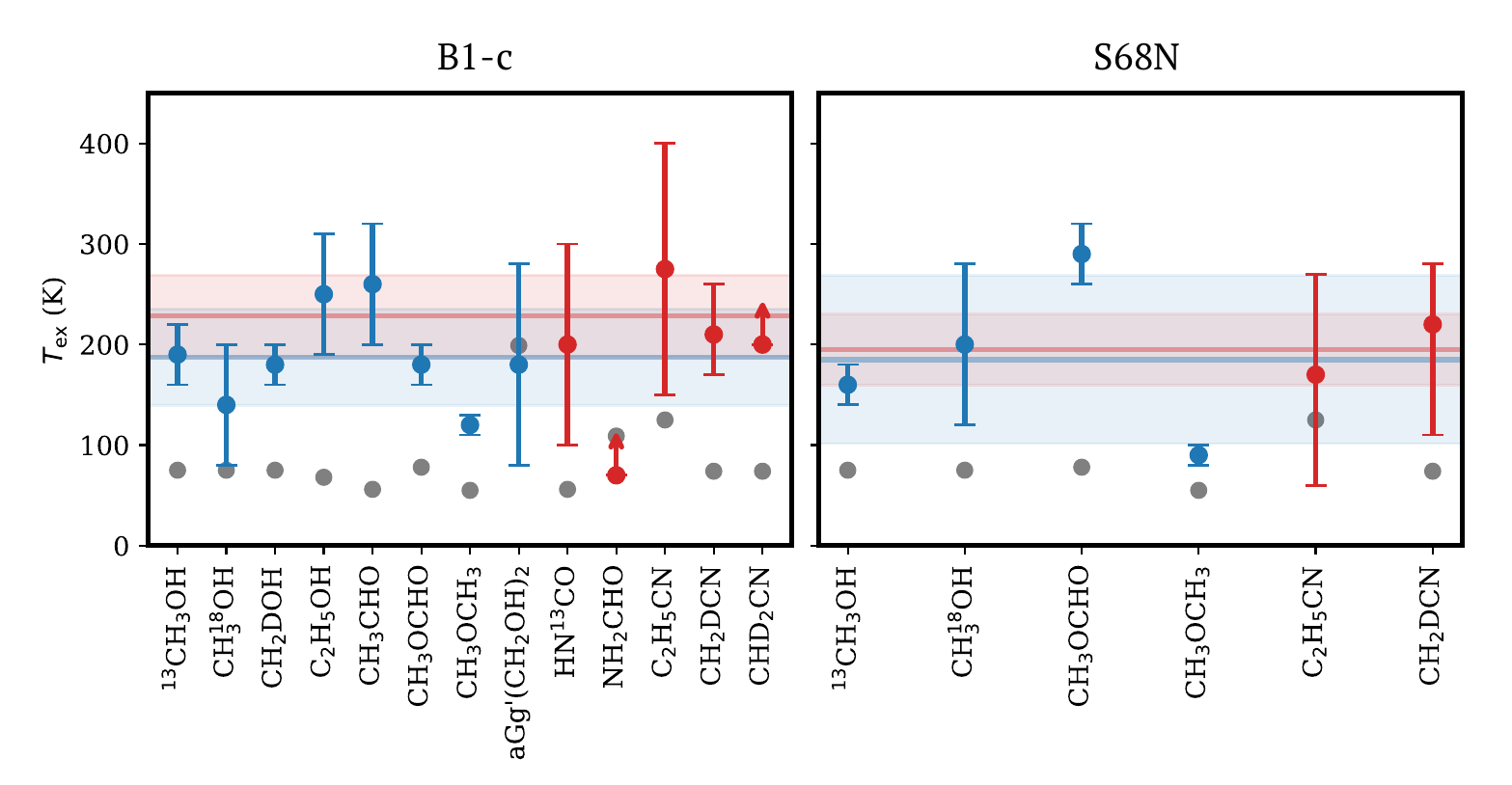}
    \caption{Excitation temperatures for N-bearing species discussed in this work (red) and O-bearing species discussed in \cite{vangelder2020} (blue) for B1-c and S68N. Only the species with derived $T_{\rm ex}$ are plotted. The solid red and blue lines show the average values for the red and blue data points, respectively. The shaded red and blue areas show the standard deviation of the data points. The grey points show the sublimation temperatures of the corresponding molecules found by the gas-grain balance model (\citealt{Hasegawa1992}) and the respective binding energies of each molecule (\citealt{Penteado2017}; \citealt{Garrod2013}).}
    \label{fig:Tex_both}
\end{figure*}

Figure \ref{fig:Tex_both} presents a comparison between excitation temperatures of the O-bearing species studied by \cite{vangelder2020} and the N-bearing species studied in this work towards B1-c and S68N. This figure shows that there is no significant difference in $T_{\rm ex}$ between the O-bearing and N-bearing species; the mean and the scatter of $T_{\rm ex}$ are very similar. Assuming that the excitation temperature is close to the kinetic temperature of the region in which a molecule resides, Fig. \ref{fig:Tex_both} suggests that N-bearing species trace a range of temperatures from ${\sim} 100\,\rm K$ to ${\sim} 300\,\rm K$. We note that at the densities considered here some molecules may be sub-thermally excited, and hence the $T_{\rm ex}$ found in this work may be a lower limit to the kinetic temperature (\citealt{Jorgensen2016}). \cite{Bisschop2007} also found that there is no difference between the excitation temperatures of N-bearing and O-bearing species in seven high-mass protostars, which is consistent with what is found here.

Each molecule can desorb at a significantly different temperature
according to its binding energy, ice, and grain environment (\citealt{Cuppen2017}), so molecules come off the ice roughly at their respective snow lines depending on their sublimation temperatures. Therefore, one would expect differences in the excitation temperatures of molecules. The idea of an onion-like temperature structure around low-mass protostars was discussed in \cite{Jorgensen2018}, in which O-bearing molecules are divided into two categories (see also \citealt{Manigand2020}): one that includes molecules associated with temperatures of 100-150\,K and another that includes molecules associated with temperatures of 250-300\,K. Figure \ref{fig:Tex_both} shows that $\rm C_{2}H_{5}CN$ towards B1-c falls under the hottest category ($>250\rm \,K$). In addition, three of the N-bearing molecules studied here (HN$^{13}$CO, CH$_2$DCN, and CHD$_2$CN) seem to trace the gas with temperatures around ${\sim} 200\,\rm K$ in B1-c. 

The potential relation between the excitation and sublimation temperatures of molecules can be further explored by comparing our results with the sublimation temperatures found from the gas-grain balance model (\citealt{Hasegawa1992}) and the binding energies of each molecule in the solid state. In this model, the number density of the solid to the gas phase of species $i$ is given by

\begin{equation}
    \frac{n_{\rm ice}}{n_{\rm gas}} = \frac{\pi a_{\rm d}^{2} n_{\rm d} S \sqrt{3k_{\rm B}T_{\rm gas}/m_{i}}}{e^{-E_{\rm b}/T_{\rm d}} \sqrt{2 k_{\rm B} n_{\rm ss} E_{\rm b}/(\pi^{2}m_{i})}},
    \label{eq:gas-grain}
\end{equation}

\noindent where $a_{\rm d}$ is the dust grain size (assumed to be $0.1\,\mu \rm m$), $n_{\rm d}=10^{-12} \times n_{\rm H}$ is the dust number density (with $n_{\rm H}$ the hydrogen number density, assumed to be $10^{7}\,\rm cm^{-3}$), $S$ is the sticking coefficient (assumed to be 1), $n_{\rm ss}$ is the number of binding sites per surface area (taken as $8\times 10^{14}\,\rm cm^{-2}$), $E_{\rm b}$ is the binding energy of species $i$ in units of kelvin, $m_{i}$ is the mass of species $i$, and $T_{\rm gas}$ and $T_{\rm d}$ are the gas and dust temperatures, respectively. Assuming that the environment is sufficiently dense, the gas would be thermally coupled with the dust and $T_{\rm d}$ would be equal to $T_{\rm gas}$. The sublimation temperature of species $i$ is the temperature for which the ice and the gas are in balance for that molecule. In other words, $n_{\rm ice}/n_{\rm gas} = 1$. 

Therefore, using Eq. \eqref{eq:gas-grain} and the binding energies for most molecules from \cite{Penteado2017} (the binding energy for $\rm (CH_{2}OH)_{2}$ is taken from \citealt{Garrod2013}), the desorption temperatures of O-bearing and N-bearing species can be calculated. The grey points in Fig. \ref{fig:Tex_both} represent these values. It can be safely assumed that the desorption temperatures for the isotopologues will barely differ from that of the  main isotopologue. However, binding energies differ for different molecule interactions in a mixed ice or on different grain surfaces (e.g. \citealt{Tielens1991}; \citealt{Collings2004}; \citealt{Ferrero2020}), and hence in principle desorption temperatures may vary (typically by several tens of kelvin). The values adopted here are mostly for pure ices; if these COMs are mixed in ice layers consisting predominantly of $\rm H_{2}O$ or $\rm CH_{3}OH$, values close to their desorption temperatures (${\sim} 100\,\rm K$) are expected.

Figure \ref{fig:Tex_both} shows that there is no significant difference between the sublimation temperatures of O-bearing and N-bearing species. This is consistent with what is seen for the excitation temperatures, but with an offset between sublimation temperatures and excitation temperatures for most molecules (except for $\rm (CH_{2}OH)_{2}$). Specifically, the excitation temperatures are typically higher than the sublimation temperatures for all molecules by a factor of ${\sim} 2-4$, except for $\rm (CH_{2}OH)_{2}$ where these two values are very similar. This can be interpreted as the excitation temperatures reflecting an average temperature of the region around the protostar, where it is hotter than the sublimation temperature of each molecule. Using the toy model for a spherically symmetric infalling envelope with power law structure in temperature and density (see Appendix \ref{sec:toy_model_avgT}), the mass-weighted average temperature can be calculated as $1.36T_{\rm sub}$. Therefore, if the excitation temperature is measuring an average temperature, it is expected to be larger than the sublimation temperature for each molecule. We note that the factor 1.36 is not large enough to explain the difference seen in Fig. \ref{fig:Tex_both}. However, this is only using a simple toy model: Developing a more complete model that takes small-scale structures  such as disks into account can improve our understanding of the difference between sublimation temperatures and excitation temperatures.

The sublimation temperatures for all species in Fig. \ref{fig:Tex_both} (except for (CH$_{2}$OH)$_{2}$) are between ${\sim} 55$ and ${\sim} 125\, \rm K$. While the sublimation temperatures of O-bearing and N-bearing groups of species do not differ significantly, these values for some molecules (i.e. CH$_{3}$OCH$_{3}$, CH$_{3}$CHO, and HNCO) are lower than the rest, and others (i.e. C$_{2}$H$_{5}$CN and (CH$_{2}$OH)$_{2}$)  are higher. Such trends are reflected in the excitation temperatures for some of the molecules with small error bars. For example, CH$_{3}$OCH$_{3}$ shows very low excitation temperatures in both sources, which agrees with its low sublimation temperature. This may be good evidence for the formation of this molecule in the ice and its desorption at its snow line. On the other hand, a molecule such as CH$_{3}$CHO shows a larger excitation temperature compared to the other species, in contradiction with its low sublimation temperature. This could be evidence for other mechanisms (e.g. gas-phase formation pathways or source structures) potentially playing a role in the formation (\citealt{Garrod2008}; \citealt{Tideswell2010}; \citealt{Vazart2020}) and emission of these species. However, given the large error bars on most of the excitation temperatures shown in Fig. \ref{fig:Tex_both}, all these arguments should be taken with care.

\begin{figure}
  \resizebox{\hsize}{!}{\includegraphics{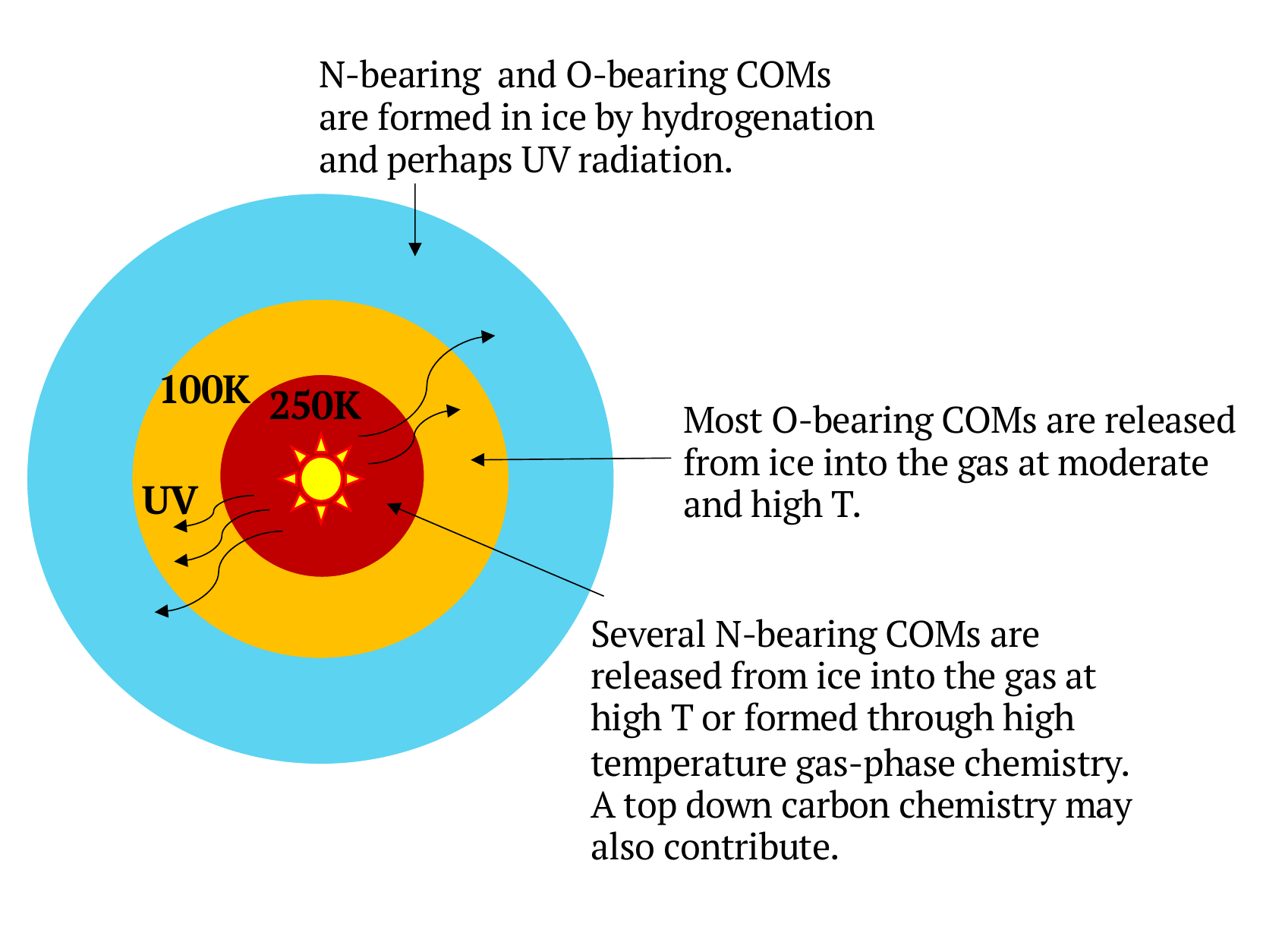}}
  \caption{Cartoon of the temperature structure of protostellar envelopes showing where O-bearing and N-bearing species are most likely to be found in the gas. This is based on the findings of \cite{Crockett2015} for massive protostars, who showed that higher excitation temperatures for N-bearing COMs are found towards several high-mass protostars, and \cite{vantHoff2020b}, who proposed a top-down carbon chemistry for nitriles. It is not yet clear whether this structure also applies to low-mass protostars.}
  \label{fig:cartoon_T}
\end{figure}

Using \textit{Herschel}-HIFI data, \cite{Crockett2015} find that N-bearing species, C$_{2}$H$_{5}$CN, NH$_{2}$CHO, and CH$_{3}$CN, trace hotter gas (${\sim} 250\,\rm K$) compared with the oxygen-bearing species, CH$_{3}$OH, CH$_{3}$OCH$_{3}$, C$_{2}$H$_{5}$OH, and CH$_{3}$OCHO, (${\sim} \rm 100\, K$), in the Orion Kleinmann-Low nebula (see Fig. \ref{fig:cartoon_T}). They argue that this could indicate either that N-bearing COMs require higher dust temperatures to desorb from the ice or that more N-bearing molecules are formed through high temperature gas-phase chemistry, reflecting the two possible COM formation scenarios: colder solid-state chemistry and warmer gas-phase chemistry.

Alternatively, at such high temperatures \cite{Crockett2015} could be seeing the carbon grain sublimation into the gas phase that leads to the formation of CH$_{3}$CN, an idea put forward by \cite{vantHoff2020b}. They propose the formation of N-rich COMs inside a `soot line' at 300 K through carbon grain sublimation and suggest that an excess of hydrocarbons and nitriles with excitation temperatures higher than those of O-bearing species can be a signature of this phenomenon. This process implies that such molecules could be formed through top-down chemistry (from the destruction of larger species) rather than bottom-up chemistry in the solid state or gas phase. \cite{vantHoff2020b} assume that no molecules with oxygen atoms are formed through carbon grain sublimation inside of the soot line at ${\sim} 300\,\rm K$ as all the oxygen is locked up in H$_2$O, CO, and CO$_2$. Thus, the only two molecules in this work that can be signatures for carbon grain sublimation are $\rm C_{2}H_{5}CN$ and CH$_{3}$CN. The toy model for a spherically symmetric infalling envelope explained in Appendix \ref{sec:toy_model_avgT} predicts the ${\sim}300\,\rm K$ radius to be significantly smaller than the spatial resolution of our data. Therefore, the fact that carbon grain sublimation is not seen for $\rm C_{2}H_{5}CN$ and CH$_{3}$CN in B1-c or S68N might be due to this effect being concealed at the present spatial resolution.

\subsection{Comparison of abundances in different sources}

In this section the column density ratios of B1-c and S68N are compared with other sources. In most studies, column density ratios are interpreted as abundances. Therefore, this section also assumes no difference between column density ratios and abundances. The validity of this assumption is further discussed in Sect. \ref{sec:abund_ratio}.   

\subsubsection{Dependence on luminosity}
\label{subsec:desorbT}

Higher luminosity implies a higher temperature, where chemical reactions can take place more efficiently thereby affecting the production rate of molecules and possibly changing their abundances. In addition, a source with a higher luminosity has more UV radiation, and hence the UV processing on the grains can alter the gas-phase abundances of species, for example through (non-)dissociative photo-desorption but also following UV-induced photochemistry in the ice (\citealt{Oberg2009b}; \citealt{Bertin2013}). Therefore, a relation between the luminosity and the abundance of a species with respect to methanol may be present. This is especially important for species with higher desorption temperatures (e.g. $\rm NH_{2}CHO$; see Sect. \ref{sec:T_ex}) that may stay mostly in the ice in sources with lower luminosities. 

\begin{figure}
  \resizebox{\hsize}{!}{\includegraphics{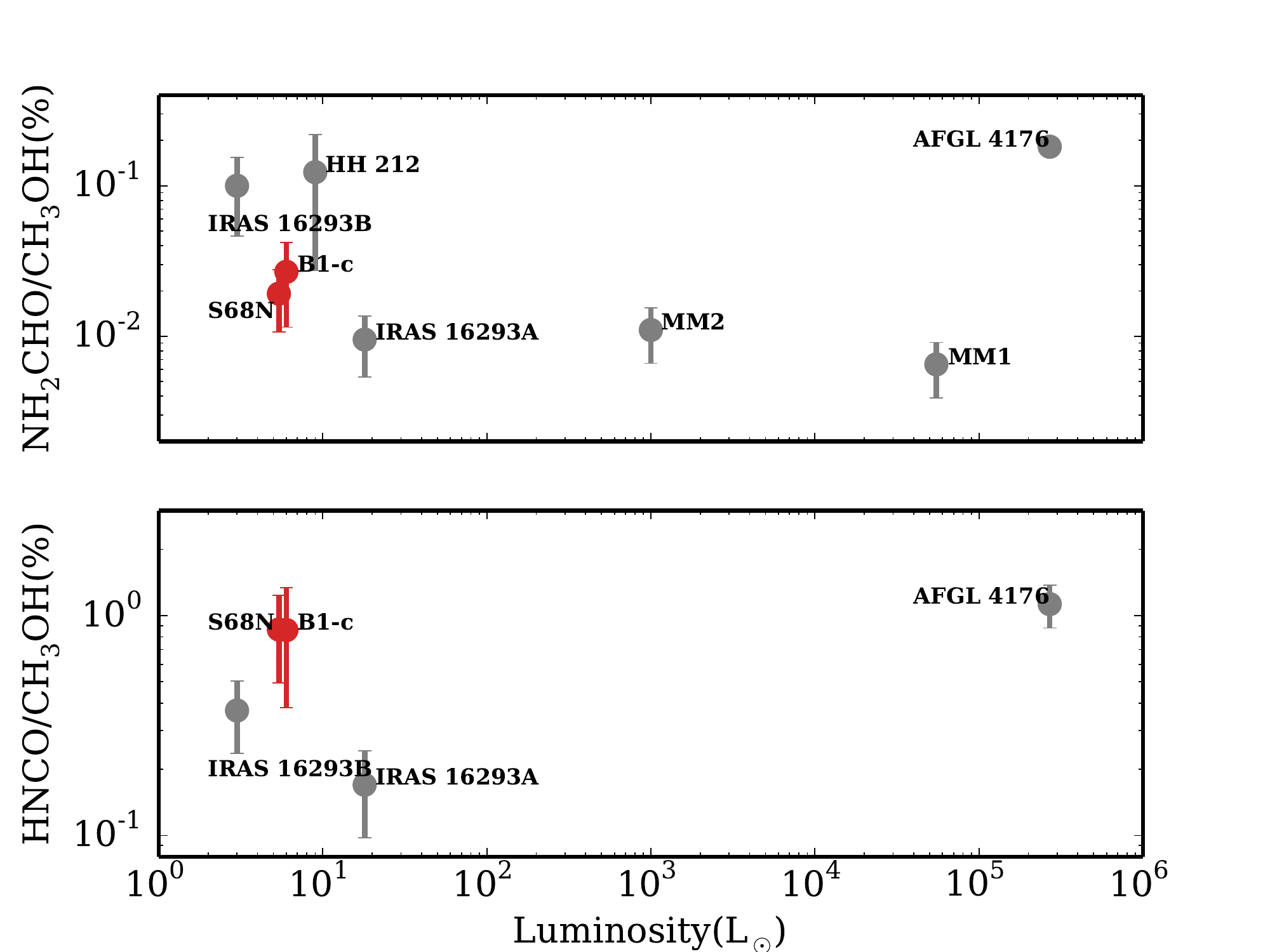}}
  \caption{Abundance ratio of $\rm NH_{2}CHO$ and HNCO with respect to $\rm CH_{3}OH$ against source luminosity. Red points show the results for the sources considered in this work. The grey points indicate other low- and high-mass sources. These sources are IRAS 16293A (\citealt{Ligterink2017}; \citealt{Manigand2020}), IRAS 16293B (\citealt{Coutens2016}; \citealt{Ligterink2017}; \citealt{Jorgensen2018}), HH 212 (\citealt{Lee2019}), AFGL 4176 (\citealt{Bogelund2019}), and NGC6334 MM1 and MM2 (\citealt{Bogelund2019ngc}).}
  \label{fig:lum}
\end{figure}

Figure \ref{fig:lum} shows the abundances of $\rm NH_{2}CHO$ and HNCO with respect to methanol for our sources and other low- and high-mass protostars against the source luminosity (see the caption of Fig. \ref{fig:lum} for the full list of sources and reference papers). There is no significant relation between abundances and source luminosity. The same conclusion holds true for the rest of the N-bearing species, which is consistent with the results of \cite{vangelder2020} for the O-bearing COMs. Moreover, our results are consistent with what \cite{Belloche2020} find for protostellar sources observed by the PdBI. Nevertheless, Fig. \ref{fig:lum} shows that $\rm NH_{2}CHO/CH_{3}OH$ ranges from ${\sim} 0.01\%$ to ${\sim} 0.1\%$ (discussed further in Sect. \ref{sec:compare}), whereas \cite{vangelder2020} find that for most O-bearing molecules the column density ratios with respect to methanol are within a factor of a few for various sources. Given that in all these sources the methanol column density is derived from one of its optically thin isotopologues, the difference seen here cannot be due to issues with optical depth.

\subsubsection{Abundances with respect to $\rm CH_{3}OH$ and HNCO}
\label{sec:compare}

We compared our abundances to values obtained for other low-mass and high-mass sources that have been studied by ALMA. The low-mass sources are IRAS 16293A (\citealt{Coutens2016}; \citealt{Ligterink2017}; \citealt{calcutt2018}; \citealt{Manigand2020}), IRAS 16293B (\citealt{Coutens2016}; \citealt{Ligterink2017}; \citealt{Jorgensen2018}; \citealt{calcutt2018}; \citealt{Coutens2018}), HH 212 (\citealt{Lee2019}), and NGC 1333 IRAS 4A2 (\citealt{Lopez-Sepulcre2017}). The high-mass sources are AFGL 4176 (\citealt{Bogelund2019}) and NGC 6334 MM1, MM2, and MM3 (\citealt{Bogelund2019ngc}). 

\begin{figure*}
    \centering
    \includegraphics[width=17cm]{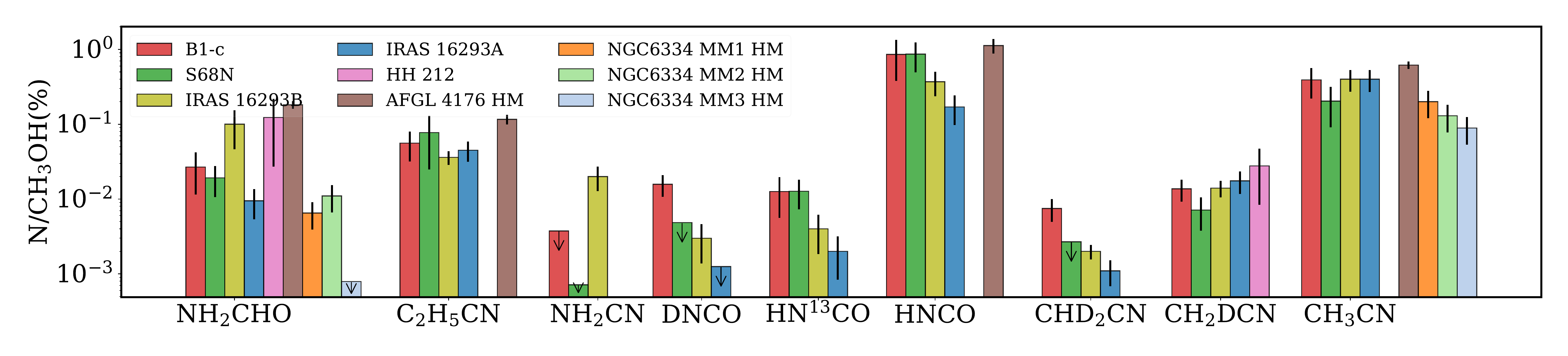}
    \caption{Column densities of the N-bearing molecules with respect to methanol for sources observed by ALMA. Column densities of methanol for B1-c and S68N are taken from \cite{vangelder2020}. Shown are IRAS 16293A (\citealt{Coutens2016}; \citealt{Ligterink2017}; \citealt{calcutt2018}; \citealt{Manigand2020}), IRAS 16293B (\citealt{Coutens2016}; \citealt{Ligterink2017}; \citealt{Jorgensen2018}; \citealt{calcutt2018}; \citealt{Coutens2018}), HH 212 (\citealt{Lee2019}), AFGL 4176 (\citealt{Bogelund2019}), and NGC 6334 MM1, MM2, and MM3 (\citealt{Bogelund2019ngc}).}
    \label{fig:CH3OH_compare}
\end{figure*}

Figure \ref{fig:CH3OH_compare} shows this comparison for the ratio of the column densities to methanol. At a glance, we can see that these values are generally lower than those of O-bearing species (Fig. 7 in \citealt{vangelder2020}), typically by an order of magnitude. Moreover, there is a larger variation (up to an order of magnitude) between the different sources for N-bearing molecules than for O-bearing ones (by a factor of a few). 

When comparing the column density ratios of the N-bearing species shown in Fig. \ref{fig:CH3OH_compare} among different sources, one can see that some molecules show similar abundance ratios. Most notably, $\rm C_{2}H_{5}CN$ is comparable in the low-mass sources and with the high-mass AFGL 4176 (within a factor of ${\sim} 5$).

\begin{figure*}
    \centering
    \includegraphics[width=17cm]{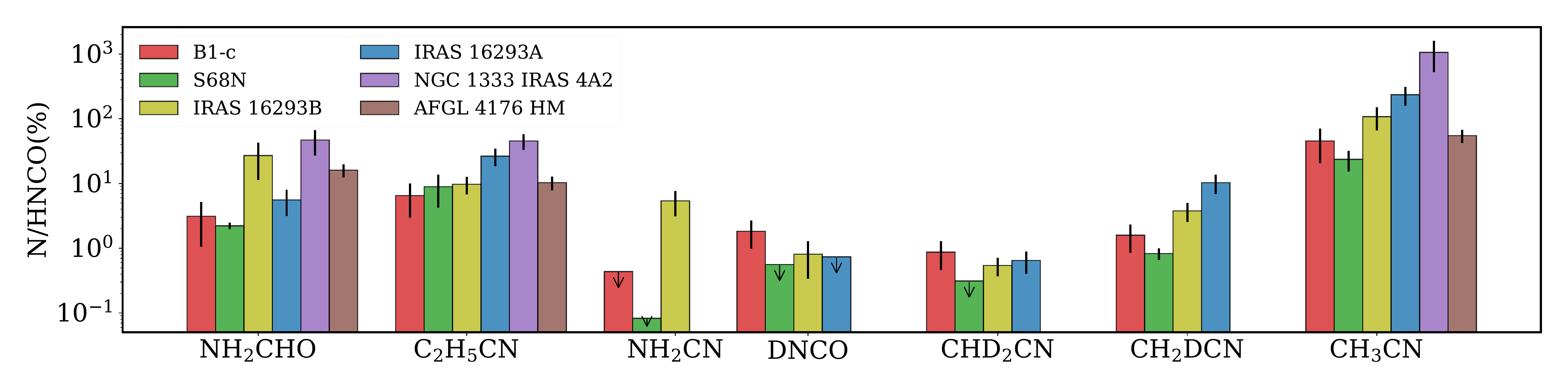}
    \caption{Column densities of N-bearing species with respect to HNCO as derived from $\rm HN^{13}CO$. Shown are IRAS 16293A (\citealt{Coutens2016}; \citealt{Ligterink2017}; \citealt{calcutt2018}; \citealt{Manigand2020}), IRAS 16293B (\citealt{Coutens2016}; \citealt{Ligterink2017}; \citealt{calcutt2018}; \citealt{Coutens2018}), NGC 1333 IRAS 4A2 (\citealt{Lopez-Sepulcre2017}), and AFGL 4176 (\citealt{Bogelund2019}).}
    \label{fig:HNCO_compare}
\end{figure*}

A larger scatter (around an order of magnitude) is seen for $\rm NH_{2}CHO$ between all sources (see also Fig. \ref{fig:lum}). Three categories seem to be present. First, two of the high-mass sources (NGC6334 MM1 and MM2) seem to agree with three of the low-mass sources (B1-c, S68N, and IRAS 16293A). Second, the high-mass source, AFGL 4176, is similar to the two other low-mass sources, IRAS 16293B and HH212. Third, the high-mass source NGC 6334 MM3 does not show a detection of this COM. \cite{Manigand2020} explain the low abundance ratio of formamide in IRAS 16293A compared with IRAS 16293B by the fact that the offset position used to extract the spectrum for the former seems to be missing the hottest gas and thus most of the emission from this molecule, while this is not the case for the latter. Moreover, in contrast with our results, they find that formamide belongs to a group of species with a high excitation temperature. Another possible explanation for the variation in formamide abundance ratios with respect to methanol across various sources is the fact that there are multiple ways to form formamide in the solid state (\citealt{Rimola2018}; \citealt{Haupa2019}; \citealt{Dulieu2019}; \citealt{Martin2020}) and gas phase (\citealt{Barone2015}; \citealt{Codella2017}). Therefore, the amount of this COM formed in the gas or ice could depend on local conditions, such as the presence of a disk, its temperature structure (\citealt{Jorgensen2002}; \citealt{schoier2002}; \citealt{Crimier2010}), or its exposure to UV (\citealt{Lopez2019}).

HNCO and its isotopologues show some scatter in their distributions with respect to methanol. B1-c ($0.9 \pm 0.5\%$), S68N ($0.9 \pm 0.4\%$), and the high-mass source AFGL 4176 ($1.1 \pm 0.3\%$) seem to have a more similar distribution of HNCO with respect to methanol, while the IRAS 16293 sources show a lower abundance ratio ($0.17 \pm 0.07\%$ and $0.37 \pm 0.13\%$ for IRAS 16293A and B, respectively). 

Figure \ref{fig:HNCO_compare} shows the abundance ratios of the molecules studied in this work with respect to HNCO. The scatter that was seen in Fig. \ref{fig:CH3OH_compare} seems to be less pronounced for most molecules. Among them, DNCO has a more uniform distribution between sources, as expected for a molecule formed directly via the deuteration of HNCO (\citealt{Noble2015}). Another molecule that shows a much more uniform distribution among sources is $\rm CHD_{2}CN$. \cite{Ligterink2020} studied amide molecules towards 12 positions in the high-mass star-forming region NGC 6334I. Their values of $\rm NH_{2}CHO/HNCO$ measured from the $\rm ^{13}C$ isotopologues of these two species span a range between ${\sim} 10\%$ and ${\sim} 100\%$. This range is ${\sim} 1-2$ orders of magnitude higher than what is found in this work towards B1-c and S68N, and more similar to IRAS 16293B (Fig. \ref{fig:HNCO_compare}).

\subsubsection{Deuteration fraction}
\label{sec:D_H}
B1-c and S68N are known to have a high deuteration fraction, as confirmed for methanol by \cite{vangelder2020} from CH$_{2}$DOH. \cite{Jorgensen2018} found that less complex molecules have similar $\rm D/H $ ratios. For example, the $\rm D/H$ ratios for HNCO (${\sim} 1\%$; \citealt{Coutens2016}) and $\rm CH_{3}OH$ ($2\%$; \citealt{Jorgensen2018}) in IRAS 16293B are comparable. Therefore, we similarly compared $\rm D/H$  for these two molecules in B1-c and S68N. The deuteration fraction found from this work for B1-c and S68N from the DNCO/HNCO ratio are $1.8\pm 0.8\%$ and $<0.6\%$. The value for B1-c is in agreement with the $2.7\pm 0.9\%$ methanol value reported in \cite{vangelder2020}, whereas the upper limit found for S68N is somewhat lower than the 1.4$\pm$0.6\% methanol value reported in the same study.

The $\rm D/H $ ratio can provide a clue as to the temperature history of the environment in which B1-c is forming. This can also be seen in Fig. \ref{fig:HNCO_compare}, where DNCO, $\rm CHD_{2}CN$, and $\rm CH_{2}DCN$ have abundance ratios with respect to HNCO that agree, within the uncertainties, between the low-mass sources. These sources are from different star-forming regions and yet show similar abundance ratios, pointing to the fact that there is a universal mechanism for increasing the $\rm D/H$ ratio prior to star formation in cold environments.

For $\rm CH_{2}DCN/CHD_{2}CN$, a column density ratio of ${\sim} 2$ is found towards B1-c, which is lower than that found for IRAS 16293A (${\sim} 16$) and IRAS 16293B (${\sim} 7$) by \cite{calcutt2018}. On the other hand, \cite{Taquet2019} find $\rm CH_{2}DOH/CHD_{2}OH \sim 1$ and ${\sim} 2$ for NGC1333-IRAS2A and NGC1333-IRAS4A, values that are more similar to our ratio for the deuterated versions of methyl cyanide. In general, these similarities and differences are interesting, but, given the small sample of hot corinos with a similar analysis at hand, we did not investigate this further.

\subsubsection{Comparison with ices}
\label{sec:ice}

B1-c is an ice-rich source, but so far CH$_3$OH is the only COM that has been identified along the line of sight towards the neighbouring B1-b source with {\it Spitzer} (\citealt{Boogert2008}). Ices have also been detected towards S68N \citep[e.g.][]{Anderson2013}, but they have not yet been analysed in detail. However, $\rm OCN^{-}$, a direct derivative of HNCO (\citealt{Broekhuizen2004}; \citealt{Fedoseev2016}), is observed towards many low- and high-mass protostars (\citealt{vanBroekhuizen2005}; \citealt{Oberg2011}). \cite{Oberg2011} find that, for the low-mass sources in their study, the median values for $\rm OCN^{-}$ and $\rm CH_{3}OH$ ice abundances with respect to water ice are $0.4^{0.4}_{0.3}\%$ and $7^{12}_{5}$\%, respectively. This gives an abundance ratio of $\rm OCN^{-}$ with respect to $\rm CH_{3}OH$ of ${\sim} 6_{2.5}^{8}\%$. This abundance ratio shows a considerable spread from source to source. Our gas-phase abundance ratios of HNCO to CH$_3$OH are ${\sim} 0.9\pm0.5$\% and ${\sim} 0.9\pm0.4$\% for B1-c and S68N, respectively (Table \ref{tab:B1c_NT}). These values are ${\sim} 6$ times lower than the median ice ratio but close to its lower limit. This comparison assumes that all OCN$^-$ comes off the grains as HNCO upon ice sublimation, but some of it may be converted into other species, such as more refractory salts (\citealt{Schutte2003}; \citealt{Boogert2008}). This is a reasonable assumption given that N-bearing salts are found to be abundant in comet 67P (\citealt{Altwegg2020}).

In addition, there are also upper limit estimates for HNCO ice itself, but mostly for high-mass protostars. Taking the upper limit of 0.7\% for HNCO/H$_2$O ice from \cite{Broekhuizen2004,vanBroekhuizen2005} and dividing it by the median of the CH$_{3}$OH/H$_{2}$O ice ratio in high-mass protostars ($8^{16}_{8}\%$) from \cite{Oberg2011} gives an upper limit estimate of 9\% for the HNCO/CH$_{3}$OH ice ratio. Again, this is higher than our observed ratios in low-mass hot corinos.

These differences are interesting, but without the direct observation of the species studied here in ices for the same sources it is not possible to investigate this further. The column density ratios discussed in this section need to be studied in a larger sample of protostars to make progress in this field. More robust conclusions will become possible with the launch of JWST and the direct observation of these molecules in ices. In addition, progress in recording the solid state features of several species will make this comparison more plausible in the near future (\citealt{Boudin1998}; \citealt{Terwisscha2018}; \citealt{Gerakines2020}).

\subsection{Effect of emitting areas of molecules on abundance ratios}
\label{sec:abund_ratio}

As the emission from our data is not spatially resolved, variations from source to source or molecule to molecule could be explained by different emitting regions that are possibly related to different binding energies. Using the spherically symmetric infalling envelope toy model (see Appendix \ref{sec:toy_model}), we can find a relation between the ratio of column densities of two molecules in the same beam with the ratio of their corresponding sublimation temperatures. This is given as

\begin{equation}
     \frac{N_{1}}{N_{2}} = \frac{X_{1}}{X_{2}} \left(\frac{T_{\rm 1,sub}}{T_{\rm 2,sub}}\right)^{-3.75},
    \label{eq:sub_T}
\end{equation}

\noindent where $X$ is the abundance of a molecule in gas phase with respect to molecular hydrogen, $N$ is the column density, and $T_{\rm sub}$ is the sublimation temperature, with subscripts 1 and 2 indicating the two species. Equation \eqref{eq:sub_T} shows that the column density ratios depend not only on the abundance ratios but also on the ratio of sublimation temperatures. This is because the difference in the sublimation temperature results in a difference in the emitting volume.

Figure \ref{fig:Tex_both} shows that a typical O-bearing or N-bearing
molecule, $\mathcal{M}$, comes off the grains at a temperature of
${\sim} 100\, \rm K$. Hence, one can rewrite Eq. \eqref{eq:sub_T}
for species 1 as

\begin{equation}
     f_{\rm sub} \equiv \frac{X_{\mathcal{M}}}{X_{1}}
     \frac{N_{1}}{N_{\mathcal{M}}} =  \left(\frac{T_{\rm 1,sub}}{100}\right)^{-3.75}.
    \label{eq:sub_T_norm_i}
\end{equation}

\noindent This can be rewritten as

\begin{equation}
    \frac{X_{1}}{X_{\mathcal{M}}} = \frac{1}{f_{\rm sub}} \frac{N_{1}}{N_{\mathcal{M}}}. 
    \label{eq:sub_T_norm}
\end{equation}

The column density ratios discussed in this work with respect to methanol, which has a $T_{\rm sub}$ of $\sim 100\, \rm K$, are thus the abundance ratios with respect to methanol times a factor ($f_{\rm sub}$). The abundance of a molecule is a local quantity that is sensitive to the emitting area, but column density is an integrated quantity within a given beam. In other words, using the expression `abundance' to refer to a column density ratio is only accurate if the emitting areas are similar.

\begin{figure}
  \resizebox{\hsize}{!}{\includegraphics{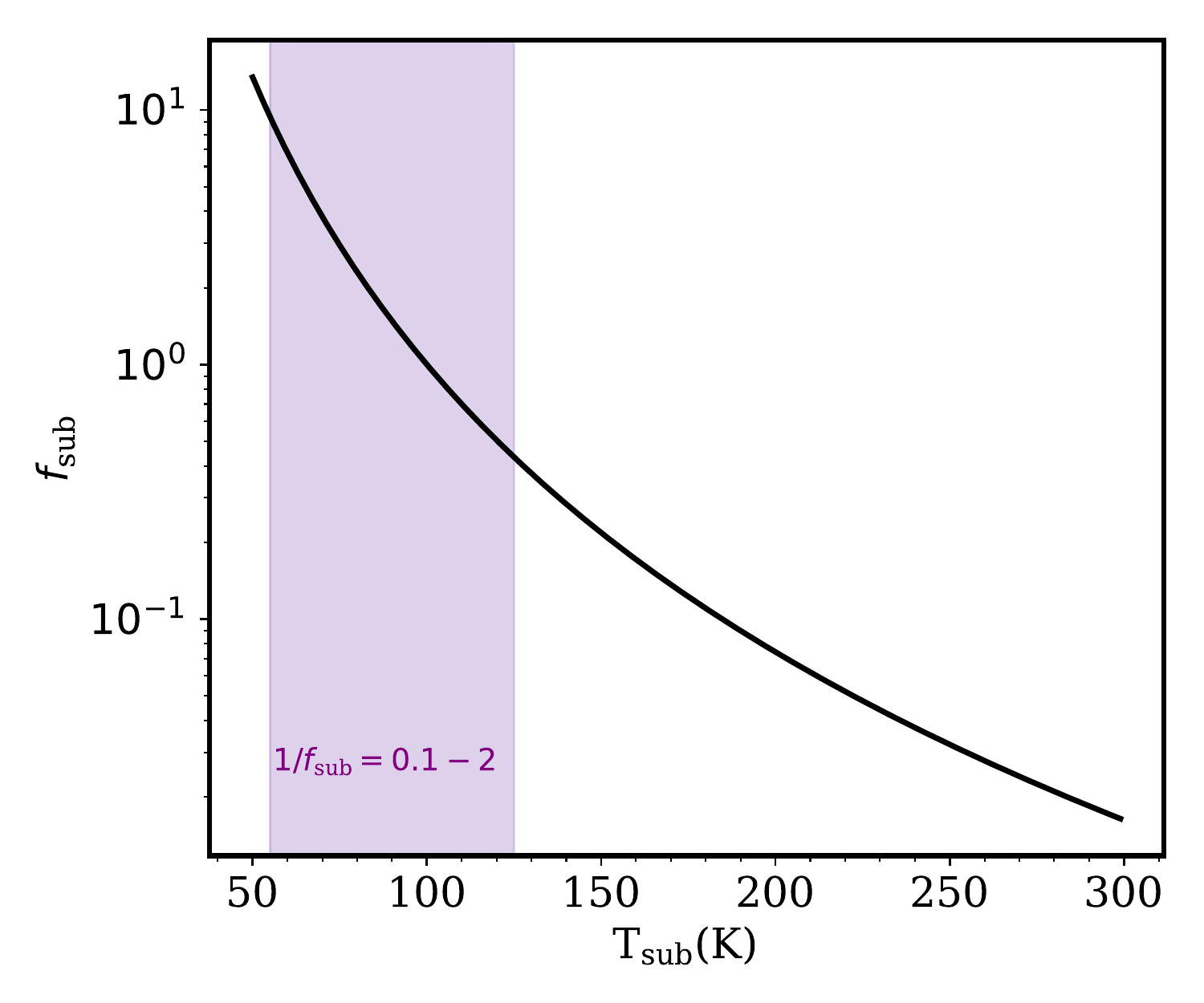}}
  \caption{Correction factor, $f_{\rm sub}$, introduced in Eq. \eqref{eq:sub_T_norm} due to different emitting areas arising from the sublimation temperature. The shaded area shows where most of the species considered in this work fall on this graph (sublimation temperatures between 55\,K and 125\,K). The abundance ratios of these species with respect to methanol are found by multiplying their column density ratios with respect to methanol by a factor of between 0.1 and 2.}
  \label{fig:toy_model}
\end{figure}

Figure \ref{fig:toy_model} shows $f_{\rm sub}$ as defined in Eq. \eqref{eq:sub_T_norm_i} against the sublimation temperature for a molecule. This figure shows that, for species with desorption temperatures between 55\,K and 125\,K (see the grey points in Fig. \ref{fig:Tex_both}), one would need to multiply the column density ratios of the species with respect to methanol by a factor of between 0.1 and 2 to get the respective abundance ratios. This factor becomes much larger as the sublimation temperature increases. For instance, $\rm (CH_{2}OH)_{2}$ has a sublimation temperature of ${\sim} 200\,\rm K$, and hence the factor becomes ${\sim} 14$ for this molecule. However, this factor is not applied here because of the simplicity of our model. In reality, one needs to build a model that takes small-scale structures such as disks into account  (\citealt{Harsono2015}; \citealt{Persson2016}), as discovered by other studies of Class 0 objects (\citealt{Tobin2012}; \citealt{Murillo2013}; \citealt{Martin2019}; \citealt{Maret2020}). This is important because the temperature profile in an envelope of a Class 0 object does not correspond to the temperature profile in the inner regions of the disk (\citealt{Murillo2018}; \citealt{vantHoff2020a}). Moreover, the opacity of the dust continuum emission can also become relevant at $\lesssim 50\,\rm au$ scales (\citealt{DeSimone2020}). 

In conclusion, some of the variations seen in abundances between various molecules and between different sources may be related to different emitting regions originating from different binding energies rather than from chemistry.

\section{Conclusions}
\label{sec:conclusion}

In this work we derived the column densities and excitation temperatures of N-bearing species in two Class 0 objects, B1-c and S68N, using the Band 6, Band 5, and Band 3 of ALMA. The main conclusions are listed below.

\begin{itemize}
    \item Four N-bearing molecules and their isotopologues are identified towards B1-c and S68N with ALMA. These species are HNCO, $\rm HN^{13}CO$, DNCO, $\rm C_{2}H_{5}CN$, $\rm NH_{2}CHO$, $\rm CH_{3}CN$, $\rm CH_{2}DCN$, and $\rm CHD_{2}CN$. None of these species are spatially resolved, but they are located within $200\,\rm au$ of the central protostar.
    \item N-bearing and O-bearing species show a similar scatter in their excitation temperatures (${\sim} 100-300\,\rm K$), and the average excitation temperature of N-bearing species is roughly the same as the average for the O-bearing species studied by \cite{vangelder2020} for B1-c and S68N. 
    \item The excitation temperatures for the N-bearing and O-bearing species are larger than the sublimation temperatures of each molecule, which span a range between ${\sim} 55$ and ${\sim} 125\, \rm K$. This can be interpreted as the excitation temperature measuring a mass-weighted average temperature of the region that is hotter than the sublimation temperature of the detected molecule.
    \item The abundances of the N-bearing species with respect to methanol and HNCO are very similar for the two sources studied in this work. Their abundances with respect to methanol are lower than those of the O-bearing species by an order of magnitude.
    \item Overall, N-bearing species show more uniform abundance ratios with respect to HNCO than relative to methanol between low- and high-mass sources.
    \item $\rm CH_{2}DCN$, $\rm CHD_{2}CN$, and DNCO show similar abundance ratios with respect to HNCO between different low-mass sources, suggesting a universal mechanism for increasing the $\rm D/H$ in cold regions prior to the star formation process.
    \item\ The $\rm NH_{2}CHO/CH_{3}OH$ varies within an order of magnitude between different sources. This can be due either to the different formation pathways suggested for formamide in the solid state or the high sublimation temperature of formamide and the different local temperatures where formamide is detected.
    \item Variation in abundance ratios between various molecules and sources could be due to the different emitting areas related to the binding energies of the species in spatially unresolved emission. 
\end{itemize}

We emphasise that the sample of COM-rich low-mass protostars that have been subjected to such analyses is still very small. Therefore, reaching robust conclusions on the chemistry of the species discussed here is difficult. An increase in the size of this sample by an order of magnitude will enhance our understanding considerably. Moreover, a broader frequency coverage can help to determine more accurate excitation temperatures and identify other molecular species. Therefore, more ALMA and NOEMA observations of COM-rich Class 0/I objects are necessary for moving this field forward. Moreover, JWST data of the sources discussed here can give a direct comparison between the ice abundances and the gas abundances found in this work. The conclusions presented here can act as a guideline for this future work.

\begin{acknowledgements}
      We thank the Allegro Team at Leiden Observatory, especially Aida Ahmadi for her invaluable help in reducing the Band 5 data for B1-c. We also thank the referee for a constructive report. This paper makes use of the following ALMA data: $\rm ADS/JAO.ALMA\# 2017.1.01174.S$ and $\rm ADS/JAO.ALMA\# 2017.1.01371.S$. ALMA is a partnership of ESO (representing its member states), NSF (USA) and NINS (Japan), together with NRC (Canada), MOST and ASIAA (Taiwan), and KASI (Republic of Korea), in cooperation with the Republic of Chile. The Joint ALMA Observatory is operated by ESO, $\rm AUI/NRAO$ and NAOJ. Astrochemistry in Leiden is supported by the Netherlands Research School for Astronomy (NOVA). M.L.G. acknowledges support from the Dutch Research Council (NWO) with project number NWO TOP-1 614.001.751. B.T. acknowledges support from the Dutch Astrochemistry Network II with project number 614.001.751, financed by the Netherlands Organisation for Scientific Research (NWO). M.L.R.H. acknowledges support from the Michigan Society of Fellows. A.C.G. has received funding from the European Research Council under the European Union’s Horizon 2020 research and innovation programme (grant agreement No. 743029). H.B. acknowledges support from the European Research Council under the Horizon 2020 Framework Program via the ERC Consolidator Grant CSF-648505. H.B. also acknowledges support from the Deutsche Forschungsgemeinschaft in the Collaborative Research Center (SFB 881) “The Milky Way System” (subproject B1). 
      
\end{acknowledgements}


\bibliographystyle{aa}
\bibliography{N-COM_script}

\begin{appendix} 

\section{Spectroscopic data}

The spectroscopic data for HNCO are taken from the CDMS (\citealt{CDMS2001}; \citealt{muller2005}), which is based on the work by \cite{Kukolich1971HNCO}, \cite{Hocking1975}, \cite{Niedenhoff1995}, and \cite{Lapinov2007}. DNCO and $\rm HN^{13}CO$ data are based on \cite{Hocking1975} and are taken from the JPL database (\citealt{JPL}).

The line list for $\rm C_{2}H_{5}CN$ is taken from the CDMS. The molecule entries are based on the work by \cite{Pearson1994}, \cite{Fukuyama1996}, and \cite{Brauer2009}. A vibrational factor of 1.8316 is used at 200\,K (\citealt{Heise1981}).

The line list for $\rm NH_{2}CHO$ is taken from the JPL database. The methods used to analyse the experimental measurements are in \cite{Kirchhoff1972}. The measurements are taken from \cite{Kurland1957}, \cite{Costain1960}, \cite{Kukolich1971NH2CHO}, \cite{Johnson1972}, and \cite{Kirchhoff1973}. An upper limit of 1.5 for the vibrational correction factor of $\rm NH_{2}CHO$ is used at 300\,K.

The spectroscopic data used for $\rm CH_{3}CN$ are taken from the CDMS entry (\citealt{Bocquet1988}; \citealt{Koivusaari1992}; \citealt{Tolonen1993}; \citealt{Anttila1993}; \citealt{Gadhi1995}; \citealt{Simeckova2004}; \citealt{Cazzoli2006}; \citealt{Muller2009}; \citealt{muller2015}). The line lists of $\rm CH_{2}DCN$ and $\rm CHD_{2}CN$ are taken from the CDMS (\citealt{Nguyen2013}). $\rm CH_{2}DCN$ has additional data from \cite{Guennec1992} and \cite{Muller2009}. $\rm CHD_{2}CN$ has additional data from \cite{Halonen1978}. The vibration factor for $\rm CH_{3}CN$ is lower than 1.1 up to a temperature of 180\,K (\citealt{muller2015}) and is thus negligible in this work. It is assumed that the difference between the vibrational factors for the main molecule and its isotopologues are small at the excitation temperatures found in this work.  

The spectroscopic data for $\rm CH_{3}NCO$ are taken from the CDMS (\citealt{Koput1986}; \citealt{Cernicharo2016}). The partition function takes vibrational states into account up to 580\,K. The line list for $\rm NH_{2}CN$ is taken from the JPL database (\citealt{Read1986}). The spectroscopic data for $\rm HOCH_{2}CN$ are taken from the CDMS (\citealt{Margules2017}). The line list for $\rm CH_{3}NH_{2}$ is taken from the JPL database (\citealt{Ilyushin2005}; \citealt{kreglewski1992}; \citealt{kreglewski1992b}; \citealt{Ohashi1987}; \citealt{Takagi1971}; \citealt{Nishikawa1957}; \citealt{Lide1954}; \citealt{Shimoda1954}; \citealt{Lide1957}).

\section{Toy model for a spherically symmetric envelope}
\label{sec:toy_model}

\subsection{Physical model}
A toy model for a spherically symmetric infalling envelope can be developed by assuming a power law structure in density and temperature. The number density of hydrogen is assumed to be 

\begin{equation}
    n_{\rm H}(R) = n_{\rm H,0} \left(\frac{R}{R_{0}}\right)^{-\alpha},
    \label{eq:nH}
\end{equation}

\noindent where $\alpha$ is 3/2 for a free-falling envelope, $R$ is the radius, and $n_{\rm H,0}$ is the number density of hydrogen at radius $R_{0}$. Moreover, assuming that the envelope is passively heated by the central protostar, the temperature profile can be written as 

\begin{equation}
    T(R) = T_{0} \left(\frac{R}{R_{0}}\right)^{-\beta},
    \label{eq:T_R}
\end{equation}

\noindent where $\beta \simeq 2/5$ (at $R \gtrsim 10\, \rm au$) from radiative transfer calculations (\citealt{Adams1985}) and $T_{0}$ is the temperature at $R_{0}$ and is proportional to $L_{\rm bol}^{1/5}$, where $L_{\rm bol}$ is the bolometric luminosity of the source. 

In this model we assume that a constant amount of material crosses the snow line and sublimates into the gas phase without further gas-phase reactions. Therefore, each COM comes off the ice at its sublimation temperature, $T_{\rm sub}$, corresponding to a sublimation radius of $R_{T_{\rm sub}}$. This radius is found by rearranging Eq. \eqref{eq:T_R}:

\begin{equation}
    R(T_{\rm sub}) = R_{0} \left(\frac{T_{\rm sub}}{T_{0}}\right)^{-1/\beta}. 
    \label{eq:R_T}
\end{equation}

\subsection{Scaling of COM emission with luminosity}

Assuming the emission is optically thin, the total number of a specific COM in the gas phase that comes off the ice at temperatures above its sublimation temperature is proportional to the line flux. The total number of a molecule of a specific COM, $\mathcal{N}_i$, is given by:
\begin{align}
    \mathcal{N}_{i} & = \int_{0}^{R_{T_{\rm sub}}} 4 \pi X_{i} n_{\rm H}(R) R^{2}dR \\
    & = 4\pi X_{i} \int_{0}^{R_{T_{\rm sub}}} n_{\rm H,0} \left(\frac{R}{R_{0}}\right)^{-\alpha} R^{2}dR \\
    & = \frac{4\pi}{3-\alpha} X_{i} n_{\rm H,0} R_{0}^{\alpha} R_{T_{\rm sub}}^{3-\alpha},
    \label{eq:COM_num}
\end{align}

\noindent where $X_{i}$ is the abundance of species $i$ with respect to hydrogen atoms. 

Dropping the factors that do not depend on the source properties, Eqs. \eqref{eq:R_T} and \eqref{eq:COM_num} yield

\begin{equation}
    \mathcal{N}_{i} \propto n_{\rm H,0} R_{T_{\rm sub}}^{3- \alpha} \propto  n_{\rm H,0} T_{0}^{(3-\alpha)/\beta} \propto  n_{\rm H,0} L_{\rm bol}^{(3 - \alpha)/(5\beta)},
    \label{eq:num_prop}
\end{equation}

\noindent and assuming $\alpha = 3/2$ and $\beta = 2/5$ gives

\begin{equation}
    \mathcal{N}_{i} \propto n_{\rm H,0} L_{\rm bol}^{3/4}.
\end{equation}

\noindent Therefore, the toy model predicts that the total number of molecules depends on the density of the envelope at a reference radius, $R_0$, and on the luminosity of the source.

\subsection{COM emission as a function of $T_{\rm sub}$}

The measured column density of a molecule depends on the assumed source size. If the emission is spatially unresolved and optically thin, one generally assumes a fixed source size for the different species. Hence, the ratio of the measured column densities of two species, 1 and 2, does not depend on the assumed source size. Moreover, the ratio of the measured column densities, assuming the same emitting region, is the same as the ratio of the number of the two molecules in the gas phase. Therefore, the ratio of column densities as found in Sect. \ref{sec:col_exT} ($N_{1}/N_{2}$) is the same as the ratio of the total number of molecules for two COMs ($\mathcal{N}_{1}/\mathcal{N}_{2}$). For the rest of this section, $N_{1}/N_{2}$ and $\mathcal{N}_{1}/\mathcal{N}_{2}$ are the same. 

From Eq. \eqref{eq:COM_num}, this ratio is given by

\begin{equation}
    \frac{N_{1}}{N_{2}} = \frac{X_{1}}{X_{2}} \left(\frac{R_{T_{\rm sub,1}}}{R_{T_{\rm sub,2}}}\right)^{3-\alpha}.
\end{equation}

\noindent Using Eq. \eqref{eq:R_T} and assuming $\alpha = 3/2$ and $\beta = 2/5$, this becomes

\begin{equation}
    \frac{N_{1}}{N_{2}} = \frac{X_{1}}{X_{2}} \left(\frac{T_{\rm sub,1}}{T_{\rm sub,2}}\right)^{-3.75}.
\end{equation}

\noindent Therefore, the measured column density ratio for two molecules is not the same as their abundance ratio: The former also depends on the ratio of their sublimation temperatures. In principle, if the thermal structure of the envelope is well constrained, one can recover the abundance ratio from the measured column density ratio by applying the correction factor defined in Eq. \ref{eq:sub_T_norm_i}.  

\subsection{Mass-weighted kinetic temperature}
\label{sec:toy_model_avgT}

Assuming that the excitation temperatures measured in this work probe the mass-weighted kinetic temperature of the environment, one can find a relation between this quantity and the sublimation temperature of each COM. 
The mass-weighted temperature is given by

\begin{equation}
    <T> = \frac{\int_{0}^{R_{T_{\rm sub}}} T(R) n_{\rm H}(R)R^{2}dR}{\int_{0}^{R_{T_{\rm sub}}} n_{\rm H}(R)R^{2}dR}.
\end{equation}

\noindent Using Eqs. \eqref{eq:nH}, \eqref{eq:T_R}, and \eqref{eq:R_T}, one can find the above average temperature as

\begin{equation}
  <T> = 1.36 T_{\rm sub}.  
\end{equation}

\noindent Therefore, the measured excitation temperature of a species would be 1.36 times the sublimation temperature of that molecule.

\section{Spectral fitting results}
\label{app:spec_fit}

\begin{figure*}
    \centering
    \includegraphics[width=17cm]{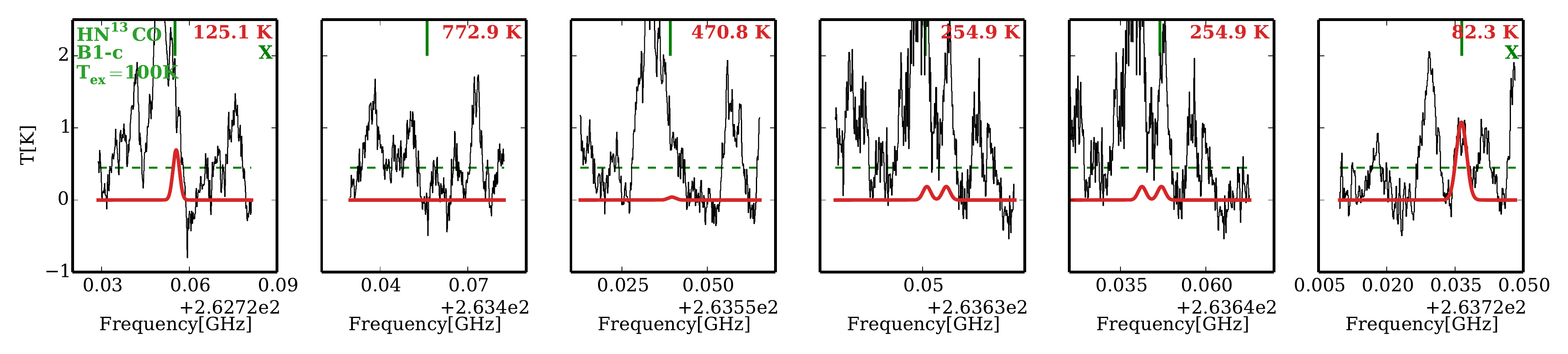}
    \caption{Model to combined Band 5 and 6 data of $\rm HN^{13}CO$ for B1-c in red and data in black. The model uses the lower limit on the excitation temperature of $\rm HN^{13}CO$. Each graph shows one line of $\rm HN^{13}CO$, indicated by the solid green line at the top of the box, along with its upper energy level, written in red at the top right. The lines with upper energy levels above 1000\,K and/or $A_{ij}$ below $10^{-5}$ are not plotted. The excitation temperature used for the figure is shown at the top left. The dashed green line shows the $3\sigma$ level. Cases where a line is seen at the $3\sigma$ level or above and is used as part of the fitting are marked with a green X at the top right corner of the box.}
    \label{fig:HN13CO_fit}
\end{figure*}

\begin{figure*}
 \includegraphics[width=17cm]{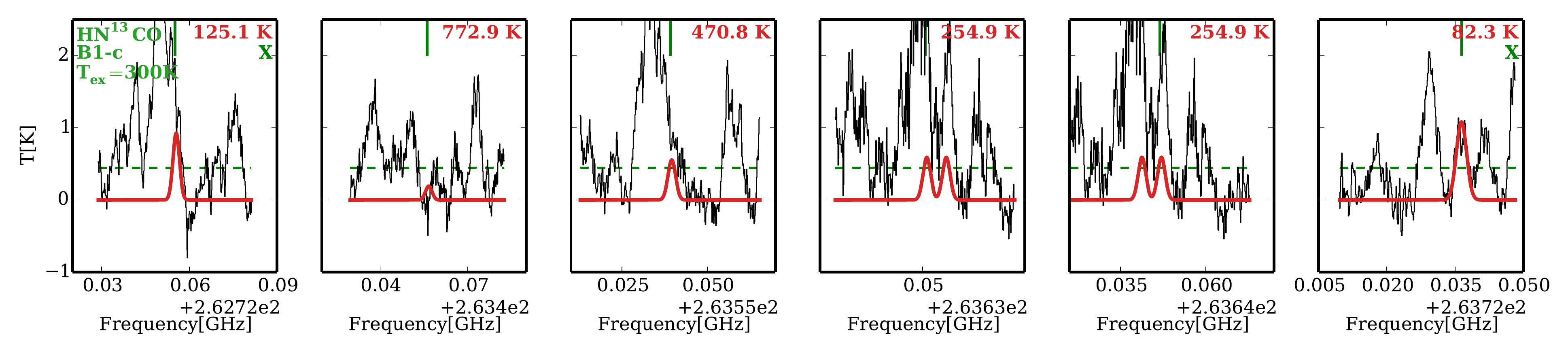}
  \caption{Same as Fig. \ref{fig:HN13CO_fit} but for its upper limit on temperature.}
  \label{fig:HN13CO_fitup}
\end{figure*}

\begin{figure}
  \resizebox{0.5\hsize}{!}{\includegraphics{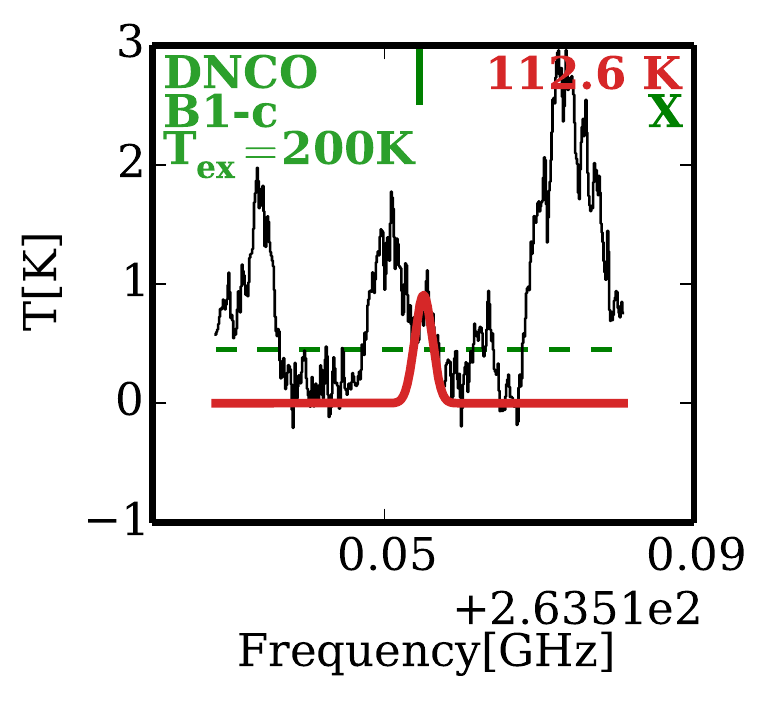}}
  \caption{Same as Fig. \ref{fig:HN13CO_fit} but for DNCO and the best-fit model where the temperature is fixed.}
  \label{fig:DNCO_fit}
\end{figure}

\begin{figure*}
    \centering
    \includegraphics[width=17cm]{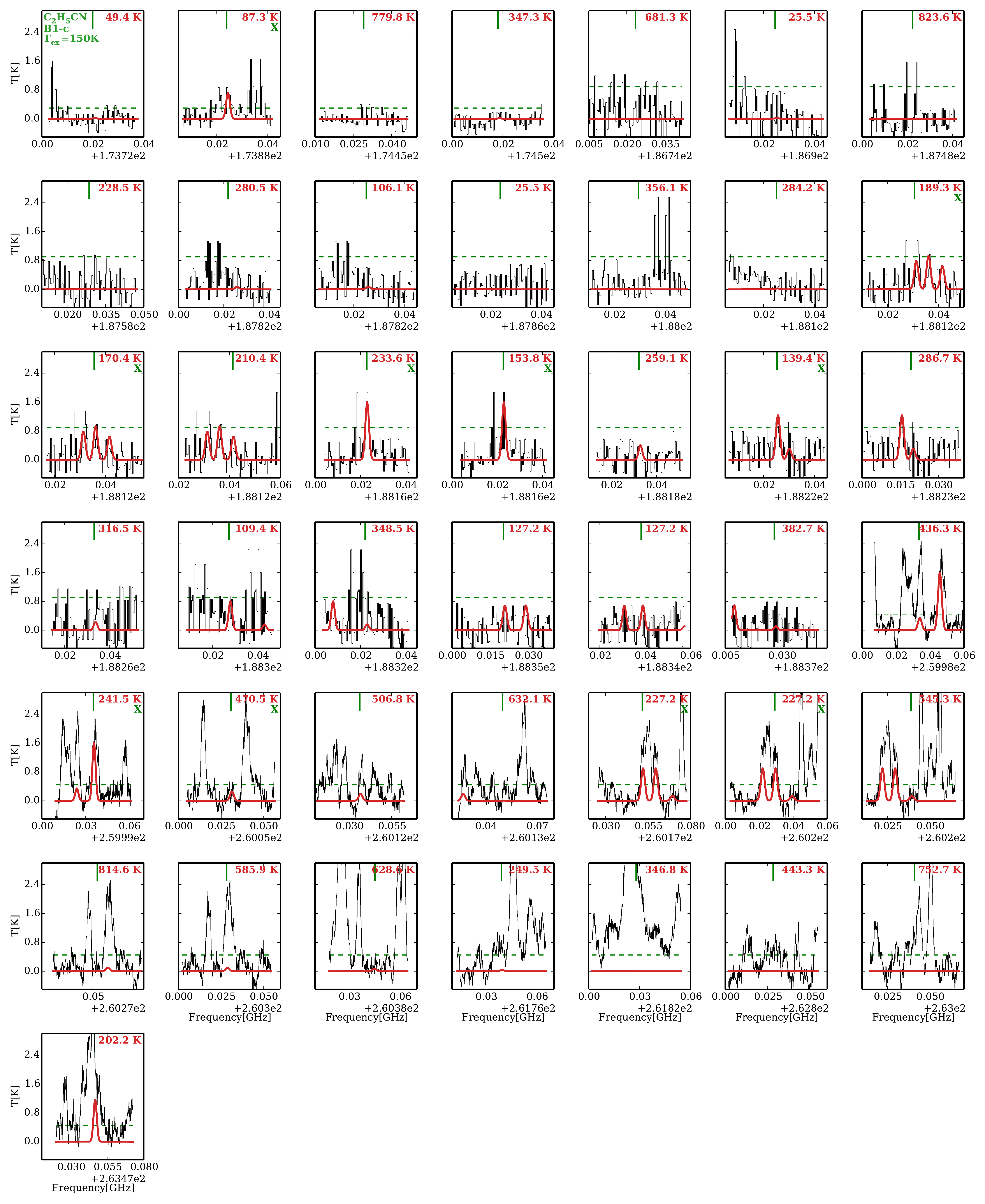}
    \caption{Same as Fig. \ref{fig:HN13CO_fit} but for $\rm C_{2}H_{5}CN$.}
    \label{fig:C2H5CN_fit}
\end{figure*}

\begin{figure*}
    \centering
    \includegraphics[width=17cm]{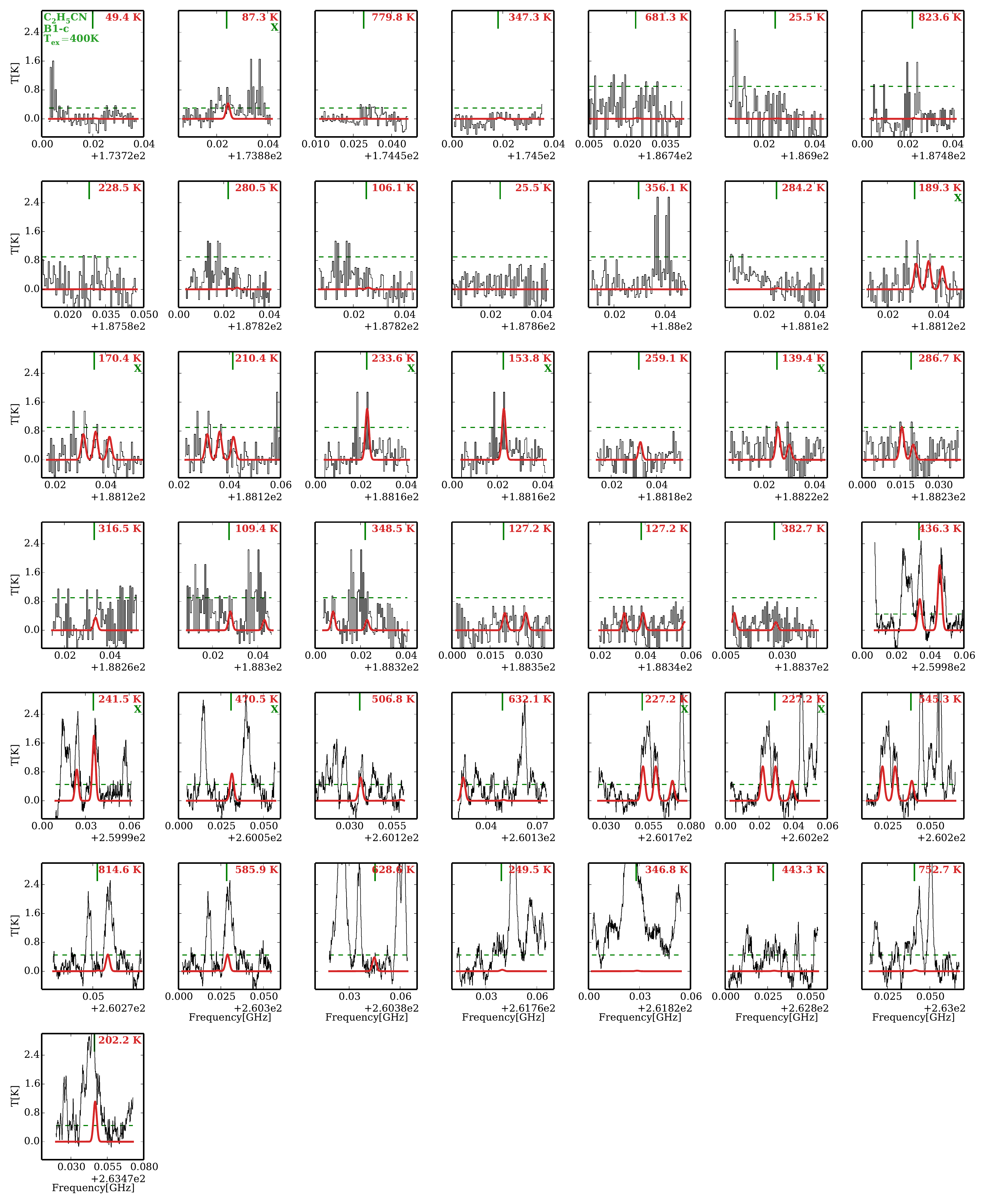}
    \caption{Same as Fig. \ref{fig:HN13CO_fit} but for $\rm C_{2}H_{5}CN$ and for its upper limit on temperature.}
    \label{fig:C2H5CN_fitup}
\end{figure*}

\begin{figure*}
    \centering
    \includegraphics[width=17cm]{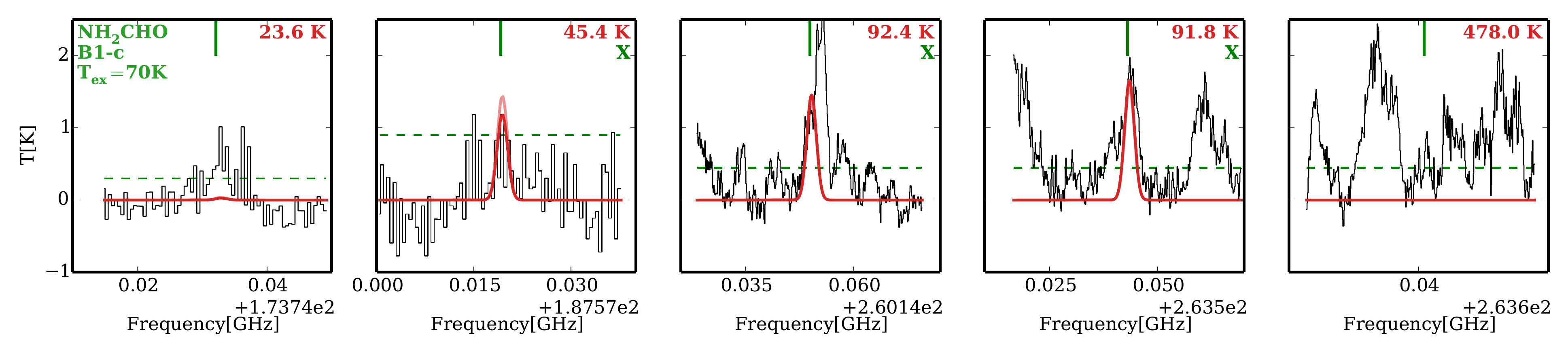}
    \caption{Same as Fig. \ref{fig:HN13CO_fit} but for $\rm NH_{2}CHO$. The more transparent red line shows the model with a $T_{\rm ex}$ of 50\,K to show that it overestimates the line with an E$_{\rm up}$ of 45.4\,K.}
    \label{fig:NH2CHO_fit_low}
\end{figure*}

\begin{figure*}
    \centering
    \includegraphics[width=17cm]{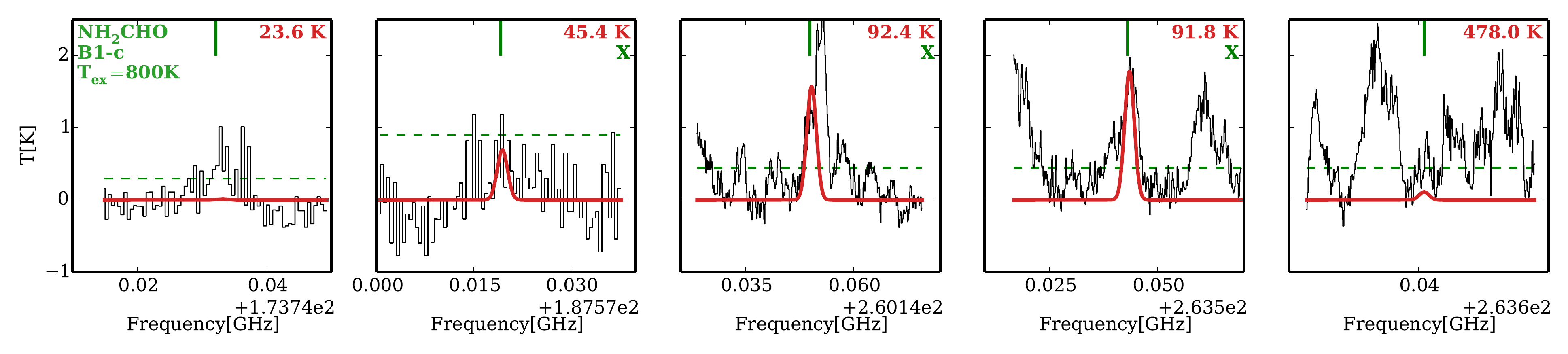}
    \caption{Same as Fig. \ref{fig:HN13CO_fit} but for $\rm NH_{2}CHO$ and a high temperature for the model. This graph is presented to demonstrate that it is not possible to derive an upper limit for the excitation temperature of this molecule because the model used in this plot is very similar to the model in Fig. \ref{fig:NH2CHO_fit_low}.}
    \label{fig:NH2CHO_fit_high}
\end{figure*}

\begin{figure*}
    \centering
    \includegraphics[width=17cm]{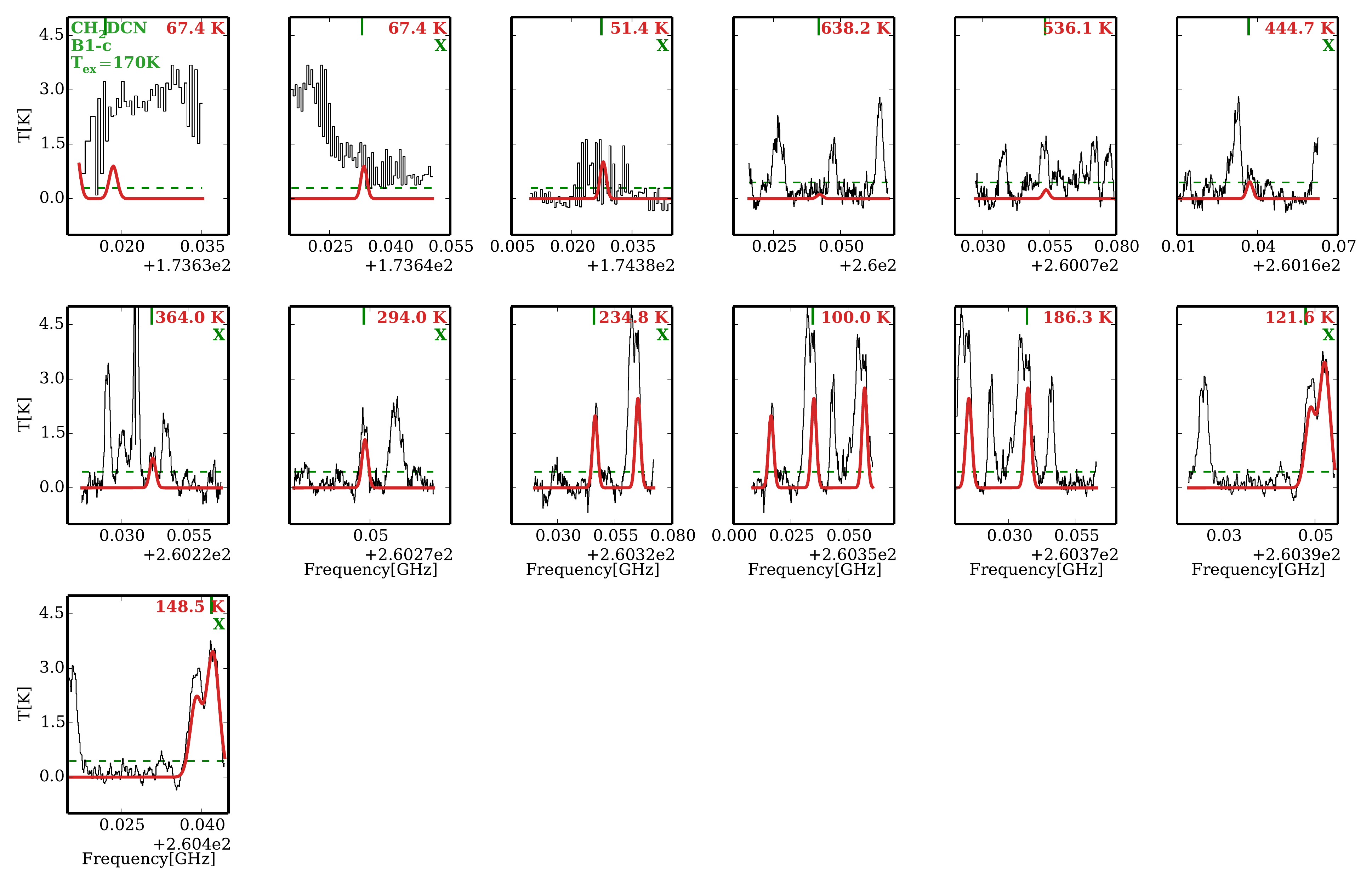}
    \caption{Same as Fig.\ \ref{fig:HN13CO_fit} but for $\rm CH_{2}DCN$.}
    \label{fig:CH2DCN_fit_low}
\end{figure*}

\begin{figure*}
    \centering
    \includegraphics[width=17cm]{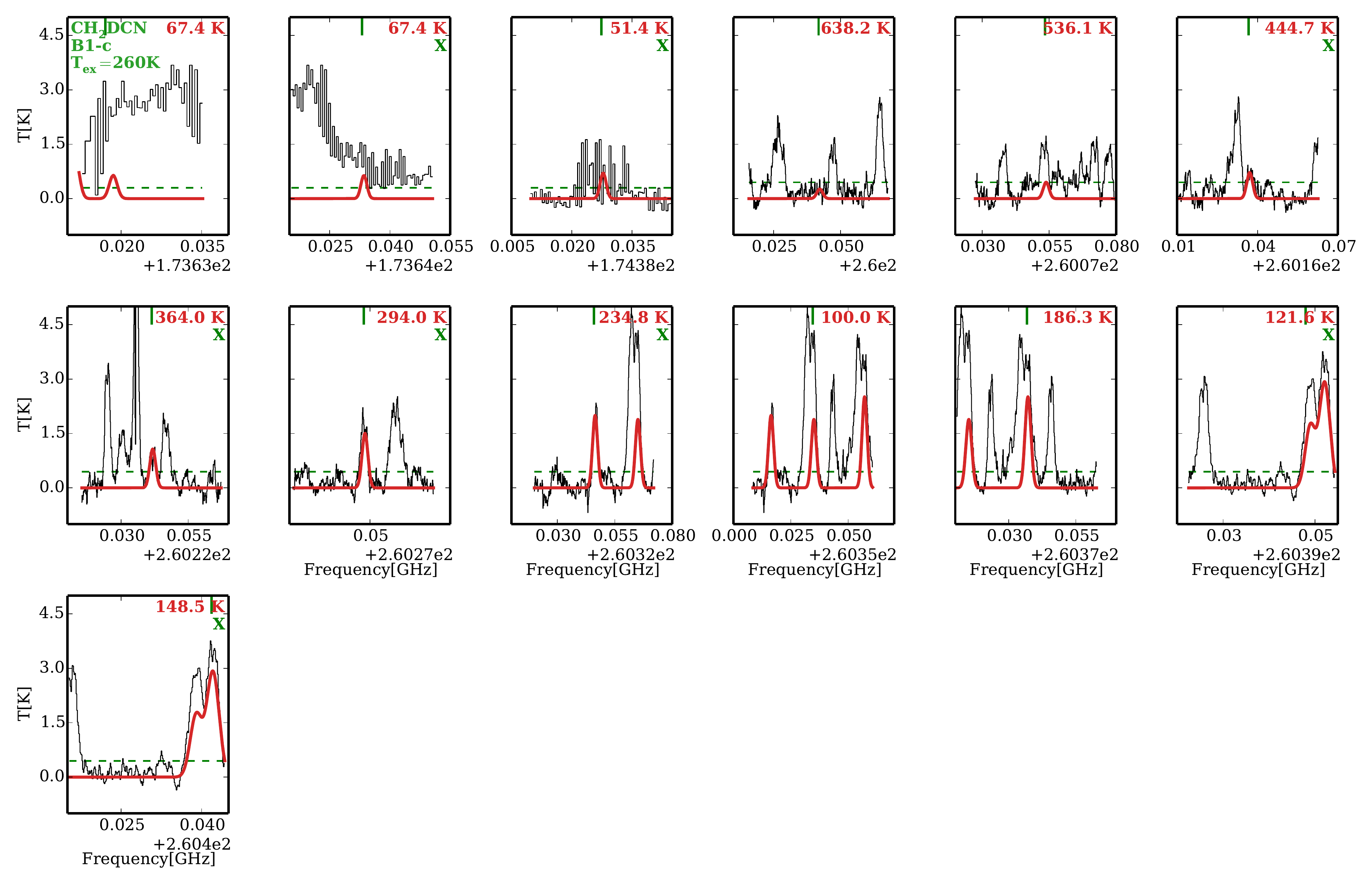}
    \caption{Same as Fig. \ref{fig:HN13CO_fit} but for $\rm CH_{2}DCN$ and for its upper limit on temperature.}
    \label{fig:CH2DCN_fit_high}
\end{figure*}

\begin{figure*}
    \centering
    \includegraphics[width=17cm]{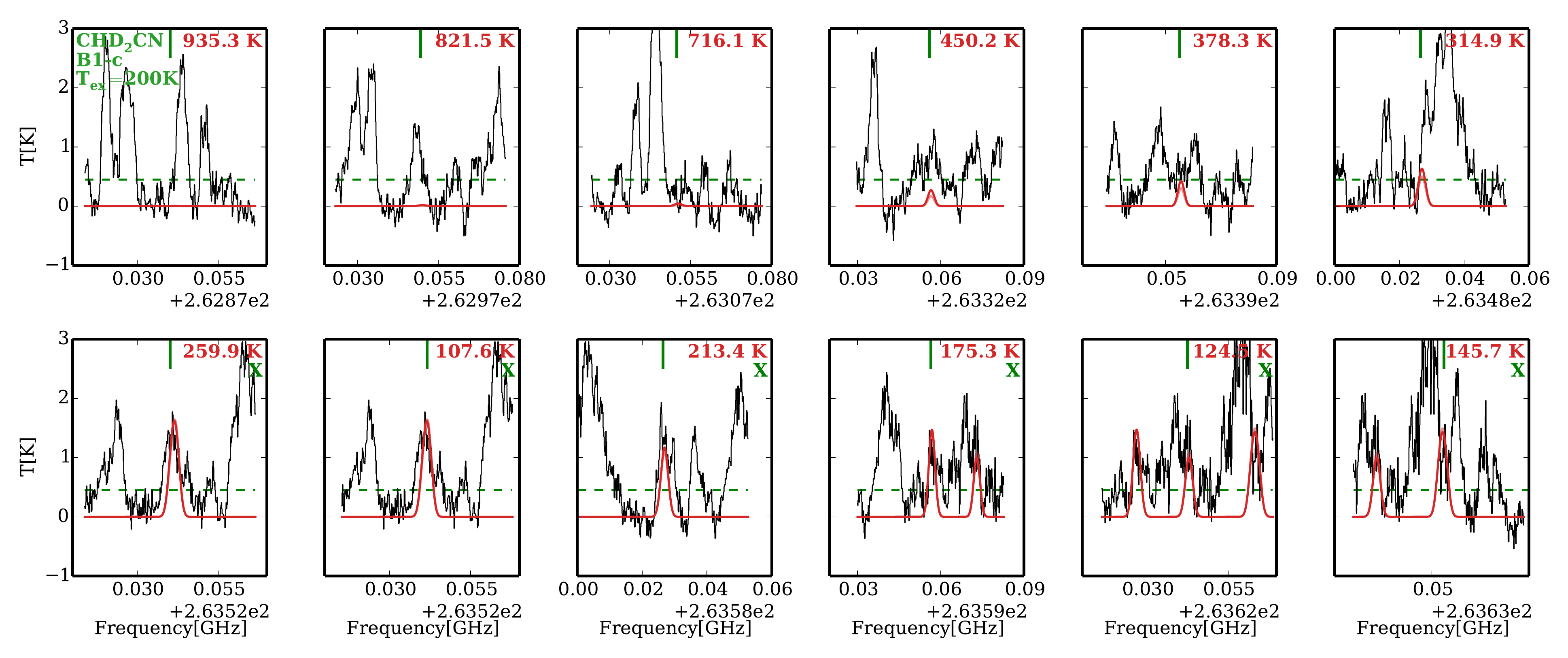}
    \caption{Same as Fig. \ref{fig:HN13CO_fit} but for $\rm CHD_{2}CN$.}
    \label{fig:CHD2CN_fit}
\end{figure*}

\begin{figure*}
    \centering
    \includegraphics[width=17cm]{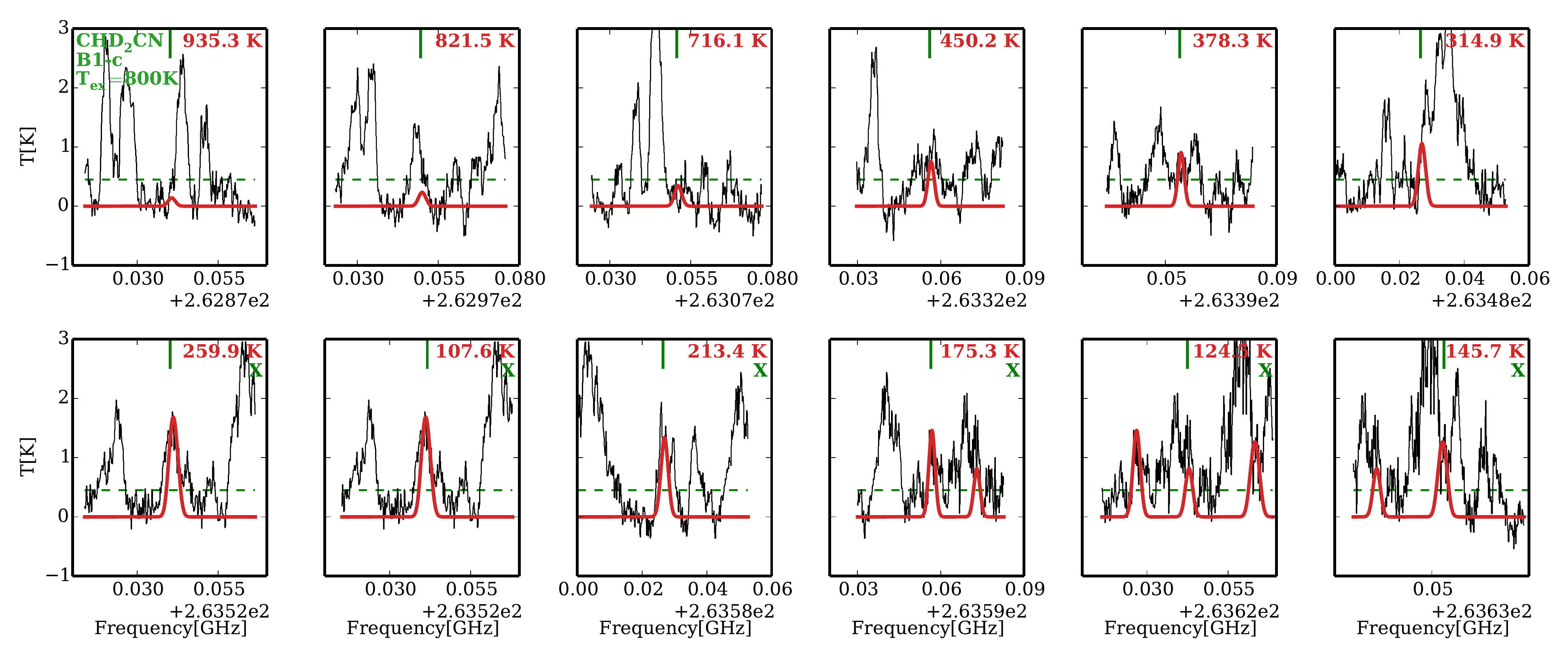}
    \caption{Same as Fig. \ref{fig:HN13CO_fit} but for $\rm CHD_{2}CN$ and a high temperature for the model. This graph is presented to demonstrate that it is not possible to derive an upper limit for the excitation temperature of this molecule because the model used in this plot is very similar to the model in Fig. \ref{fig:CHD2CN_fit}.}
    \label{fig:CHD2CN_fit_err}
\end{figure*}

\begin{figure}
  \resizebox{\hsize}{!}{\includegraphics{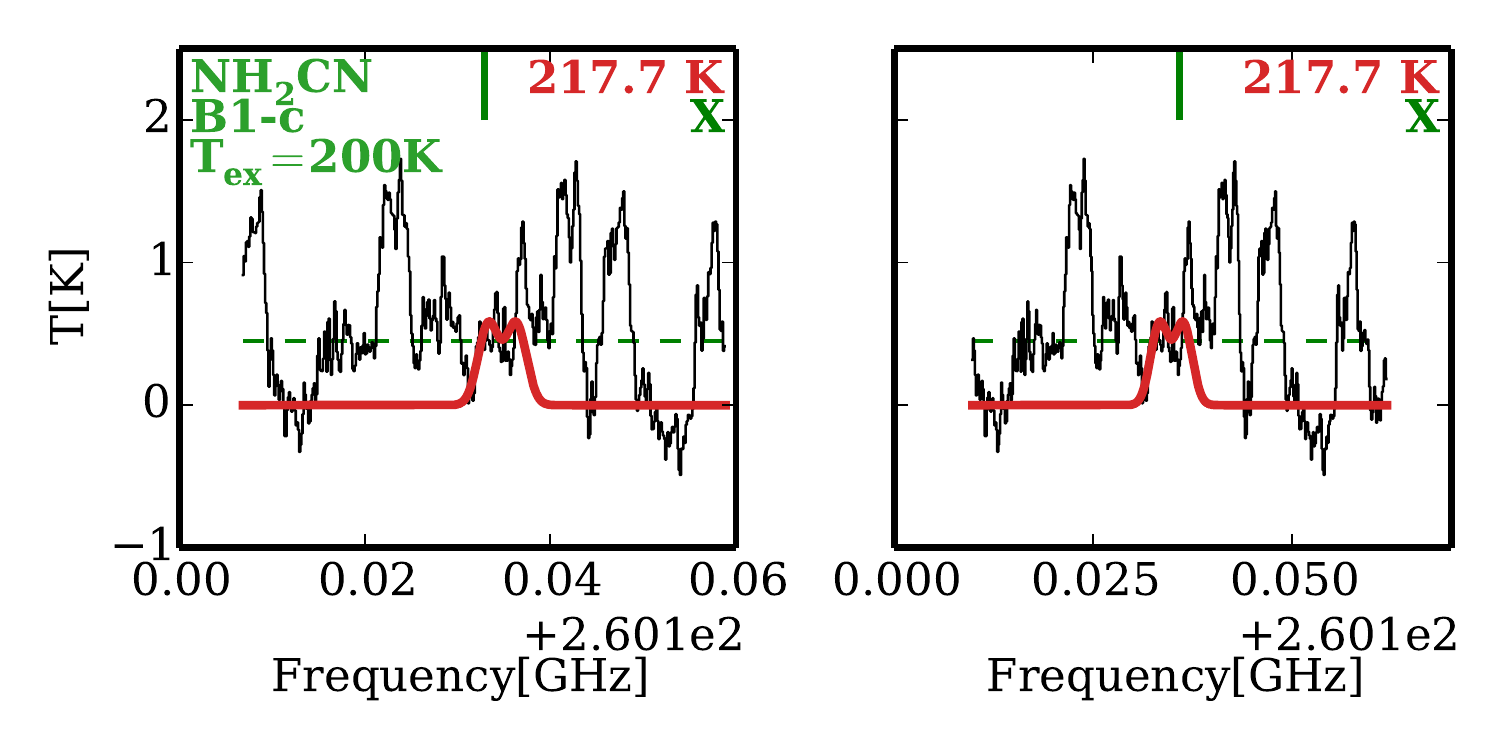}}
  \caption{Same as Fig. \ref{fig:HN13CO_fit} but for $\rm NH_{2}CN$ and the best-fit model where the temperature is fixed.}
  \label{fig:NH2CN_fit}
\end{figure}

\begin{figure*}
    \centering
    \includegraphics[width=17cm]{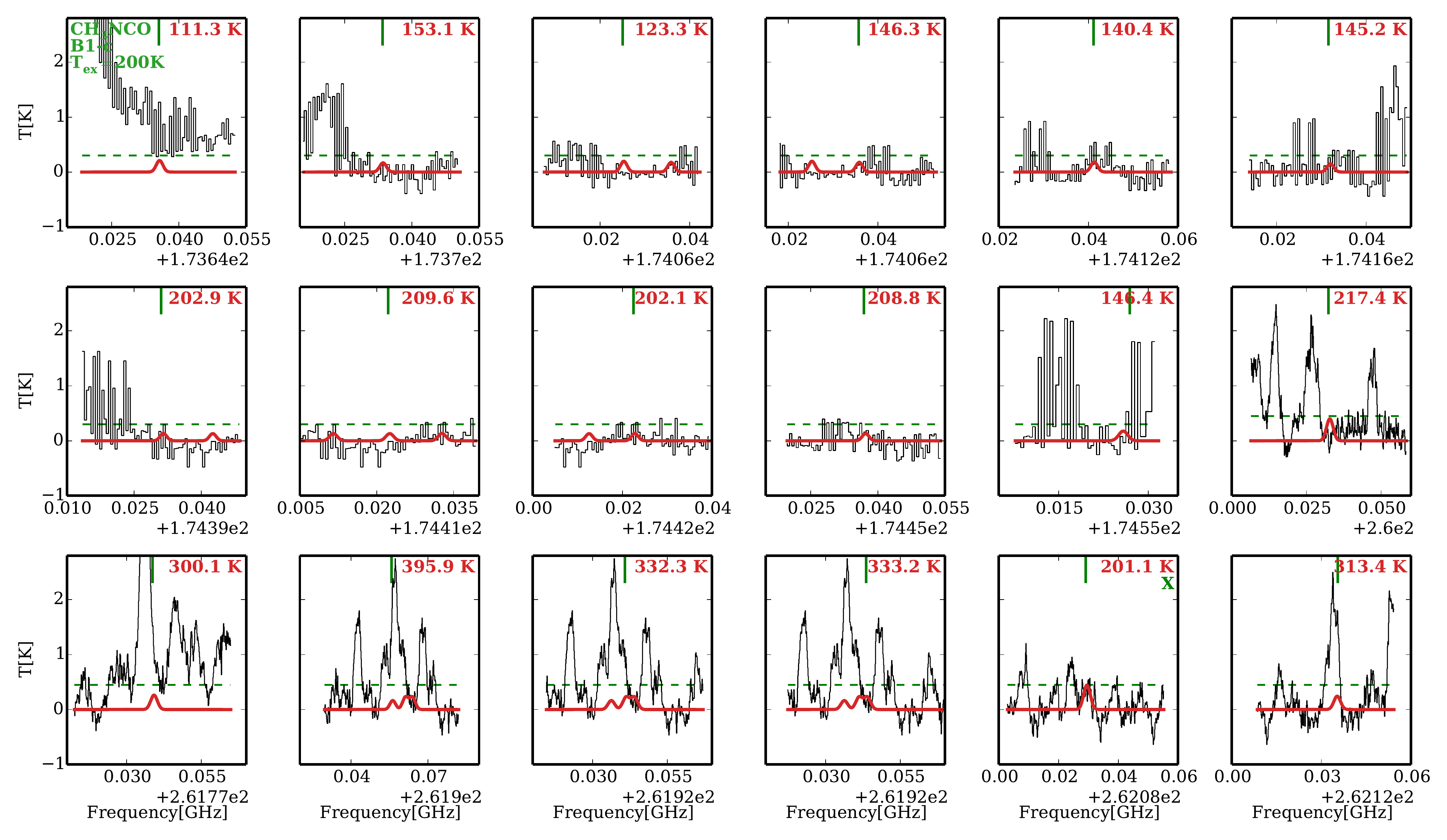}
    \caption{Same as Fig. \ref{fig:HN13CO_fit} but for $\rm CH_{3}NCO$ and the best-fit model where the temperature is fixed.}
    \label{fig:CH3NCO_fit}
\end{figure*}

\begin{figure*}
    \centering
    \includegraphics[width=17cm]{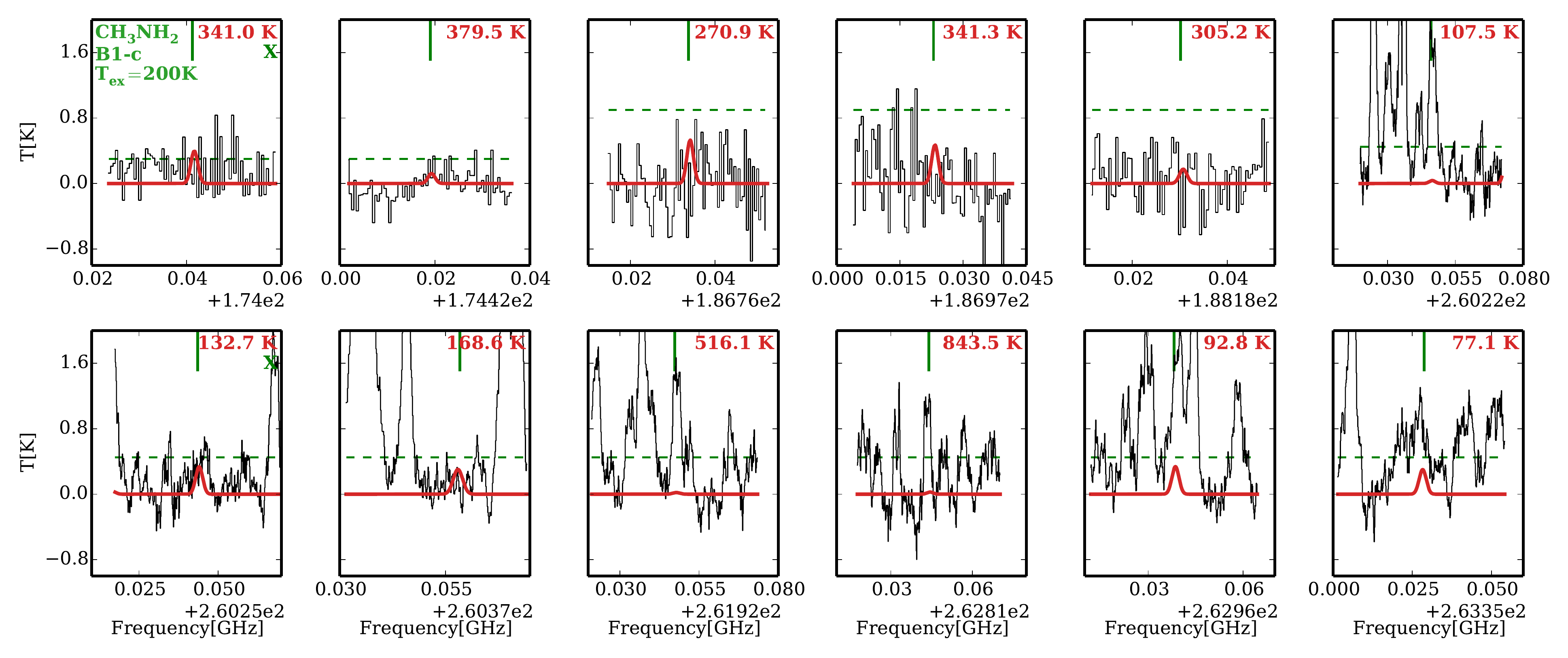}
    \caption{Same as Fig. \ref{fig:HN13CO_fit} but for $\rm CH_{3}NH_{2}$ and the best-fit model where the temperature is fixed.}
    \label{fig:CH3NH2_fit}
\end{figure*}

\begin{figure*}
    \centering
    \includegraphics[width=17cm]{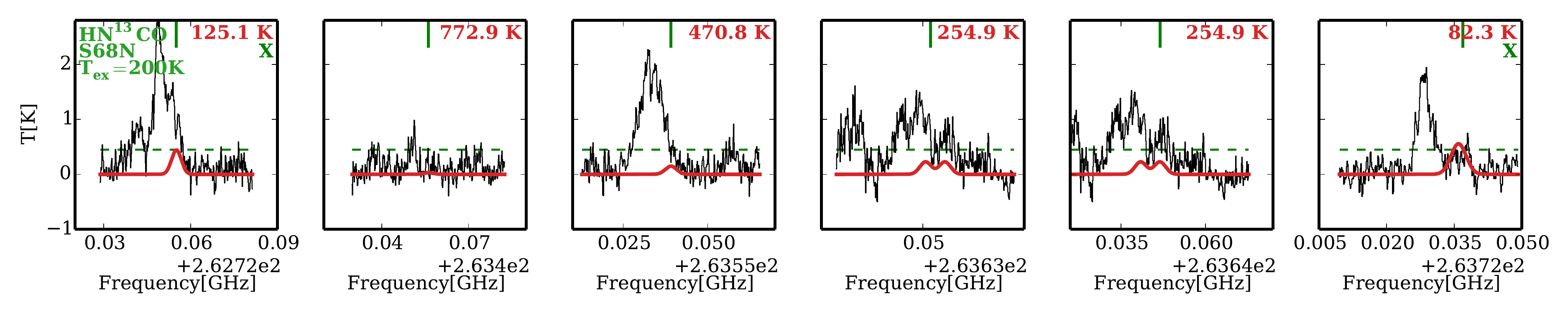}
    \caption{Model to Band 6 data of $\rm HN^{13}CO$ for S68N in red and data in black. The model uses an excitation temperature that is fixed for $\rm HN^{13}CO$. Each graph shows one line of $\rm HN^{13}CO$, indicated by the solid green line at the top middle along with its upper energy level at the top right. The lines with upper energy levels above 1000\,K and/or $A_{ij}$ below $10^{-5}$ are not plotted. The excitation temperature used for this figure is shown at the top left. The dashed green line shows the $3\sigma$ level. Cases where a line is seen at the $3\sigma$ level or above and is used as part of the fitting are marked with a green X at the top right corner of the box.}
    \label{fig:HN13CO_fit_S68N}
\end{figure*}

\begin{figure}
  \resizebox{0.5\hsize}{!}{\includegraphics{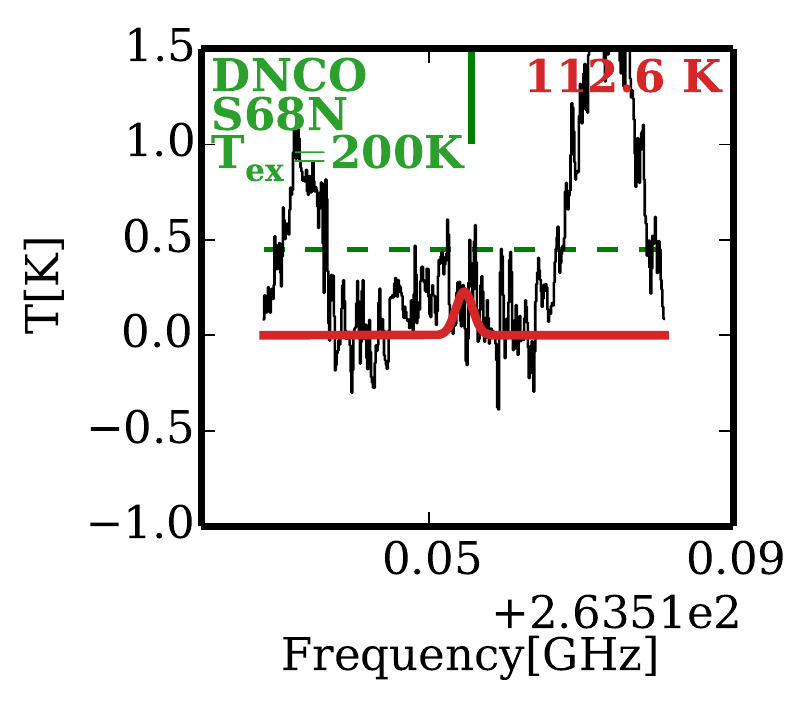}}
  \caption{Same as Fig.\ \ref{fig:HN13CO_fit_S68N} but for DNCO.}
  \label{fig:DNCO_fit_S68N}
\end{figure}

\begin{figure*}
    \centering
    \includegraphics[width=17cm]{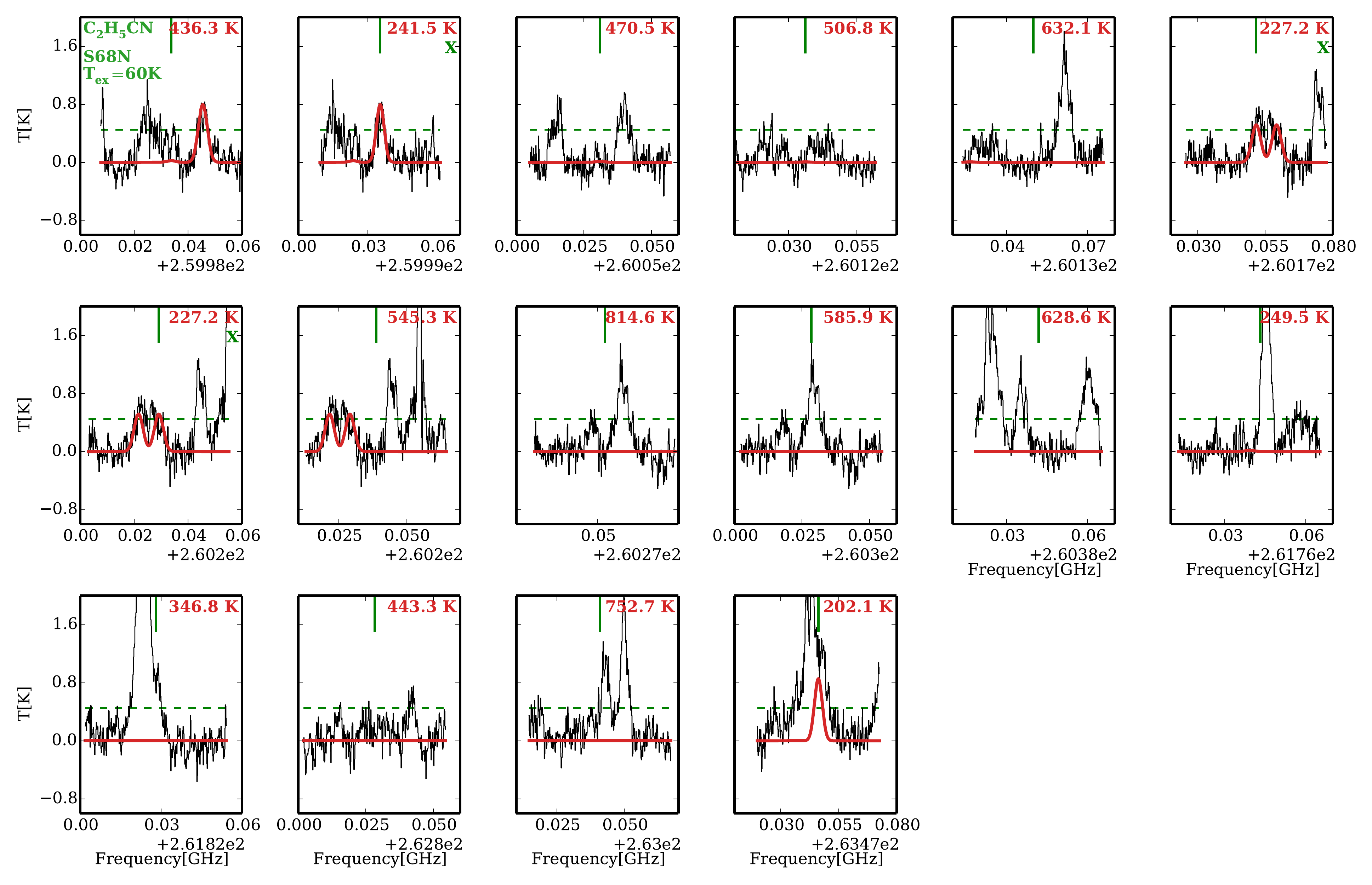}
    \caption{Same as Fig.\ \ref{fig:HN13CO_fit_S68N} but for $\rm C_{2}H_{5}CN$ and the excitation temperature used is the lower limit for $\rm C_{2}H_{5}CN$.}
    \label{fig:C2H5CN_fit_S68N_low}
\end{figure*}

\begin{figure*}
    \centering
    \includegraphics[width=17cm]{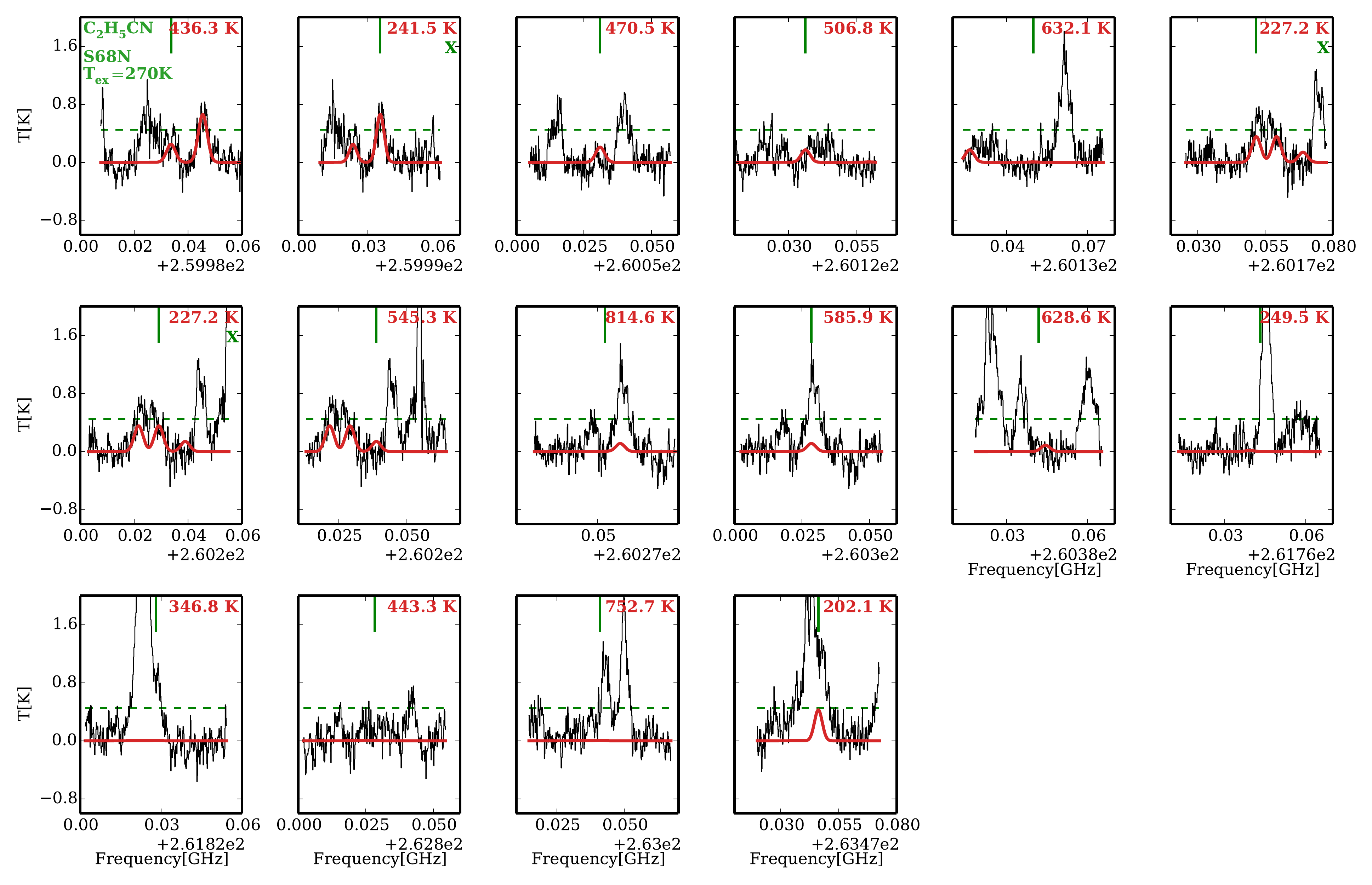}
    \caption{Same as Fig. \ref{fig:HN13CO_fit_S68N} but for $\rm C_{2}H_{5}CN$ and the excitation temperature used is the upper limit for $\rm C_{2}H_{5}CN$.}
    \label{fig:C2H5CN_fit_S68N_high}
\end{figure*}

\begin{figure*}
\centering
  \includegraphics[width=17cm]{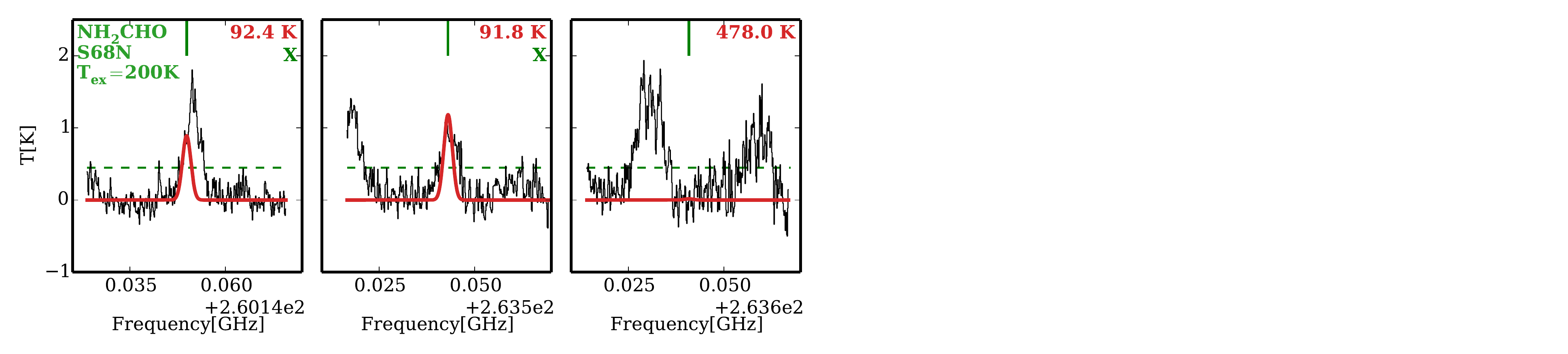}
  \caption{Same as Fig.\ \ref{fig:HN13CO_fit_S68N} but for $\rm NH_{2}CHO$.}
  \label{fig:NH2CHO_fit_S68N}
\end{figure*}

\begin{figure*}
    \centering
    \includegraphics[width=17cm]{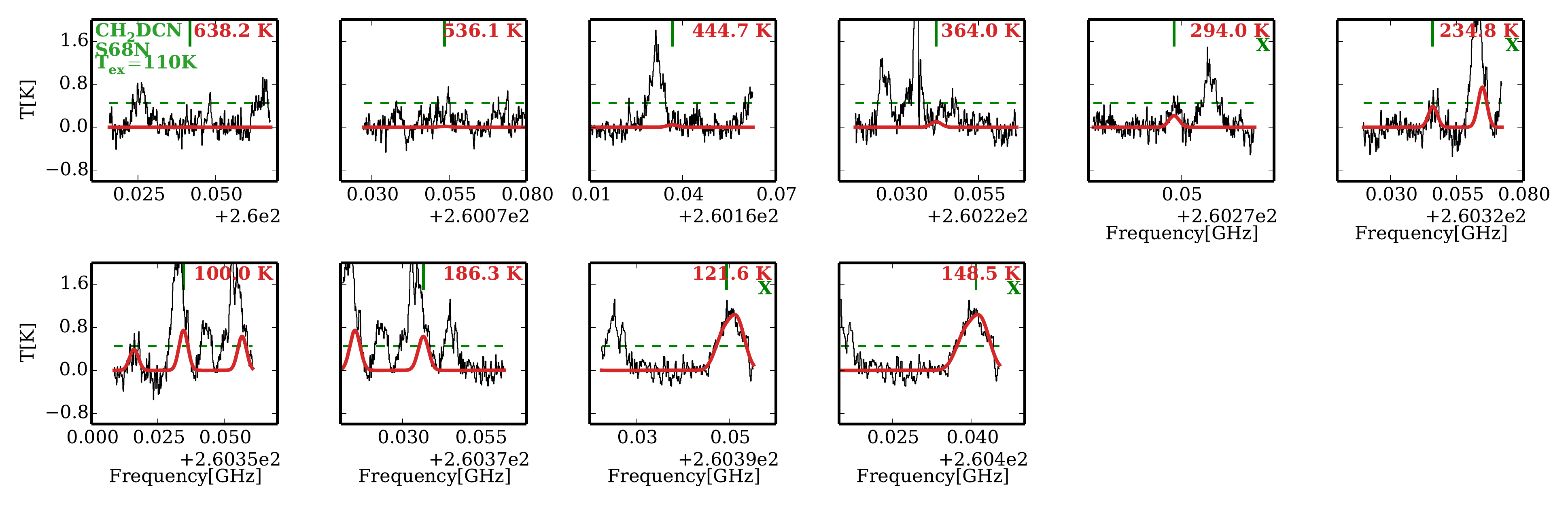}
    \caption{Same as Fig. \ref{fig:HN13CO_fit_S68N} but for $\rm CH_{2}DCN$ and the excitation temperature used is the lower limit for $\rm CH_{2}DCN$.}
    \label{fig:CH2DCN_fit_S68N_low}
\end{figure*}

\begin{figure*}
    \centering
    \includegraphics[width=17cm]{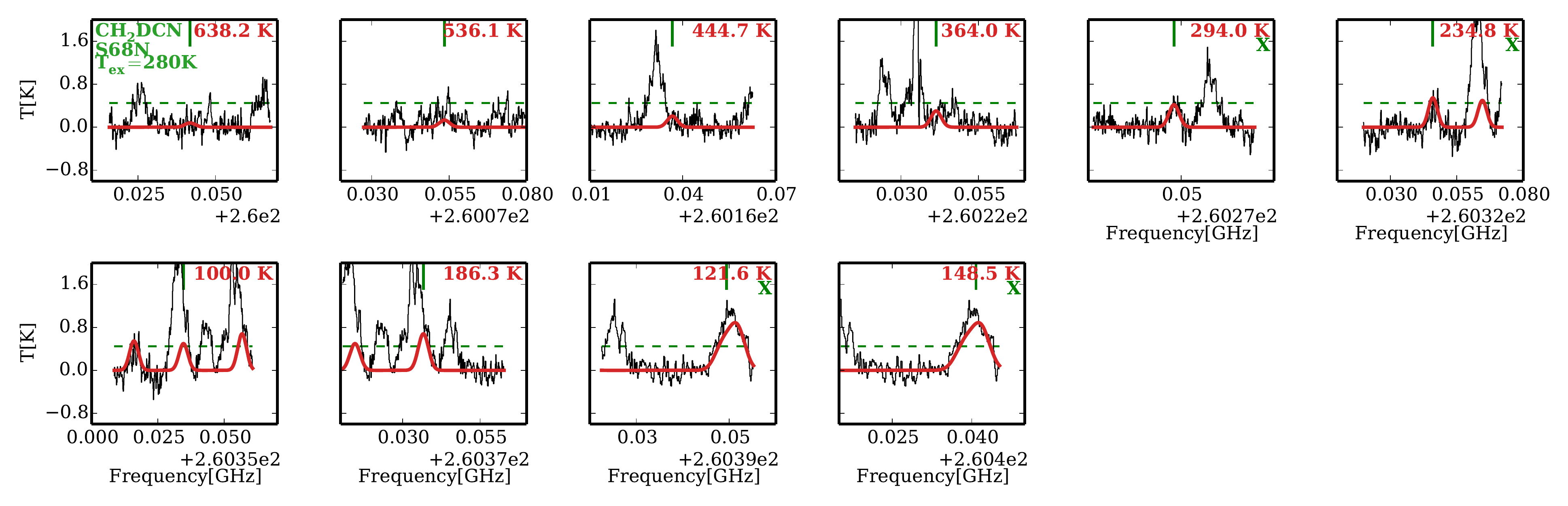}
    \caption{Same as Fig.\ \ref{fig:HN13CO_fit_S68N} but for $\rm CH_{2}DCN$ and the excitation temperature used is the upper limit for $\rm CH_{2}DCN$.}
    \label{fig:CH2DCN_fit_S68N_high}
\end{figure*}

\begin{figure*}
    \centering
    \includegraphics[width=17cm]{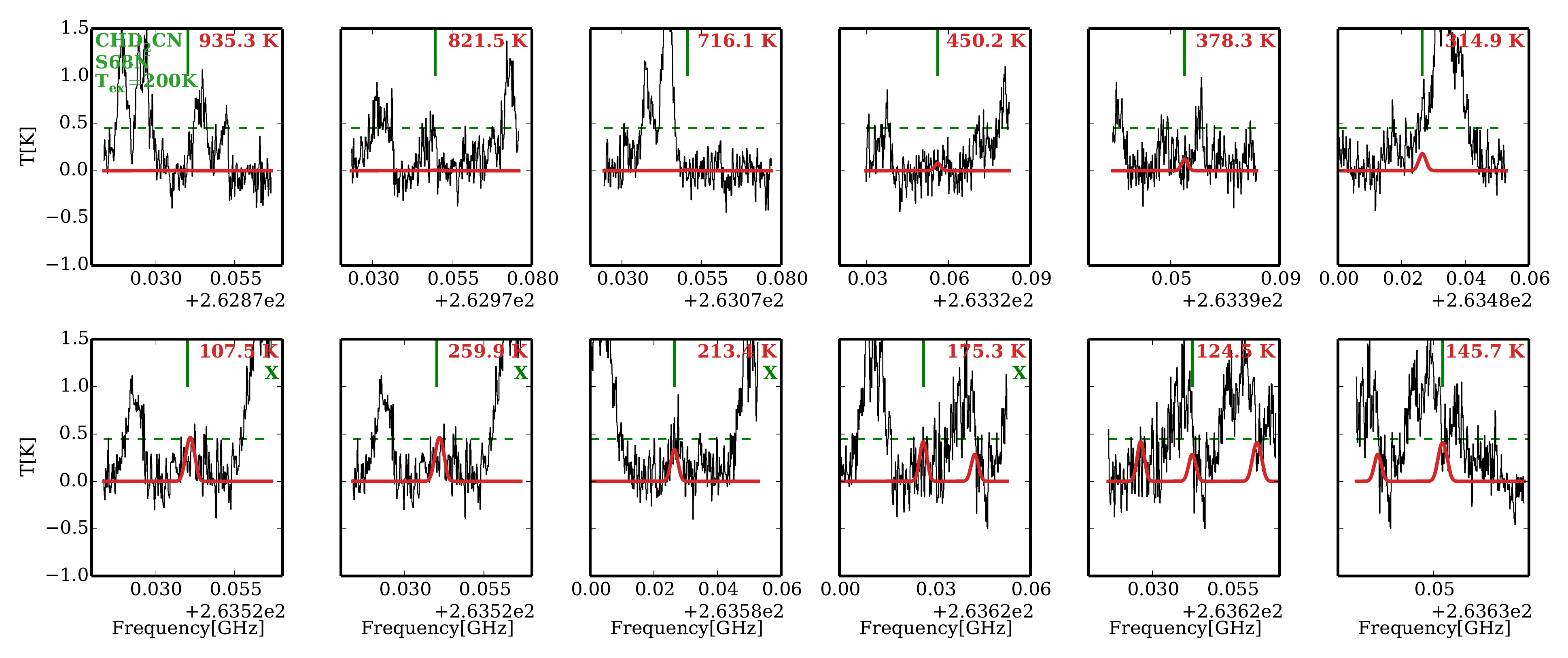}
    \caption{Same as Fig.\ \ref{fig:HN13CO_fit_S68N} but for $\rm CHD_{2}CN$.}
    \label{fig:CHD2CN_fit_S68N}
\end{figure*}

\begin{figure}
  \resizebox{\hsize}{!}{\includegraphics{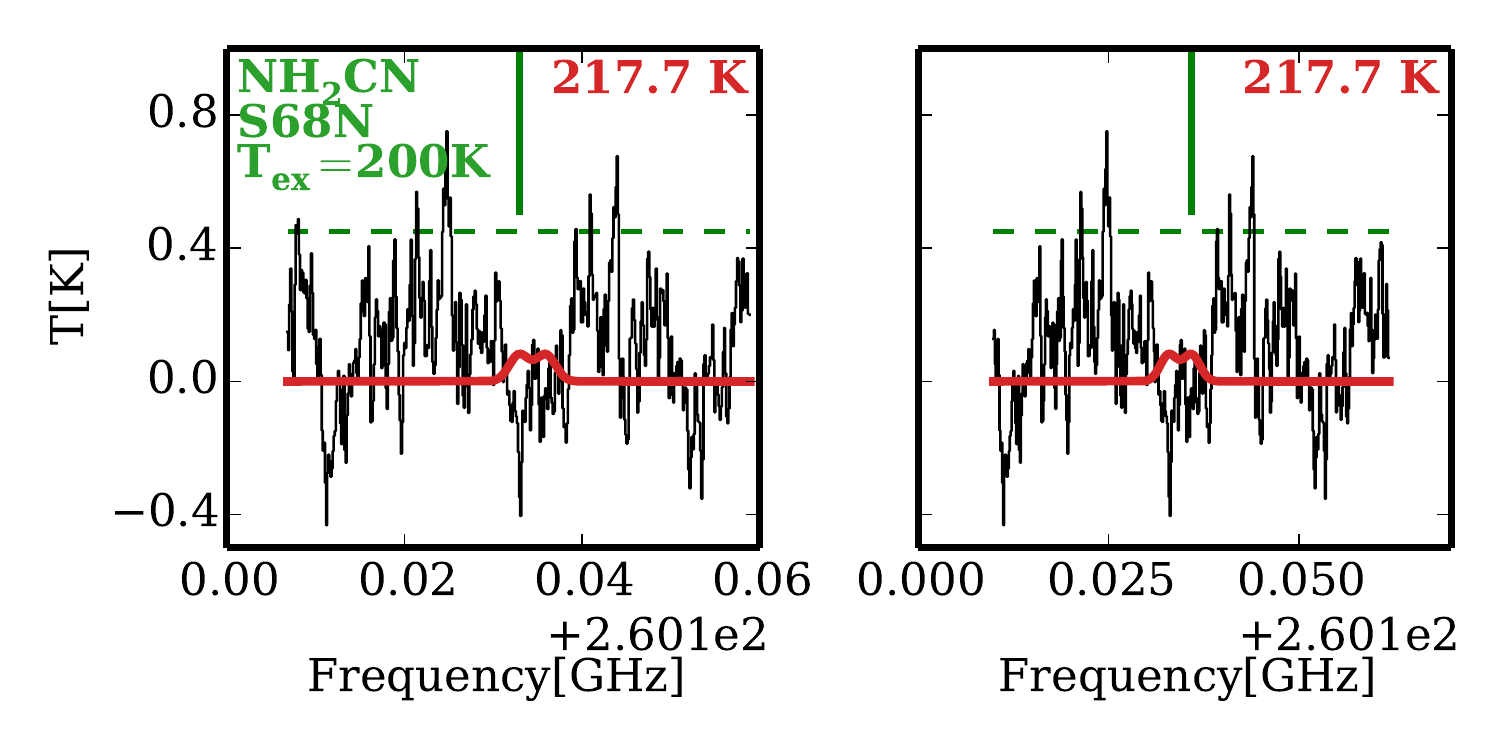}}
  \caption{Same as Fig. \ref{fig:HN13CO_fit_S68N} but for $\rm NH_{2}CN$.}
  \label{fig:NH2CN_fit_S68N}
\end{figure}

\begin{figure*}
    \centering
    \includegraphics[width=17cm]{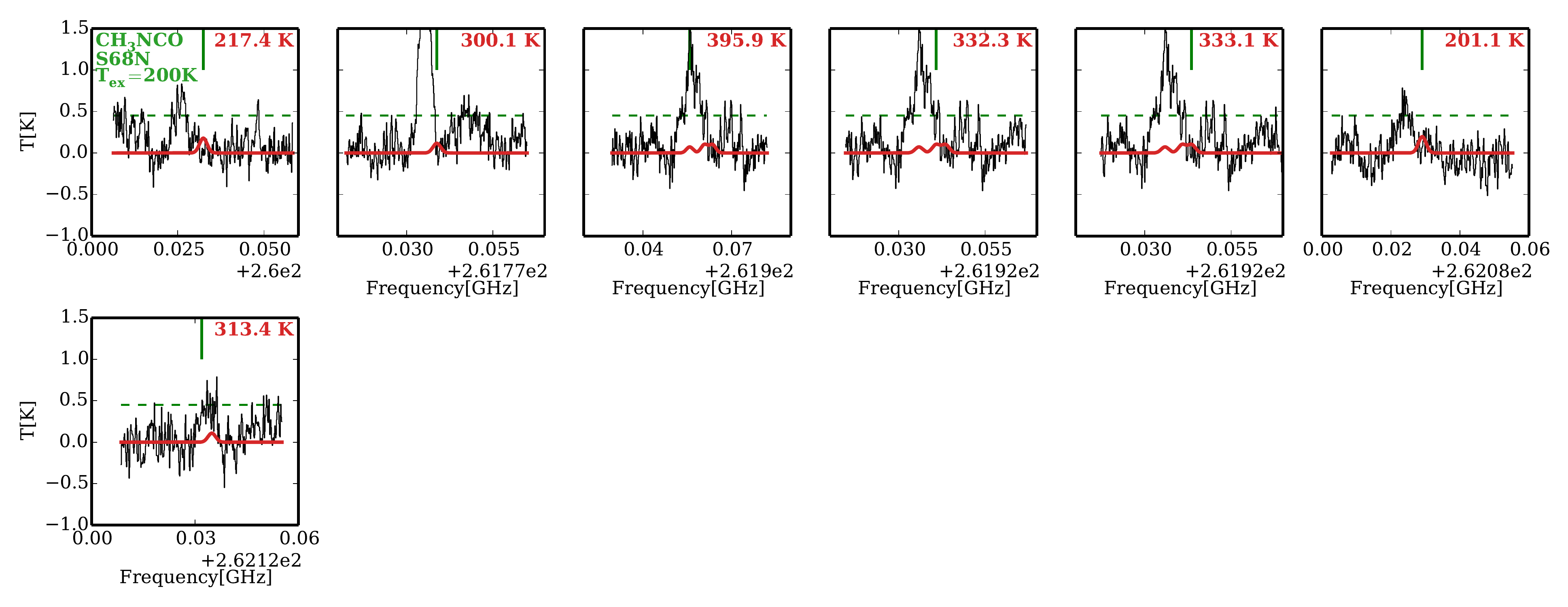}
    \caption{Same as Fig.\ \ref{fig:HN13CO_fit_S68N} but for $\rm CH_{3}NCO$.}
    \label{fig:CH3NCO_fit_S68N}
\end{figure*}

\begin{figure*}
    \centering
    \includegraphics[width=17cm]{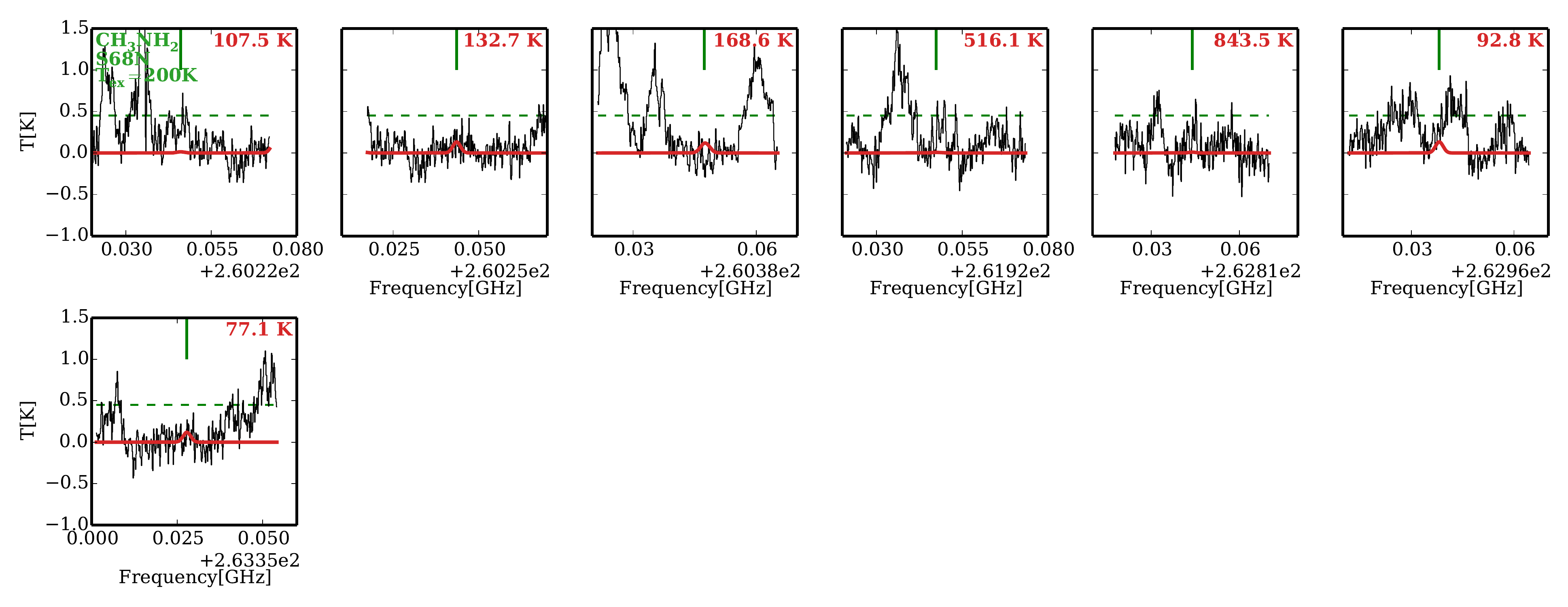}
    \caption{Same as Fig.\ \ref{fig:HN13CO_fit_S68N} but for $\rm CH_{3}NH_{2}$.}
    \label{fig:CH3NH2_fit_S68N}
\end{figure*}

The fitted models are presented in Figs. \ref{fig:HN13CO_fit} to \ref{fig:CH3NH2_fit_S68N} for B1-c and S68N. To derive the most accurate excitation temperatures, the $\chi^{2}$ plot for each molecule is examined. Figure \ref{fig:chi_squared} shows the $\chi^{2}$ plot for one of the molecules (C$_{2}$H$_{5}$CN) where the $\chi^{2}$ calculation is most constraining. When the $\chi^{2}$ method was not constraining or the results from fitting the spectrum by eye were not consistent with the $\chi^{2}$ results, the spectrum was fitted by eye. The fit by eye was done by fixing the column density to the best fitted value or changing it slightly around the best fitted value while varying the temperature in steps of 10\,K to find the best fit. Errors of ${\sim} 50$\,K to ${\sim} 100$\,K were typically found. All the excitation temperatures that were not fixed were fitted by eye due to line blending. Figures \ref{fig:HN13CO_fit}-\ref{fig:CH3NH2_fit} and \ref{fig:HN13CO_fit_S68N}-\ref{fig:CH3NH2_fit_S68N} present the range of fitted models for B1-c and S68N, including the upper and lower limit temperature fit for each molecule where the fit-by-eye method was used ($\rm NH_{2}CHO$, C$_2$H$_5$CN, HN$^{13}$CO, $\rm CH_{2}DCN$, and $\rm CHD_{2}CN$ towards B1-c and $\rm C_{2}H_{5}CN$ and $\rm CH_{2}DCN$ towards S68N).

In general, the model, as expected, does not predict a line for transitions with low $A_{ij}$. This can be seen for the NH$_2$CHO line with an E$_{\rm up}$ of 23.6\,K and an $A_{ij}$ of $1.2 \times 10^{-5}$ (Figs. \ref{fig:NH2CHO_fit_low} and \ref{fig:NH2CHO_fit_high}). Moreover, the line with an E$_{\rm up}$ of 478.0\,K has a large error in its frequency in the JPL catalogue and hence is not used in the fit.

In the specific case of C$_2$H$_5$CN towards B1-c (Figs. \ref{fig:C2H5CN_fit} and \ref{fig:C2H5CN_fitup}), the two lines with upper energy levels of $153.7$\,K and $233.6$\,K have the same frequencies and plausible $A_{ij}$ of $4.99 \times 10^{-4}$ and $4.07 \times 10^{-4}$, respectively. Hence it is not possible to say which of them is responsible for the emission line. We produced two models with the best-fit parameters for this molecule, once limiting the $E_{\rm up}$ of lines included in the model to below 160\,K and once to above this value to make sure that these two models do not include the two lines at the same time. We found that the two models show very similar line peak intensities, and thus the final best-fit model most likely includes emission from both of these lines.

\begin{figure*}
    \centering
    \includegraphics[width=17cm]{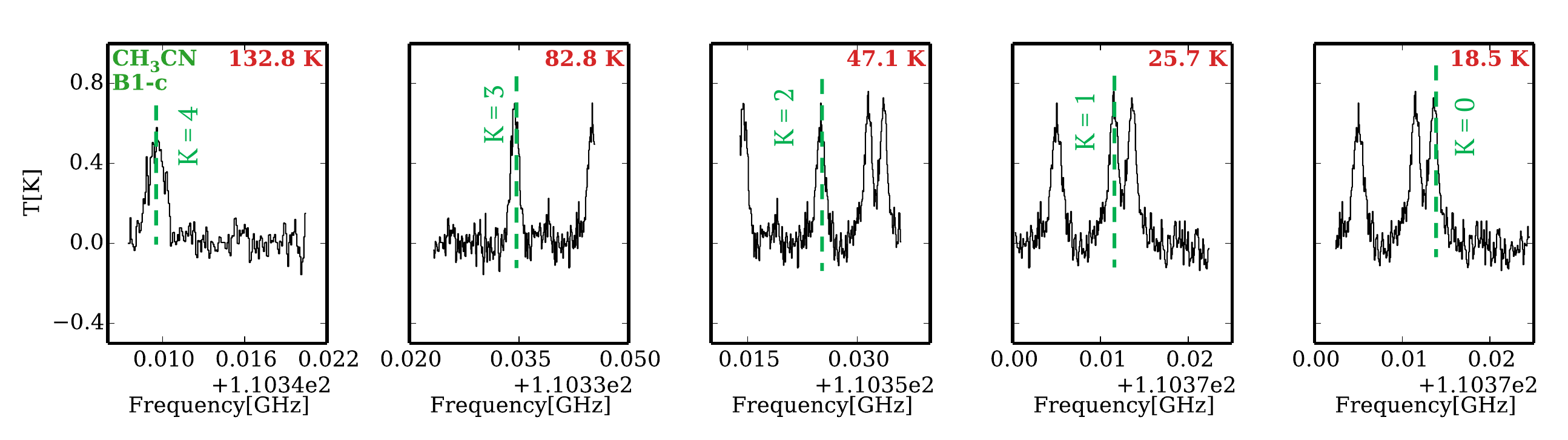}
    \caption{Spectrum for $\rm CH_{3}CN$ towards B1-c in the Band 3 data. The upper energy levels of each line are shown at the top right of each panel.}
    \label{fig:CH3CN}
\end{figure*}

\begin{figure*}
    \centering
    \includegraphics[width=17cm]{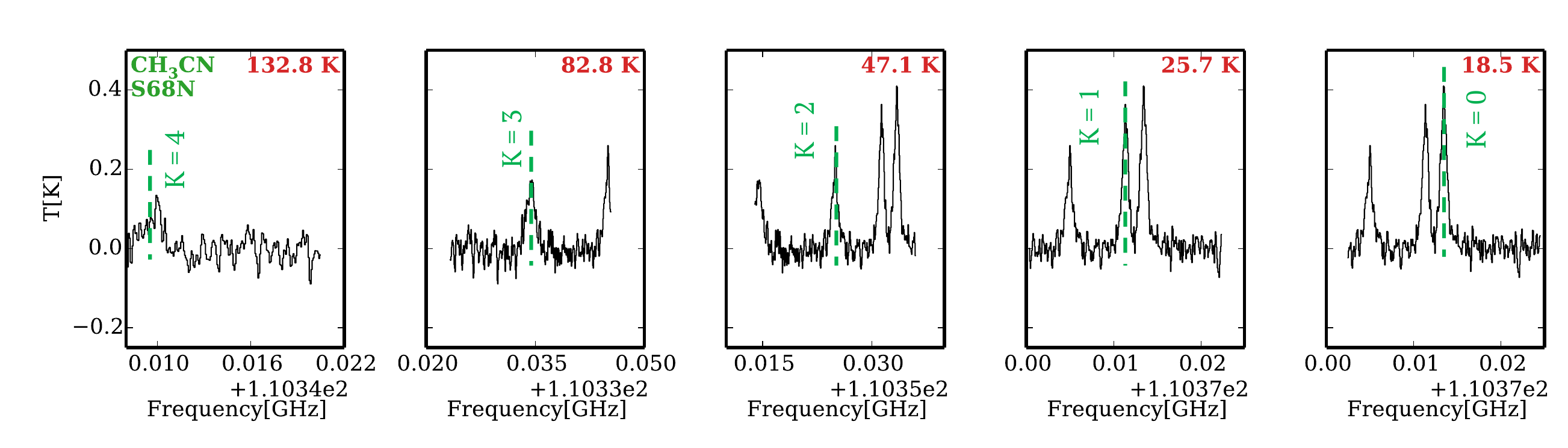}
    \caption{Spectrum for $\rm CH_{3}CN$ towards S68N in the Band 3 data. The upper energy levels of each line are shown at the top right of each panel.}
    \label{fig:CH3CN_S68N}
\end{figure*}

\begin{figure*}
\centering 
  \includegraphics[width=17cm]{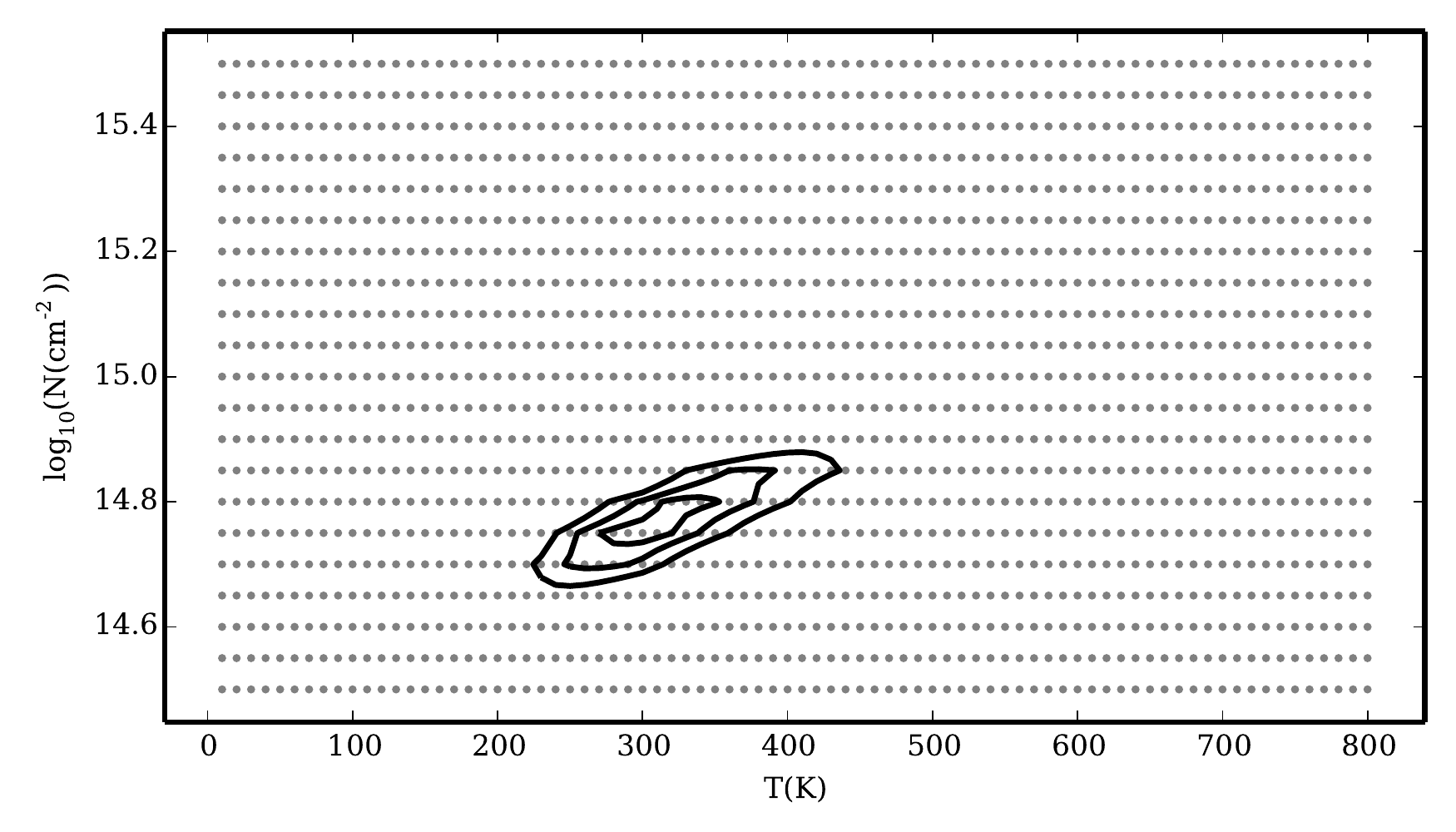}
  \caption{$\chi^{2}$ plot for $\rm C_{2}H_{5}CN$ for B1-c in the Band 5 and 6 data. The dots in the background show the grid used for this molecule, and the contours show the $3\sigma$, $2\sigma$, and $\sigma$ levels of the best fit. It is seen that T$_{\rm ex}$ is constrained to between ${\sim} 200$\,K and ${\sim} 400$\,K, whereas the column density is well determined at $(5.8 \pm 1.7) \times 10^{14}$. We note that the vibrational correction factor is not taken into account for this plot.}
  \label{fig:chi_squared}
\end{figure*}

\section{Additional tables}
\label{app:lines_tabs}

\begin{table*}
    \caption{Observational parameters of the data from observing programme 2017.1.01174.S.}
    \label{tab:obs_pars}
    \resizebox{\textwidth}{!}{\begin{tabular}{@{\extracolsep{1cm}}*{6}{c}}
          \toprule    
          & \multicolumn{2}{c}{B1-c}&\multicolumn{2}{c}{S68N}\\
          \cmidrule{2-3} \cmidrule{4-5}
                Parameters & Band 3 & Band 6 & Band 3 & Band 6 \\
                \midrule     
Configuration & C43-3 & C43-4 & C43-2 & C43-4\\
Beam (\arcsec) & $2.1 \times 1.4$ & $0.58 \times 0.39$ & $2.8 \times 1.8$ & $0.46 \times 0.42$\\
LAS (\arcsec) & 16 & 6 & 22 & 6 \\
$\rm \Delta V \, \rm (km \, s^{-1})$ & 0.2 / 0.4 & 0.14 & 0.17 / 0.32 & 0.14\\
Continuum rms (mJy) & 14.4 & 24.4 & 14.6 & 9.2\\
Line rms (K) & 0.06 & 0.14 & 0.04 & 0.15\\

\bottomrule
        \end{tabular}}
\end{table*}
\begin{table}
\caption{Frequency ranges covered in the Band 5 and 6 data.}              
\label{tab:freq}      
\centering                                      
\begin{tabular}{l l}          
\hline\hline                        
 ALMA Band & Frequency range (GHz) \\    
\hline                                   
    Band 5: & ${\sim} 172.65 {\sim} 172.71 $\\      
    & ${\sim} 173.48 {\sim} 173.54 $ \\
    & ${\sim} 173.65 {\sim} 174.58 $\\
    & ${\sim} 186.55 {\sim} 188.43 $\\
    Band 6: & ${\sim} 259.98 {\sim} 260.45 $ \\
    & ${\sim} 261.71 {\sim} 262.18 $\\
    & ${\sim} 262.70 {\sim} 263.19 $\\
    & ${\sim} 263.30 {\sim} 263.77 $\\
\hline   
\end{tabular}
\end{table}


\begin{table}
\caption{Characteristics of the present lines towards B1-c.}              
\label{tab:B1c_Nfitted}      
\centering                                      
\begin{tabular}{l l l l l}          
\hline\hline                        
Species & $E_\mathrm{up}$ (K) range\tablefootmark{a} & Transitions\tablefootmark{b} & FWHM (km/s) \tablefootmark{c}  \\    
\hline                                   
    HNCO & 82.3-732.9 & 5 & --\\      
    HN$^{13}$CO & 82.3-772.9 & 6 & 3.0 \\
    DNCO & 112.6-112.6 &  1 & 3.0\\
    NH$_{2}$CHO & 23.6-478.0 & 6 & 3.0\\
    C$_{2}$H$_{5}$CN & 25.5-939.2 & 65 & 3.0 \\
    CH$_{3}$CN & -- & --  & 3.0 \\
    CH$_{2}$DCN & 51.4-638.2 & 15 &  3.0\\
    CHD$_{2}$CN & 28.8-976.3 & 22 & 3.0 \\
    NH$_{2}$CN & 217.7-217.7 & 2 & 3.0 \\
    CH$_{3}$NCO & 111.3-395.9 & 18 & 3.0\\
    CH$_{3}$NH$_{2}$ & 48.6-843.5 & 26 & 3.0\\
\hline  
\end{tabular}
\tablefoot{\tablefoottext{a}{Upper energy level ranges below 1000\,K for the lines present in the less noisy spectral windows of the Band 5 and 6 data.}
\tablefoottext{b}{Number of transitions with upper energy levels below 1000\,K in the less noisy spectral windows of the Band 5 and 6 data.}
\tablefoottext{c}{The FWHM used in the final best-fit model for each molecule.}}
\end{table}

\begin{table}
\caption{Characteristics of the present lines towards S68N.}             
\label{tab:S68N_Nfitted}      
\centering                                      
\begin{tabular}{l l l l l}          
\hline\hline                        
Species & $E_\mathrm{up}$ (K) range \tablefootmark{a}  & Transitions \tablefootmark{b} & FWHM (km/s) \tablefootmark{c} \\    
\hline                                   
    HNCO & 82.3-732.9 & 5 & -- \\      
    HN$^{13}$CO & 82.3-772.9 & 6 & 4.5 \\
    DNCO & 112.6-112.6 &  1 & 3.0 \\
    NH$_{2}$CHO & 91.8-478.0 & 3 & 3.0 \\
    C$_{2}$H$_{5}$CN & 56.2-814.6 & 22 & 4.5 \\
    CH$_{3}$CN & -- &  -- & 3.0 \\
    CH$_{2}$DCN & 100-638.2 &  10 & 4.5 \\
    CHD$_{2}$CN & 28.8-935.3 & 17 & 3.0\\
    NH$_{2}$CN & 217.7-217.7 & 2 & 3.0\\
    CH$_{3}$NCO & 201.1-395.9  & 7 & 3.0\\
    CH$_{3}$NH$_{2}$ & 48.6-843.5 & 10 & 3.0 \\
\hline                                             
\end{tabular}
\tablefoot{\tablefoottext{a}{Upper energy level ranges below 1000\,K for the lines present in the Band 6 data.}
\tablefoottext{b}{Number of transitions with upper energy levels below 1000\,K in the Band 6 data.}
\tablefoottext{c}{The FWHM used in the final best-fit model for each molecule.}}
\end{table}

\clearpage
\onecolumn

\setcounter{table}{2}
\begin{longtable}{lllll}
\caption{Lines of the species in the Band 3 data.} \\
\hline \hline
Species & Transition & Frequency & $A_{ij}$ & $E_\mathrm{up}$ \\
 & J K L (M) & (MHz) & (s$^{-1}$) & (K) \\
\hline
\endfirsthead
\caption{continued.}\\
\hline\hline
Species & Transition & Frequency & $A_{ij}$ & $E_\mathrm{up}$ \\
 & J K L (M) & (GHz) & (s$^{-1}$) & (K) \\
\hline
\endhead
\hline
\endfoot

HNCO & 5 4 1 - 4 4 0 & 109\,778.752 & $5.20 \times 10^{-6}$ & 666.6\\ 
  & 5 4 2 - 4 4 1 & 109\,778.752 & $5.20 \times 10^{-6}$ & 666.6\\ 
  & 5 0 5 - 4 0 4 & 109\,905.749 & $1.75 \times 10^{-5}$ & 15.8\\ 
\hline 
$\rm CH_{3}CN$ & 6 4 0 - 5 4 0 & 110\,349.471 & $6.17 \times 10^{-5}$ & 132.8\\ 
  & 6 3 0 - 5 -3 0 & 110\,364.354 & $8.33 \times 10^{-5}$ & 82.8\\ 
  & 6 -3 0 - 5 3 0 & 110\,364.354 & $8.33 \times 10^{-5}$ & 82.8\\ 
  & 6 2 0 - 5 2 0 & 110\,374.989 & $9.88 \times 10^{-5}$ & 47.1\\ 
  & 6 1 0 - 5 1 0 & 110\,381.372 & $1.08 \times 10^{-4}$ & 25.7\\ 
  & 6 0 0 - 5 0 0 & 110\,383.5 & $1.11 \times 10^{-4}$ & 18.5
\label{tab:lines_band3}
\end{longtable}

\setcounter{table}{2}
\begin{longtable}{lllll}
\caption{Lines of the species in the Band 5 and 6 data.} \\
\hline \hline
Species & Transition & Frequency & $A_{ij}$ & $E_\mathrm{up}$ \\
 & J K L (M) & (MHz) & (s$^{-1}$) & (K) \\
\hline
\endfirsthead
\caption{continued.}\\
\hline\hline
Species & Transition & Frequency & $A_{ij}$ & $E_\mathrm{up}$ \\
 & J K L (M) & (GHz) & (s$^{-1}$) & (K) \\
\hline
\endhead
\hline
\endfoot

HNCO & 12 1 12 - 11 1 11 & 262\,769.477 & $2.48 \times 10^{-4}$ & 125.3\\ 
  & 12 4 8 - 11 4 7 & 263\,449.24 & $1.87 \times 10^{-4}$ & 732.9\\ 
  & 12 4 9 - 11 4 8 & 263\,449.24 & $1.87 \times 10^{-4}$ & 732.9\\ 
  & 12 3 10 - 11 3 9 & 263\,580.924 & $2.16 \times 10^{-4}$ & 457.2\\ 
  & 12 3 9 - 11 3 8 & 263\,580.928 & $2.16 \times 10^{-4}$ & 457.2\\ 
  & 12 2 11 - 11 2 10 & 263\,672.912 & $2.37 \times 10^{-4}$ & 252.5\\ 
  & 12 2 10 - 11 2 9 & 263\,678.709 & $2.37 \times 10^{-4}$ & 252.5\\ 
  & 12 0 12 - 11 0 11 & 263\,748.625 & $2.56 \times 10^{-4}$ & 82.3\\ 
\hline 
$\rm HN^{13}CO$ & 12 1 12 11 - 11 1 11 11 & 262\,774.075 & $1.95 \times 10^{-6}$ & 125.1\\ 
  & 12 1 12 13 - 11 1 11 12 & 262\,775.057 & $2.58 \times 10^{-4}$ & 125.1\\ 
  & 12 1 12 12 - 11 1 11 11 & 262\,775.062 & $2.57 \times 10^{-4}$ & 125.1\\ 
  & 12 1 12 11 - 11 1 11 10 & 262\,775.064 & $2.56 \times 10^{-4}$ & 125.1\\ 
  & 12 1 12 12 - 11 1 11 12 & 262\,775.968 & $1.79 \times 10^{-6}$ & 125.1\\ 
  & 12 4 8 11 - 11 4 7 11 & 263\,455.652 & $1.76 \times 10^{-6}$ & 772.9\\ 
  & 12 4 9 11 - 11 4 8 11 & 263\,455.652 & $1.76 \times 10^{-6}$ & 772.9\\ 
  & 12 4 9 11 - 11 4 8 10 & 263\,456.155 & $2.31 \times 10^{-4}$ & 772.9\\ 
  & 12 4 8 11 - 11 4 7 10 & 263\,456.155 & $2.31 \times 10^{-4}$ & 772.9\\ 
  & 12 4 9 13 - 11 4 8 12 & 263\,456.155 & $2.33 \times 10^{-4}$ & 772.9\\ 
  & 12 4 8 13 - 11 4 7 12 & 263\,456.155 & $2.33 \times 10^{-4}$ & 772.9\\ 
  & 12 4 8 12 - 11 4 7 11 & 263\,456.198 & $2.31 \times 10^{-4}$ & 772.9\\ 
  & 12 4 9 12 - 11 4 8 11 & 263\,456.198 & $2.31 \times 10^{-4}$ & 772.9\\ 
  & 12 4 8 12 - 11 4 7 12 & 263\,456.659 & $1.62 \times 10^{-6}$ & 772.9\\ 
  & 12 4 9 12 - 11 4 8 12 & 263\,456.659 & $1.62 \times 10^{-6}$ & 772.9\\ 
  & 12 3 10 11 - 11 3 9 11 & 263\,588.502 & $1.86 \times 10^{-6}$ & 470.8\\ 
  & 12 3 9 11 - 11 3 8 11 & 263\,588.505 & $1.86 \times 10^{-6}$ & 470.8\\ 
  & 12 3 10 13 - 11 3 9 12 & 263\,589.13 & $2.46 \times 10^{-4}$ & 470.8\\ 
  & 12 3 10 11 - 11 3 9 10 & 263\,589.132 & $2.44 \times 10^{-4}$ & 470.8\\ 
  & 12 3 9 13 - 11 3 8 12 & 263\,589.132 & $2.46 \times 10^{-4}$ & 470.8\\ 
  & 12 3 9 11 - 11 3 8 10 & 263\,589.135 & $2.44 \times 10^{-4}$ & 470.8\\ 
  & 12 3 10 12 - 11 3 9 11 & 263\,589.154 & $2.44 \times 10^{-4}$ & 470.8\\ 
  & 12 3 9 12 - 11 3 8 11 & 263\,589.157 & $2.44 \times 10^{-4}$ & 470.8\\ 
  & 12 3 10 12 - 11 3 9 12 & 263\,589.731 & $1.71 \times 10^{-6}$ & 470.8\\ 
  & 12 3 9 12 - 11 3 8 12 & 263\,589.734 & $1.71 \times 10^{-6}$ & 470.8\\ 
  & 12 2 11 11 - 11 2 10 11 & 263\,680.119 & $1.93 \times 10^{-6}$ & 254.9\\ 
  & 12 2 11 13 - 11 2 10 12 & 263\,680.835 & $2.56 \times 10^{-4}$ & 254.9\\ 
  & 12 2 11 11 - 11 2 10 10 & 263\,680.839 & $2.54 \times 10^{-4}$ & 254.9\\ 
  & 12 2 11 12 - 11 2 10 11 & 263\,680.847 & $2.54 \times 10^{-4}$ & 254.9\\ 
  & 12 2 11 12 - 11 2 10 12 & 263\,681.508 & $1.77 \times 10^{-6}$ & 254.9\\ 
  & 12 2 10 11 - 11 2 9 11 & 263\,685.825 & $1.93 \times 10^{-6}$ & 254.9\\ 
  & 12 2 10 13 - 11 2 9 12 & 263\,686.54 & $2.56 \times 10^{-4}$ & 254.9\\ 
  & 12 2 10 11 - 11 2 9 10 & 263\,686.544 & $2.54 \times 10^{-4}$ & 254.9\\ 
  & 12 2 10 12 - 11 2 9 11 & 263\,686.552 & $2.54 \times 10^{-4}$ & 254.9\\ 
  & 12 2 10 12 - 11 2 9 12 & 263\,687.21 & $1.77 \times 10^{-6}$ & 254.9\\ 
  & 12 0 12 11 - 11 0 11 11 & 263\,755.182 & $1.99 \times 10^{-6}$ & 82.3\\ 
  & 12 0 12 11 - 11 0 11 12 & 263\,755.909 & $3.45 \times 10^{-9}$ & 82.3\\ 
  & 12 0 12 13 - 11 0 11 12 & 263\,755.97 & $2.63 \times 10^{-4}$ & 82.3\\ 
  & 12 0 12 12 - 11 0 11 11 & 263\,755.972 & $2.61 \times 10^{-4}$ & 82.3\\ 
  & 12 0 12 11 - 11 0 11 10 & 263\,755.975 & $2.61 \times 10^{-4}$ & 82.3\\ 
  & 12 0 12 12 - 11 0 11 12 & 263\,756.699 & $1.83 \times 10^{-6}$ & 82.3\\ 
\hline 
DNCO & 13 1 13 12 - 12 1 12 12 & 263\,563.661 & $1.67 \times 10^{-6}$ & 112.6\\ 
  & 13 1 13 14 - 12 1 12 13 & 263\,564.594 & $2.62 \times 10^{-4}$ & 112.6\\ 
  & 13 1 13 13 - 12 1 12 12 & 263\,564.598 & $2.60 \times 10^{-4}$ & 112.6\\ 
  & 13 1 13 12 - 12 1 12 11 & 263\,564.599 & $2.60 \times 10^{-4}$ & 112.6\\ 
  & 13 1 13 13 - 12 1 12 13 & 263\,565.464 & $1.55 \times 10^{-6}$ & 112.6\\ 
\hline 
$\rm NH_{2}CHO$ & 6 1 6 5 - 5 0 5 4 & 173\,772.286 & $1.22 \times 10^{-5}$ & 23.6\\ 
  & 6 1 6 7 - 5 0 5 6 & 173\,772.398 & $1.27 \times 10^{-5}$ & 23.6\\ 
  & 6 1 6 6 - 5 0 5 5 & 173\,773.239 & $1.23 \times 10^{-5}$ & 23.6\\ 
  & 19 2 17 20 - 18 3 16 19 & 174\,482.455 & $4.46 \times 10^{-6}$ & 209.7\\ 
  & 9 0 9 10 - 8 0 8 9 & 187\,589.055 & $4.75 \times 10^{-4}$ & 45.4\\ 
  & 9 0 9 8 - 8 0 8 7 & 187\,589.059 & $4.68 \times 10^{-4}$ & 45.4\\ 
  & 9 0 9 9 - 8 0 8 8 & 187\,589.158 & $4.69 \times 10^{-4}$ & 45.4\\ 
  & 12 2 10 - 11 2 9 & 260\,189.848 & $1.25 \times 10^{-3}$ & 92.4\\ 
  & 13 1 13 - 12 1 12 & 263\,543.025 & $1.33 \times 10^{-3}$ & 91.8\\ 
  & 29 3 26 29 - 29 2 27 29 & 263\,640.492 & $6.36 \times 10^{-5}$ & 478.0\\ 
  & 29 3 26 30 - 29 2 27 30 & 263\,640.985 & $6.37 \times 10^{-5}$ & 478.0\\ 
  & 29 3 26 28 - 29 2 27 28 & 263\,641.002 & $6.37 \times 10^{-5}$ & 478.0\\ 
\hline 
$\rm C_{2}H_{5}CN$ & 67 4 63 - 66 6 60 & 173\,479.725 & $3.59 \times 10^{-7}$ & 1006.9\\ 
  & 68 10 58 - 69 8 61 & 173\,481.563 & $1.22 \times 10^{-7}$ & 1119.8\\ 
  & 53 3 51 - 54 0 54 & 173\,496.547 & $8.35 \times 10^{-8}$ & 619.7\\ 
  & 44 13 31 - 45 12 34 & 173\,505.22 & $6.02 \times 10^{-6}$ & 613.1\\ 
  & 44 13 32 - 45 12 33 & 173\,505.22 & $6.02 \times 10^{-6}$ & 613.1\\ 
  & 10 3 8 - 10 1 9 & 173\,739.951 & $1.99 \times 10^{-7}$ & 33.7\\ 
  & 14 2 13 - 13 1 12 & 173\,740.541 & $1.64 \times 10^{-5}$ & 49.4\\ 
  & 19 2 17 - 18 2 16 & 173\,904.151 & $4.37 \times 10^{-4}$ & 87.3\\ 
  & 8 6 3 - 9 5 4 & 174\,114.528 & $1.84 \times 10^{-6}$ & 55.5\\ 
  & 8 6 2 - 9 5 5 & 174\,114.533 & $1.84 \times 10^{-6}$ & 55.5\\ 
  & 57 4 53 - 56 6 50 & 174\,241.737 & $5.62 \times 10^{-7}$ & 738.0\\ 
  & 58 6 52 - 58 5 53 & 174\,479.058 & $3.53 \times 10^{-5}$ & 779.8\\ 
  & 38 5 33 - 38 4 34 & 174\,518.373 & $2.90 \times 10^{-5}$ & 347.4\\ 
  & 64 4 60 - 63 6 57 & 186\,317.795 & $5.52 \times 10^{-7}$ & 922.2\\ 
  & 38 7 31 - 39 5 34 & 186\,652.738 & $1.10 \times 10^{-7}$ & 373.3\\ 
  & 54 6 48 - 54 5 49 & 186\,763.733 & $4.01 \times 10^{-5}$ & 681.3\\ 
  & 22 9 14 - 23 8 15 & 186\,905.217 & $5.85 \times 10^{-6}$ & 198.8\\ 
  & 22 9 13 - 23 8 16 & 186\,905.217 & $5.85 \times 10^{-6}$ & 198.8\\ 
  & 8 3 6 - 7 2 5 & 186\,924.942 & $2.24 \times 10^{-5}$ & 25.5\\ 
  & 60 4 56 - 59 6 53 & 187\,192.928 & $6.78 \times 10^{-7}$ & 814.6\\ 
  & 58 7 52 - 57 8 49 & 187\,379.431 & $9.51 \times 10^{-6}$ & 791.6\\ 
  & 32 1 31 - 32 1 32 & 187\,437.247 & $2.18 \times 10^{-6}$ & 228.5\\ 
  & 60 5 55 - 60 4 56 & 187\,502.511 & $3.90 \times 10^{-5}$ & 823.6\\ 
  & 32 1 31 - 32 0 32 & 187\,608.572 & $1.25 \times 10^{-5}$ & 228.5\\ 
  & 56 2 54 - 57 1 57 & 187\,643.268 & $9.43 \times 10^{-8}$ & 689.0\\ 
  & 10 2 8 - 9 0 9 & 187\,663.517 & $6.94 \times 10^{-7}$ & 28.2\\ 
  & 35 3 33 - 35 2 34 & 187\,842.025 & $2.26 \times 10^{-5}$ & 280.5\\ 
  & 22 0 22 - 21 1 21 & 187\,845.577 & $4.68 \times 10^{-5}$ & 106.1\\ 
  & 90 13 78 - 89 14 75 & 187\,874.315 & $9.72 \times 10^{-6}$ & 1945.3\\ 
  & 90 13 77 - 89 14 76 & 187\,880.086 & $9.72 \times 10^{-6}$ & 1945.3\\ 
  & 8 3 5 - 7 2 6 & 187\,884.105 & $2.26 \times 10^{-5}$ & 25.5\\ 
  & 63 8 55 - 62 9 54 & 187\,931.933 & $9.70 \times 10^{-6}$ & 939.2\\ 
  & 39 4 35 - 38 5 34 & 188\,030.172 & $1.01 \times 10^{-5}$ & 356.1\\ 
  & 27 10 18 - 28 9 19 & 188\,061.915 & $6.51 \times 10^{-6}$ & 273.6\\ 
  & 27 10 17 - 28 9 20 & 188\,061.915 & $6.51 \times 10^{-6}$ & 273.6\\ 
  & 34 5 29 - 34 4 30 & 188\,125.196 & $3.39 \times 10^{-5}$ & 284.2\\ 
  & 21 9 12 - 20 9 11 & 188\,151.024 & $4.58 \times 10^{-4}$ & 189.3\\ 
  & 21 9 13 - 20 9 12 & 188\,151.024 & $4.58 \times 10^{-4}$ & 189.3\\ 
  & 21 8 13 - 20 8 12 & 188\,155.826 & $4.80 \times 10^{-4}$ & 170.4\\ 
  & 21 8 14 - 20 8 13 & 188\,155.826 & $4.80 \times 10^{-4}$ & 170.4\\ 
  & 21 10 11 - 20 10 10 & 188\,161.248 & $4.34 \times 10^{-4}$ & 210.4\\ 
  & 21 10 12 - 20 10 11 & 188\,161.248 & $4.34 \times 10^{-4}$ & 210.4\\ 
  & 21 11 10 - 20 11 9 & 188\,182.647 & $4.07 \times 10^{-4}$ & 233.6\\ 
  & 21 11 11 - 20 11 10 & 188\,182.647 & $4.07 \times 10^{-4}$ & 233.6\\ 
  & 21 7 15 - 20 7 14 & 188\,182.647 & $4.99 \times 10^{-4}$ & 153.8\\ 
  & 21 7 14 - 20 7 13 & 188\,182.647 & $4.99 \times 10^{-4}$ & 153.8\\ 
  & 21 12 9 - 20 12 8 & 188\,212.59 & $3.78 \times 10^{-4}$ & 259.1\\ 
  & 21 12 10 - 20 12 9 & 188\,212.59 & $3.78 \times 10^{-4}$ & 259.1\\ 
  & 21 6 16 - 20 6 15 & 188\,245.402 & $5.16 \times 10^{-4}$ & 139.4\\ 
  & 21 6 15 - 20 6 14 & 188\,245.402 & $5.16 \times 10^{-4}$ & 139.4\\ 
  & 56 3 54 - 57 0 57 & 188\,247.954 & $9.52 \times 10^{-8}$ & 689.0\\ 
  & 21 13 8 - 20 13 7 & 188\,249.82 & $3.47 \times 10^{-4}$ & 286.7\\ 
  & 21 13 9 - 20 13 8 & 188\,249.82 & $3.47 \times 10^{-4}$ & 286.7\\ 
  & 21 14 7 - 20 14 6 & 188\,293.433 & $3.13 \times 10^{-4}$ & 316.5\\ 
  & 21 14 8 - 20 14 7 & 188\,293.433 & $3.13 \times 10^{-4}$ & 316.5\\ 
  & 21 3 19 - 20 3 18 & 188\,327.62 & $5.51 \times 10^{-4}$ & 109.4\\ 
  & 21 15 6 - 20 15 5 & 188\,342.625 & $2.76 \times 10^{-4}$ & 348.5\\ 
  & 21 15 7 - 20 15 6 & 188\,342.625 & $2.76 \times 10^{-4}$ & 348.5\\ 
  & 29 6 24 - 30 4 27 & 188\,354.152 & $9.34 \times 10^{-8}$ & 227.2\\ 
  & 21 5 17 - 20 5 16 & 188\,370.494 & $5.31 \times 10^{-4}$ & 127.2\\ 
  & 21 5 16 - 20 5 15 & 188\,378.708 & $5.31 \times 10^{-4}$ & 127.2\\ 
  & 21 16 5 - 20 16 4 & 188\,397.003 & $2.36 \times 10^{-4}$ & 382.7\\ 
  & 21 16 6 - 20 16 5 & 188\,397.003 & $2.36 \times 10^{-4}$ & 382.7\\ 
  & 63 4 59 - 62 6 56 & 188\,398.285 & $6.07 \times 10^{-7}$ & 894.7\\ 
  & 84 4 80 - 85 3 83 & 188\,416.325 & $2.91 \times 10^{-7}$ & 1554.4\\ 
  & 26 5 22 - 27 2 25 & 188\,422.769 & $1.20 \times 10^{-7}$ & 178.8\\ 
  & 68 18 50 - 69 17 53 & 188\,436.437 & $8.17 \times 10^{-6}$ & 1366.8\\ 
  & 68 18 51 - 69 17 52 & 188\,436.437 & $8.17 \times 10^{-6}$ & 1366.8\\ 
  & 29 15 14 - 28 15 13 & 260\,013.644 & $1.09 \times 10^{-3}$ & 436.3\\ 
  & 29 15 15 - 28 15 14 & 260\,013.644 & $1.09 \times 10^{-3}$ & 436.3\\ 
  & 29 7 23 - 28 7 22 & 260\,025.426 & $1.40 \times 10^{-3}$ & 241.5\\ 
  & 29 7 22 - 28 7 21 & 260\,025.426 & $1.40 \times 10^{-3}$ & 241.5\\ 
  & 29 16 13 - 28 16 12 & 260\,081.02 & $1.04 \times 10^{-3}$ & 470.5\\ 
  & 29 16 14 - 28 16 13 & 260\,081.02 & $1.04 \times 10^{-3}$ & 470.5\\ 
  & 29 17 12 - 28 17 11 & 260\,156.297 & $9.80 \times 10^{-4}$ & 506.8\\ 
  & 29 17 13 - 28 17 12 & 260\,156.297 & $9.80 \times 10^{-4}$ & 506.8\\ 
  & 53 3 50 - 53 2 51 & 260\,179.828 & $6.18 \times 10^{-5}$ & 632.1\\ 
  & 29 6 24 - 28 6 23 & 260\,221.658 & $1.43 \times 10^{-3}$ & 227.2\\ 
  & 29 6 23 - 28 6 22 & 260\,229.158 & $1.43 \times 10^{-3}$ & 227.2\\ 
  & 44 7 38 - 45 4 41 & 260\,232.74 & $3.29 \times 10^{-7}$ & 480.6\\ 
  & 29 18 11 - 28 18 10 & 260\,238.936 & $9.19 \times 10^{-4}$ & 545.3\\ 
  & 29 18 12 - 28 18 11 & 260\,238.936 & $9.19 \times 10^{-4}$ & 545.3\\ 
  & 60 4 56 - 60 3 57 & 260\,322.752 & $7.43 \times 10^{-5}$ & 814.6\\ 
  & 29 19 10 - 28 19 9 & 260\,328.45 & $8.54 \times 10^{-4}$ & 585.9\\ 
  & 29 19 11 - 28 19 10 & 260\,328.45 & $8.54 \times 10^{-4}$ & 585.9\\ 
  & 29 20 9 - 28 20 8 & 260\,424.396 & $7.86 \times 10^{-4}$ & 628.6\\ 
  & 29 20 10 - 28 20 9 & 260\,424.396 & $7.86 \times 10^{-4}$ & 628.6\\ 
  & 31 7 24 - 32 5 27 & 261\,749.397 & $1.87 \times 10^{-7}$ & 267.8\\ 
  & 33 2 31 - 32 3 30 & 261\,799.173 & $5.81 \times 10^{-5}$ & 249.5\\ 
  & 29 12 17 - 30 11 20 & 261\,848.139 & $1.53 \times 10^{-5}$ & 346.8\\ 
  & 29 12 18 - 30 11 19 & 261\,848.139 & $1.53 \times 10^{-5}$ & 346.8\\ 
  & 75 21 54 - 76 20 57 & 261\,866.013 & $2.11 \times 10^{-5}$ & 1711.9\\ 
  & 75 21 55 - 76 20 56 & 261\,866.013 & $2.11 \times 10^{-5}$ & 1711.9\\ 
  & 53 5 49 - 53 3 50 & 262\,101.305 & $8.70 \times 10^{-6}$ & 644.7\\ 
  & 34 13 21 - 35 12 24 & 262\,828.348 & $1.68 \times 10^{-5}$ & 443.3\\ 
  & 34 13 22 - 35 12 23 & 262\,828.348 & $1.68 \times 10^{-5}$ & 443.3\\ 
  & 87 11 77 - 86 12 74 & 262\,862.696 & $2.67 \times 10^{-5}$ & 1779.1\\ 
  & 11 5 7 - 12 3 10 & 262\,873.8 & $3.81 \times 10^{-8}$ & 56.2\\ 
  & 98 13 86 - 97 14 83 & 262\,947.313 & $2.68 \times 10^{-5}$ & 2268.9\\ 
  & 21 6 15 - 22 4 18 & 262\,953.486 & $1.16 \times 10^{-7}$ & 139.4\\ 
  & 98 8 90 - 98 8 91 & 263\,004.792 & $1.02 \times 10^{-5}$ & 2170.5\\ 
  & 80 22 58 - 81 21 61 & 263\,007.909 & $2.16 \times 10^{-5}$ & 1926.5\\ 
  & 80 22 59 - 81 21 60 & 263\,007.909 & $2.16 \times 10^{-5}$ & 1926.5\\ 
  & 31 7 25 - 32 5 28 & 263\,026.766 & $1.89 \times 10^{-7}$ & 267.8\\ 
  & 57 6 52 - 57 5 53 & 263\,040.95 & $9.13 \times 10^{-5}$ & 752.7\\ 
  & 30 2 29 - 29 2 28 & 263\,516.223 & $1.54 \times 10^{-3}$ & 202.1\\ 
\hline 
$\rm CH_{2}DCN$ & 10 8 2 - 9 8 1 & 173\,480.998 & $1.60 \times 10^{-4}$ & 390.6\\ 
  & 10 8 3 - 9 8 2 & 173\,480.998 & $1.60 \times 10^{-4}$ & 390.6\\ 
  & 10 7 3 - 9 7 2 & 173\,524.046 & $2.27 \times 10^{-4}$ & 309.9\\ 
  & 10 7 4 - 9 7 3 & 173\,524.046 & $2.27 \times 10^{-4}$ & 309.9\\ 
  & 10 0 10 - 9 0 9 & 173\,638.56 & $4.46 \times 10^{-4}$ & 45.8\\ 
  & 10 3 8 - 9 3 7 & 173\,641.17 & $4.06 \times 10^{-4}$ & 94.4\\ 
  & 10 3 7 - 9 3 6 & 173\,641.17 & $4.06 \times 10^{-4}$ & 94.4\\ 
  & 10 2 9 - 9 2 8 & 173\,648.22 & $4.28 \times 10^{-4}$ & 67.4\\ 
  & 10 2 8 - 9 2 7 & 173\,673.22 & $4.28 \times 10^{-4}$ & 67.4\\ 
  & 37 1 36 - 37 0 37 & 173\,837.52 & $3.54 \times 10^{-7}$ & 593.4\\ 
  & 16 0 16 - 15 1 15 & 174\,131.5 & $2.20 \times 10^{-7}$ & 113.3\\ 
  & 10 1 9 - 9 1 8 & 174\,407.53 & $4.47 \times 10^{-4}$ & 51.4\\ 
  & 15 10 5 - 14 10 4 & 260\,041.811 & $8.46 \times 10^{-4}$ & 638.2\\ 
  & 15 10 6 - 14 10 5 & 260\,041.811 & $8.46 \times 10^{-4}$ & 638.2\\ 
  & 15 9 6 - 14 9 5 & 260\,123.472 & $9.75 \times 10^{-4}$ & 536.1\\ 
  & 15 9 7 - 14 9 6 & 260\,123.472 & $9.75 \times 10^{-4}$ & 536.1\\ 
  & 15 8 7 - 14 8 6 & 260\,196.625 & $1.09 \times 10^{-3}$ & 444.7\\ 
  & 15 8 8 - 14 8 7 & 260\,196.625 & $1.09 \times 10^{-3}$ & 444.7\\ 
  & 15 7 8 - 14 7 7 & 260\,261.337 & $1.19 \times 10^{-3}$ & 364.0\\ 
  & 15 7 9 - 14 7 8 & 260\,261.337 & $1.19 \times 10^{-3}$ & 364.0\\ 
  & 15 6 9 - 14 6 8 & 260\,317.716 & $1.28 \times 10^{-3}$ & 294.0\\ 
  & 15 6 10 - 14 6 9 & 260\,317.716 & $1.28 \times 10^{-3}$ & 294.0\\ 
  & 15 5 10 - 14 5 9 & 260\,365.984 & $1.36 \times 10^{-3}$ & 234.8\\ 
  & 15 5 11 - 14 5 10 & 260\,365.984 & $1.36 \times 10^{-3}$ & 234.8\\ 
  & 15 0 15 - 14 0 14 & 260\,384.626 & $1.53 \times 10^{-3}$ & 100.0\\ 
  & 15 4 12 - 14 4 11 & 260\,406.697 & $1.42 \times 10^{-3}$ & 186.3\\ 
  & 15 4 11 - 14 4 10 & 260\,406.699 & $1.42 \times 10^{-3}$ & 186.3\\ 
  & 15 2 14 - 14 2 13 & 260\,438.549 & $1.50 \times 10^{-3}$ & 121.6\\ 
  & 15 3 13 - 14 3 12 & 260\,441.387 & $1.47 \times 10^{-3}$ & 148.5\\ 
  & 15 3 12 - 14 3 11 & 260\,441.972 & $1.47 \times 10^{-3}$ & 148.5\\ 
\hline 
$\rm CHD_{2}CN$ & 49 1 48 - 49 1 49 & 186\,743.167 & $5.09 \times 10^{-7}$ & 976.3\\ 
  & 34 0 34 - 33 2 31 & 187\,059.069 & $4.89 \times 10^{-8}$ & 469.9\\ 
  & 26 1 25 - 25 2 23 & 187\,083.112 & $1.29 \times 10^{-7}$ & 283.1\\ 
  & 45 2 44 - 45 1 44 & 187\,826.836 & $3.12 \times 10^{-7}$ & 834.8\\ 
  & 25 4 21 - 26 3 23 & 188\,372.962 & $9.57 \times 10^{-8}$ & 324.8\\ 
  & 10 2 9 - 10 1 9 & 260\,444.239 & $6.53 \times 10^{-7}$ & 60.4\\ 
  & 8 2 7 - 8 1 7 & 261\,931.581 & $6.55 \times 10^{-7}$ & 45.4\\ 
  & 35 1 35 - 34 2 33 & 261\,979.861 & $2.60 \times 10^{-7}$ & 499.9\\ 
  & 16 14 2 - 15 14 1 & 262\,910.198 & $3.69 \times 10^{-4}$ & 935.3\\ 
  & 16 14 3 - 15 14 2 & 262\,910.198 & $3.69 \times 10^{-4}$ & 935.3\\ 
  & 16 13 3 - 15 13 2 & 263\,019.578 & $5.36 \times 10^{-4}$ & 821.5\\ 
  & 16 13 4 - 15 13 3 & 263\,019.578 & $5.36 \times 10^{-4}$ & 821.5\\ 
  & 6 2 5 - 6 1 5 & 263\,107.386 & $6.48 \times 10^{-7}$ & 33.6\\ 
  & 16 12 4 - 15 12 3 & 263\,120.686 & $6.91 \times 10^{-4}$ & 716.1\\ 
  & 16 12 5 - 15 12 4 & 263\,120.686 & $6.91 \times 10^{-4}$ & 716.1\\ 
  & 16 9 7 - 15 9 6 & 263\,376.04 & $1.08 \times 10^{-3}$ & 450.2\\ 
  & 16 9 8 - 15 9 7 & 263\,376.04 & $1.08 \times 10^{-3}$ & 450.2\\ 
  & 16 8 8 - 15 8 7 & 263\,445.162 & $1.19 \times 10^{-3}$ & 378.3\\ 
  & 16 8 9 - 15 8 8 & 263\,445.162 & $1.19 \times 10^{-3}$ & 378.3\\ 
  & 16 7 9 - 15 7 8 & 263\,506.504 & $1.28 \times 10^{-3}$ & 314.9\\ 
  & 16 7 10 - 15 7 9 & 263\,506.504 & $1.28 \times 10^{-3}$ & 314.9\\ 
  & 16 6 10 - 15 6 9 & 263\,560.166 & $1.36 \times 10^{-3}$ & 259.9\\ 
  & 16 6 11 - 15 6 10 & 263\,560.166 & $1.36 \times 10^{-3}$ & 259.9\\ 
  & 16 0 16 - 15 0 15 & 263\,561.61 & $1.59 \times 10^{-3}$ & 107.6\\ 
  & 5 2 4 - 5 1 4 & 263\,578.106 & $6.38 \times 10^{-7}$ & 28.8\\ 
  & 16 5 11 - 15 5 10 & 263\,606.473 & $1.43 \times 10^{-3}$ & 213.4\\ 
  & 16 5 12 - 15 5 11 & 263\,606.473 & $1.43 \times 10^{-3}$ & 213.4\\ 
  & 53 3 50 - 52 4 48 & 263\,614.777 & $3.08 \times 10^{-7}$ & 1169.4\\ 
  & 16 4 13 - 15 4 12 & 263\,646.356 & $1.49 \times 10^{-3}$ & 175.3\\ 
  & 16 4 12 - 15 4 11 & 263\,646.356 & $1.49 \times 10^{-3}$ & 175.3\\ 
  & 16 2 15 - 15 2 14 & 263\,662.527 & $1.57 \times 10^{-3}$ & 124.5\\ 
  & 16 3 14 - 15 3 13 & 263\,682.153 & $1.53 \times 10^{-3}$ & 145.7\\ 
  & 16 3 13 - 15 3 12 & 263\,683.596 & $1.53 \times 10^{-3}$ & 145.7\\ 
\hline 
$\rm NH_{2}CN$ & 13 3 11 0 - 12 3 10 0 & 260\,132.895 & $1.77 \times 10^{-3}$ & 217.7\\ 
  & 13 3 10 0 - 12 3 9 0 & 260\,135.878 & $1.77 \times 10^{-3}$ & 217.7\\ 
\hline 
$\rm CH_{3}NCO$ & 20 2 19 0 - 19 2 18 0 & 173\,483.466 & $2.44 \times 10^{-4}$ & 111.3\\ 
  & 20 2 18 0 - 19 2 17 0 & 173\,675.392 & $2.45 \times 10^{-4}$ & 111.3\\ 
  & 20 -3 0 1 - 19 -3 0 1 & 173\,733.277 & $2.41 \times 10^{-4}$ & 153.1\\ 
  & 20 2 0 1 - 19 2 0 1 & 174\,085.007 & $2.44 \times 10^{-4}$ & 123.3\\ 
  & 20 -1 0 2 - 19 -1 0 2 & 174\,095.57 & $2.46 \times 10^{-4}$ & 146.3\\ 
  & 20 0 0 2 - 19 0 0 2 & 174\,161.002 & $2.47 \times 10^{-4}$ & 140.4\\ 
  & 19 3 0 1 - 18 3 0 1 & 174\,191.731 & $2.17 \times 10^{-4}$ & 145.2\\ 
  & 20 0 0 -3 - 19 0 0 -3 & 174\,421.255 & $2.47 \times 10^{-4}$ & 202.9\\ 
  & 20 1 0 -3 - 19 1 0 -3 & 174\,432.265 & $2.51 \times 10^{-4}$ & 209.6\\ 
  & 20 0 0 3 - 19 0 0 3 & 174\,442.476 & $2.47 \times 10^{-4}$ & 202.1\\ 
  & 20 1 0 3 - 19 1 0 3 & 174\,486.974 & $2.51 \times 10^{-4}$ & 208.8\\ 
  & 20 1 0 2 - 19 1 0 2 & 174\,575.538 & $2.47 \times 10^{-4}$ & 146.4\\ 
  & 30 2 29 0 - 29 2 28 0 & 260\,032.391 & $8.32 \times 10^{-4}$ & 217.4\\ 
  & 30 -3 0 2 - 29 -3 0 2 & 261\,808.735 & $8.34 \times 10^{-4}$ & 300.1\\ 
  & 30 0 0 4 - 29 0 0 4 & 261\,955.843 & $8.43 \times 10^{-4}$ & 395.9\\ 
  & 30 2 0 3 - 29 2 0 3 & 261\,960.792 & $8.43 \times 10^{-4}$ & 332.3\\ 
  & 30 2 0 -3 - 29 2 0 -3 & 261\,963.534 & $8.43 \times 10^{-4}$ & 333.1\\ 
  & 30 1 29 0 - 29 1 28 0 & 262\,109.072 & $8.55 \times 10^{-4}$ & 201.1\\ 
  & 30 -1 0 -3 - 29 -1 0 -3 & 262\,154.856 & $8.26 \times 10^{-4}$ & 313.4\\ 
\hline 
$\rm CH_{3}NH_{2}$ & 10 1 7 - 9 2 7 & 173\,654.159 & $1.41 \times 10^{-6}$ & 120.0\\ 
  & 13 2 5 - 12 3 4 & 173\,922.286 & $1.37 \times 10^{-6}$ & 209.9\\ 
  & 24 5 4 - 23 6 4 & 173\,958.808 & $7.10 \times 10^{-6}$ & 735.2\\ 
  & 17 2 4 - 17 1 5 & 174\,041.351 & $2.90 \times 10^{-5}$ & 341.0\\ 
  & 10 1 4 - 9 2 5 & 174\,135.742 & $2.33 \times 10^{-6}$ & 120.2\\ 
  & 22 0 7 - 21 3 7 & 174\,146.048 & $2.49 \times 10^{-7}$ & 535.3\\ 
  & 40 -3 1 - 40 2 0 & 174\,158.953 & $3.19 \times 10^{-5}$ & 1776.0\\ 
  & 18 1 6 - 18 0 7 & 174\,198.267 & $1.07 \times 10^{-7}$ & 371.0\\ 
  & 24 5 6 - 23 6 6 & 174\,397.684 & $7.13 \times 10^{-6}$ & 735.6\\ 
  & 18 -2 1 - 18 1 0 & 174\,438.993 & $3.08 \times 10^{-5}$ & 379.5\\ 
  & 17 3 4 - 16 4 4 & 186\,769.592 & $9.24 \times 10^{-6}$ & 361.1\\ 
  & 15 2 4 - 15 1 5 & 186\,793.84 & $3.29 \times 10^{-5}$ & 270.9\\ 
  & 39 9 5 - 38 10 5 & 186\,802.979 & $8.44 \times 10^{-6}$ & 1969.6\\ 
  & 19 1 6 - 19 0 7 & 186\,980.617 & $1.10 \times 10^{-7}$ & 411.7\\ 
  & 17 2 2 - 17 -1 3 & 186\,993.005 & $3.68 \times 10^{-5}$ & 341.3\\ 
  & 22 8 0 - 23 -7 1 & 187\,323.266 & $6.99 \times 10^{-6}$ & 786.6\\ 
  & 22 -8 1 - 23 7 0 & 187\,323.267 & $6.99 \times 10^{-6}$ & 786.6\\ 
  & 22 8 5 - 23 7 5 & 187\,339.428 & $7.00 \times 10^{-6}$ & 786.3\\ 
  & 16 -2 1 - 16 1 0 & 188\,210.397 & $3.67 \times 10^{-5}$ & 305.2\\ 
  & 6 4 7 - 7 3 7 & 260\,266.034 & $1.07 \times 10^{-5}$ & 107.5\\ 
  & 10 2 4 - 10 1 4 & 260\,293.536 & $2.26 \times 10^{-5}$ & 132.7\\ 
  & 25 0 5 - 24 3 4 & 260\,303.993 & $8.54 \times 10^{-7}$ & 686.4\\ 
  & 12 1 4 - 11 2 4 & 260\,427.523 & $2.05 \times 10^{-5}$ & 168.6\\ 
  & 17 7 7 - 18 6 7 & 261\,967.364 & $1.73 \times 10^{-5}$ & 516.1\\ 
  & 9 2 6 - 9 1 7 & 262\,121.321 & $1.37 \times 10^{-6}$ & 111.5\\ 
  & 26 5 4 - 25 6 4 & 262\,853.982 & $2.58 \times 10^{-5}$ & 843.5\\ 
  & 37 8 4 - 36 9 4 & 262\,858.032 & $2.44 \times 10^{-5}$ & 1741.1\\ 
  & 8 2 0 - 8 -1 1 & 262\,998.165 & $6.98 \times 10^{-5}$ & 92.8\\ 
  & 31 -1 1 - 30 4 0 & 263\,356.252 & $7.76 \times 10^{-7}$ & 1063.8\\ 
  & 8 0 7 - 7 1 6 & 263\,377.826 & $5.77 \times 10^{-5}$ & 77.1\\ 
  & 6 1 4 - 5 1 4 & 263\,431.705 & $8.98 \times 10^{-6}$ & 48.6\\ 
  & 37 8 0 - 36 -9 1 & 263\,436.944 & $2.47 \times 10^{-5}$ & 1740.8\\ 
  & 37 -8 1 - 36 9 0 & 263\,436.963 & $2.47 \times 10^{-5}$ & 1740.8\\ 
\hline 
$\rm HOCH_{2}CN$ & 58 6 53 0 - 57 7 50 0 & 172\,653.345 & $8.21 \times 10^{-6}$ & 809.1\\ 
  & 3 3 1 1 - 2 2 0 1 & 172\,657.358 & $3.66 \times 10^{-5}$ & 20.6\\ 
  & 46 11 36 1 - 47 10 37 1 & 172\,661.411 & $7.70 \times 10^{-6}$ & 651.7\\ 
  & 46 11 35 1 - 47 10 38 1 & 172\,661.415 & $7.70 \times 10^{-6}$ & 651.7\\ 
  & 60 8 53 1 - 59 9 51 0 & 172\,661.689 & $8.77 \times 10^{-6}$ & 904.5\\ 
  & 3 3 0 1 - 2 2 1 1 & 172\,662.849 & $3.66 \times 10^{-5}$ & 20.6\\ 
  & 77 9 69 1 - 76 10 66 1 & 172\,681.141 & $8.84 \times 10^{-6}$ & 1446.6\\ 
  & 54 4 51 0 - 53 5 48 0 & 172\,689.37 & $3.60 \times 10^{-6}$ & 680.1\\ 
  & 21 7 15 0 - 22 6 16 0 & 172\,698.2 & $6.49 \times 10^{-6}$ & 170.3\\ 
  & 21 7 14 0 - 22 6 17 0 & 172\,698.426 & $6.49 \times 10^{-6}$ & 170.3\\ 
  & 17 2 16 0 - 16 0 16 1 & 172\,716.049 & $2.20 \times 10^{-6}$ & 73.1\\ 
  & 55 4 52 1 - 54 5 49 1 & 173\,518.066 & $3.03 \times 10^{-6}$ & 709.3\\ 
  & 72 7 65 0 - 71 8 63 1 & 173\,520.841 & $9.70 \times 10^{-6}$ & 1234.4\\ 
  & 36 2 34 1 - 35 4 32 0 & 173\,629.603 & $1.09 \times 10^{-6}$ & 310.1\\ 
  & 27 3 24 1 - 26 4 22 0 & 173\,647.439 & $9.29 \times 10^{-6}$ & 186.2\\ 
  & 56 4 52 0 - 56 4 53 0 & 173\,712.043 & $3.10 \times 10^{-6}$ & 737.3\\ 
  & 35 3 32 0 - 35 1 34 1 & 173\,722.216 & $2.26 \times 10^{-6}$ & 294.7\\ 
  & 16 3 13 0 - 15 2 13 1 & 173\,742.253 & $1.52 \times 10^{-5}$ & 72.8\\ 
  & 62 5 58 1 - 62 4 58 0 & 173\,760.542 & $1.15 \times 10^{-5}$ & 905.9\\ 
  & 20 0 20 1 - 19 1 19 1 & 173\,836.595 & $4.12 \times 10^{-5}$ & 96.7\\ 
  & 54 4 51 1 - 53 5 48 1 & 173\,873.415 & $3.06 \times 10^{-6}$ & 685.1\\ 
  & 52 12 41 0 - 53 11 42 0 & 173\,931.489 & $7.96 \times 10^{-6}$ & 809.7\\ 
  & 52 12 40 0 - 53 11 43 0 & 173\,931.491 & $7.96 \times 10^{-6}$ & 809.7\\ 
  & 19 2 18 1 - 18 2 17 1 & 173\,962.325 & $1.61 \times 10^{-4}$ & 94.7\\ 
  & 74 7 68 0 - 73 8 66 1 & 173\,987.188 & $1.08 \times 10^{-5}$ & 1298.5\\ 
  & 46 3 43 0 - 46 3 44 0 & 173\,996.613 & $3.03 \times 10^{-6}$ & 497.9\\ 
  & 19 2 18 0 - 18 2 17 0 & 174\,009.939 & $1.55 \times 10^{-4}$ & 89.4\\ 
  & 77 9 69 0 - 76 10 66 0 & 174\,038.606 & $8.93 \times 10^{-6}$ & 1441.9\\ 
  & 50 4 46 0 - 49 5 44 1 & 174\,040.287 & $1.91 \times 10^{-5}$ & 592.8\\ 
  & 80 8 73 0 - 79 9 71 1 & 174\,056.834 & $1.04 \times 10^{-5}$ & 1524.3\\ 
  & 15 3 13 0 - 14 2 13 1 & 174\,177.206 & $1.55 \times 10^{-5}$ & 65.7\\ 
  & 18 3 16 0 - 18 0 18 1 & 174\,309.545 & $1.30 \times 10^{-6}$ & 88.3\\ 
  & 52 4 48 0 - 52 3 49 0 & 174\,312.887 & $5.08 \times 10^{-5}$ & 639.1\\ 
  & 12 0 12 1 - 11 1 10 0 & 174\,442.231 & $1.66 \times 10^{-5}$ & 39.6\\ 
  & 47 3 44 0 - 46 4 42 1 & 186\,298.941 & $1.88 \times 10^{-5}$ & 519.0\\ 
  & 20 1 19 0 - 19 1 18 0 & 186\,316.378 & $1.93 \times 10^{-4}$ & 96.0\\ 
  & 43 4 39 0 - 42 5 38 0 & 186\,329.612 & $1.16 \times 10^{-5}$ & 444.3\\ 
  & 68 9 60 1 - 67 10 58 0 & 186\,335.61 & $1.11 \times 10^{-5}$ & 1156.0\\ 
  & 55 7 49 1 - 54 8 47 0 & 186\,729.585 & $1.13 \times 10^{-5}$ & 755.6\\ 
  & 7 5 3 0 - 8 4 4 0 & 186\,743.896 & $3.27 \times 10^{-6}$ & 47.1\\ 
  & 7 5 2 0 - 8 4 5 0 & 186\,744.026 & $3.27 \times 10^{-6}$ & 47.1\\ 
  & 32 9 24 1 - 33 8 25 1 & 186\,825.606 & $9.03 \times 10^{-6}$ & 351.6\\ 
  & 32 9 23 1 - 33 8 26 1 & 186\,825.623 & $9.03 \times 10^{-6}$ & 351.6\\ 
  & 62 12 50 1 - 63 11 52 0 & 186\,844.69 & $9.73 \times 10^{-6}$ & 1069.1\\ 
  & 62 12 51 1 - 63 11 53 0 & 186\,844.733 & $9.73 \times 10^{-6}$ & 1069.1\\ 
  & 29 4 25 0 - 29 2 27 1 & 186\,942.085 & $1.50 \times 10^{-5}$ & 215.3\\ 
  & 35 4 31 1 - 34 5 29 0 & 186\,942.311 & $1.17 \times 10^{-5}$ & 307.5\\ 
  & 73 7 66 0 - 72 8 64 1 & 186\,964.419 & $1.21 \times 10^{-5}$ & 1267.0\\ 
  & 54 4 50 0 - 54 3 51 0 & 186\,998.042 & $5.38 \times 10^{-5}$ & 687.4\\ 
  & 10 2 9 1 - 10 1 9 0 & 187\,020.922 & $3.09 \times 10^{-5}$ & 35.3\\ 
  & 27 4 23 1 - 27 3 24 1 & 187\,029.983 & $3.65 \times 10^{-5}$ & 195.2\\ 
  & 7 5 3 1 - 8 4 4 1 & 187\,053.965 & $3.29 \times 10^{-6}$ & 52.6\\ 
  & 7 5 2 1 - 8 4 5 1 & 187\,054.088 & $3.29 \times 10^{-6}$ & 52.6\\ 
  & 21 1 21 0 - 20 1 20 0 & 187\,114.251 & $2.00 \times 10^{-4}$ & 100.5\\ 
  & 19 5 14 1 - 20 4 16 0 & 187\,135.775 & $8.93 \times 10^{-6}$ & 124.3\\ 
  & 30 3 28 1 - 30 2 28 0 & 187\,174.006 & $4.29 \times 10^{-6}$ & 223.8\\ 
  & 27 2 26 1 - 26 3 24 0 & 187\,227.832 & $8.46 \times 10^{-6}$ & 177.1\\ 
  & 21 1 21 1 - 20 1 20 1 & 187\,229.632 & $1.87 \times 10^{-4}$ & 105.9\\ 
  & 75 4 72 0 - 74 5 70 1 & 187\,235.57 & $9.23 \times 10^{-6}$ & 1277.0\\ 
  & 19 5 15 1 - 20 4 17 0 & 187\,320.422 & $8.95 \times 10^{-6}$ & 124.3\\ 
  & 58 6 52 1 - 57 7 51 1 & 187\,325.576 & $1.09 \times 10^{-5}$ & 814.7\\ 
  & 51 5 46 0 - 51 4 47 0 & 187\,343.315 & $2.59 \times 10^{-5}$ & 624.7\\ 
  & 29 2 27 0 - 28 3 26 0 & 187\,359.604 & $1.55 \times 10^{-5}$ & 201.4\\ 
  & 55 7 48 1 - 54 8 46 0 & 187\,446.036 & $1.14 \times 10^{-5}$ & 755.6\\ 
  & 31 1 30 1 - 31 0 31 1 & 187\,524.072 & $1.51 \times 10^{-5}$ & 227.8\\ 
  & 63 14 49 0 - 64 13 52 0 & 187\,664.457 & $1.01 \times 10^{-5}$ & 1163.7\\ 
  & 63 14 50 0 - 64 13 51 0 & 187\,664.457 & $1.01 \times 10^{-5}$ & 1163.7\\ 
  & 51 14 37 0 - 52 13 39 1 & 187\,682.587 & $8.07 \times 10^{-6}$ & 858.5\\ 
  & 51 14 38 0 - 52 13 40 1 & 187\,682.587 & $8.07 \times 10^{-6}$ & 858.5\\ 
  & 20 2 18 1 - 19 2 17 1 & 187\,815.959 & $2.11 \times 10^{-4}$ & 105.0\\ 
  & 21 0 21 0 - 20 0 20 0 & 187\,877.118 & $2.03 \times 10^{-4}$ & 100.3\\ 
  & 52 5 47 1 - 52 4 48 1 & 187\,921.804 & $4.73 \times 10^{-5}$ & 653.0\\ 
  & 15 0 15 1 - 14 1 13 0 & 187\,985.123 & $1.80 \times 10^{-5}$ & 57.9\\ 
  & 31 3 29 0 - 31 2 30 0 & 188\,047.392 & $3.27 \times 10^{-5}$ & 232.2\\ 
  & 20 2 18 0 - 19 2 17 0 & 188\,060.556 & $2.02 \times 10^{-4}$ & 99.7\\ 
  & 38 10 29 0 - 39 9 30 0 & 188\,170.704 & $9.51 \times 10^{-6}$ & 466.9\\ 
  & 38 10 28 0 - 39 9 31 0 & 188\,170.709 & $9.51 \times 10^{-6}$ & 466.9\\ 
  & 36 3 33 1 - 35 4 32 1 & 188\,171.403 & $1.24 \times 10^{-5}$ & 316.0\\ 
  & 68 13 55 1 - 69 12 57 0 & 188\,176.019 & $9.95 \times 10^{-6}$ & 1277.2\\ 
  & 68 13 56 1 - 69 12 58 0 & 188\,176.029 & $9.95 \times 10^{-6}$ & 1277.2\\ 
  & 30 1 30 1 - 29 2 28 0 & 188\,209.049 & $5.83 \times 10^{-6}$ & 205.7\\ 
  & 21 0 21 1 - 20 0 20 1 & 188\,283.514 & $1.59 \times 10^{-4}$ & 105.7\\ 
  & 31 3 29 1 - 30 4 27 0 & 188\,303.563 & $1.12 \times 10^{-5}$ & 237.5\\ 
  & 31 3 29 1 - 31 2 30 1 & 188\,394.528 & $1.02 \times 10^{-5}$ & 237.5\\ 
  & 76 5 72 0 - 75 6 69 0 & 188\,399.299 & $5.32 \times 10^{-6}$ & 1329.6\\ 
  & 46 3 43 0 - 46 2 44 0 & 188\,408.722 & $4.36 \times 10^{-5}$ & 497.9\\ 
  & 50 5 45 0 - 50 4 46 0 & 188\,418.654 & $7.90 \times 10^{-6}$ & 601.9\\ 
  & 46 2 44 0 - 46 2 45 0 & 259\,949.966 & $5.25 \times 10^{-6}$ & 488.9\\ 
  & 78 7 71 1 - 78 6 72 1 & 259\,964.069 & $1.25 \times 10^{-4}$ & 1441.1\\ 
  & 21 5 17 1 - 21 4 18 1 & 259\,967.536 & $8.82 \times 10^{-5}$ & 142.4\\ 
  & 77 7 70 0 - 77 6 71 0 & 259\,970.889 & $8.69 \times 10^{-5}$ & 1401.8\\ 
  & 18 5 13 0 - 18 4 14 0 & 259\,977.031 & $8.61 \times 10^{-5}$ & 110.5\\ 
  & 28 25 3 0 - 27 25 2 0 & 260\,045.93 & $1.09 \times 10^{-4}$ & 1033.3\\ 
  & 28 25 4 0 - 27 25 3 0 & 260\,045.93 & $1.09 \times 10^{-4}$ & 1033.3\\ 
  & 28 26 2 1 - 27 26 1 1 & 260\,047.45 & $7.53 \times 10^{-5}$ & 1107.7\\ 
  & 28 26 3 1 - 27 26 2 1 & 260\,047.45 & $7.53 \times 10^{-5}$ & 1107.7\\ 
  & 23 3 21 0 - 22 2 21 1 & 260\,050.159 & $4.54 \times 10^{-5}$ & 134.8\\ 
  & 18 5 14 0 - 18 4 15 0 & 260\,057.55 & $8.61 \times 10^{-5}$ & 110.5\\ 
  & 24 9 16 0 - 25 8 17 0 & 260\,073.689 & $1.95 \times 10^{-5}$ & 245.3\\ 
  & 24 9 15 0 - 25 8 18 0 & 260\,073.69 & $1.95 \times 10^{-5}$ & 245.3\\ 
  & 17 5 12 0 - 17 4 13 0 & 260\,110.383 & $8.53 \times 10^{-5}$ & 102.5\\ 
  & 20 5 16 1 - 20 4 17 1 & 260\,116.183 & $8.76 \times 10^{-5}$ & 133.1\\ 
  & 19 5 14 1 - 19 4 15 1 & 260\,131.715 & $8.68 \times 10^{-5}$ & 124.3\\ 
  & 17 5 13 0 - 17 4 14 0 & 260\,161.917 & $8.53 \times 10^{-5}$ & 102.5\\ 
  & 42 4 39 0 - 42 2 40 0 & 260\,162.5 & $5.40 \times 10^{-6}$ & 422.9\\ 
  & 45 4 42 1 - 44 5 40 0 & 260\,194.78 & $2.87 \times 10^{-5}$ & 486.5\\ 
  & 83 7 77 0 - 82 8 75 1 & 260\,208.253 & $3.98 \times 10^{-5}$ & 1613.5\\ 
  & 28 26 2 0 - 27 26 1 0 & 260\,213.351 & $7.44 \times 10^{-5}$ & 1101.5\\ 
  & 28 26 3 0 - 27 26 2 0 & 260\,213.351 & $7.44 \times 10^{-5}$ & 1101.5\\ 
  & 16 5 11 0 - 16 4 12 0 & 260\,218.203 & $8.43 \times 10^{-5}$ & 95.0\\ 
  & 28 27 1 1 - 27 27 0 1 & 260\,220.963 & $3.84 \times 10^{-5}$ & 1178.5\\ 
  & 28 27 2 1 - 27 27 1 1 & 260\,220.963 & $3.84 \times 10^{-5}$ & 1178.5\\ 
  & 19 5 15 1 - 19 4 16 1 & 260\,247.467 & $8.69 \times 10^{-5}$ & 124.3\\ 
  & 16 5 12 0 - 16 4 13 0 & 260\,250.329 & $8.44 \times 10^{-5}$ & 95.0\\ 
  & 18 5 13 1 - 18 4 14 1 & 260\,285.695 & $8.61 \times 10^{-5}$ & 115.9\\ 
  & 13 3 10 1 - 13 2 12 0 & 260\,291.934 & $5.81 \times 10^{-5}$ & 58.2\\ 
  & 6 6 0 0 - 5 5 0 1 & 260\,295.415 & $1.29 \times 10^{-4}$ & 59.3\\ 
  & 6 6 1 0 - 5 5 1 1 & 260\,295.415 & $1.29 \times 10^{-4}$ & 59.3\\ 
  & 40 1 39 1 - 40 0 40 1 & 260\,297.087 & $3.26 \times 10^{-5}$ & 368.9\\ 
  & 15 3 12 1 - 15 2 14 0 & 260\,297.601 & $5.67 \times 10^{-5}$ & 71.1\\ 
  & 39 4 36 0 - 39 2 37 0 & 260\,302.383 & $4.94 \times 10^{-6}$ & 368.3\\ 
  & 15 5 10 0 - 15 4 11 0 & 260\,304.408 & $8.33 \times 10^{-5}$ & 87.9\\ 
  & 15 5 11 0 - 15 4 12 0 & 260\,323.792 & $8.33 \times 10^{-5}$ & 87.9\\ 
  & 18 5 14 1 - 18 4 15 1 & 260\,361.653 & $8.62 \times 10^{-5}$ & 115.9\\ 
  & 48 7 42 0 - 48 6 42 1 & 260\,362.51 & $6.58 \times 10^{-5}$ & 589.1\\ 
  & 34 4 31 1 - 34 3 31 0 & 260\,370.704 & $5.21 \times 10^{-5}$ & 291.4\\ 
  & 14 5 9 0 - 14 4 10 0 & 260\,372.309 & $8.20 \times 10^{-5}$ & 81.2\\ 
  & 14 5 10 0 - 14 4 11 0 & 260\,383.545 & $8.20 \times 10^{-5}$ & 81.2\\ 
  & 28 27 1 0 - 27 27 0 0 & 260\,386.776 & $3.80 \times 10^{-5}$ & 1172.2\\ 
  & 28 27 2 0 - 27 27 1 0 & 260\,386.776 & $3.80 \times 10^{-5}$ & 1172.2\\ 
  & 5 4 1 1 - 6 3 3 0 & 260\,408.16 & $6.97 \times 10^{-6}$ & 34.3\\ 
  & 17 5 12 1 - 17 4 13 1 & 260\,410.819 & $8.53 \times 10^{-5}$ & 107.9\\ 
  & 5 4 2 1 - 6 3 4 0 & 260\,411.948 & $6.97 \times 10^{-6}$ & 34.3\\ 
  & 13 5 8 0 - 13 4 9 0 & 260\,424.892 & $8.05 \times 10^{-5}$ & 75.0\\ 
  & 13 5 9 0 - 13 4 10 0 & 260\,431.158 & $8.05 \times 10^{-5}$ & 75.0\\ 
  & 17 5 13 1 - 17 4 14 1 & 260\,459.444 & $8.54 \times 10^{-5}$ & 107.9\\ 
  & 44 5 40 1 - 44 4 41 1 & 261\,739.468 & $1.08 \times 10^{-4}$ & 479.1\\ 
  & 18 2 16 1 - 18 1 18 0 & 261\,767.085 & $2.02 \times 10^{-5}$ & 87.4\\ 
  & 29 2 27 1 - 28 3 25 0 & 261\,797.89 & $5.47 \times 10^{-5}$ & 206.3\\ 
  & 82 7 76 1 - 82 6 76 0 & 261\,833.407 & $1.81 \times 10^{-4}$ & 1581.4\\ 
  & 40 1 39 0 - 40 1 40 0 & 261\,835.122 & $3.23 \times 10^{-6}$ & 363.6\\ 
  & 16 3 13 1 - 16 2 15 0 & 261\,835.599 & $6.36 \times 10^{-5}$ & 78.2\\ 
  & 43 5 38 0 - 43 3 40 1 & 261\,888.854 & $1.74 \times 10^{-6}$ & 454.8\\ 
  & 50 7 43 0 - 50 6 45 1 & 261\,922.423 & $6.56 \times 10^{-5}$ & 633.1\\ 
  & 58 4 54 0 - 57 5 52 1 & 261\,937.931 & $5.31 \times 10^{-5}$ & 789.0\\ 
  & 40 1 39 0 - 40 0 40 0 & 261\,939.987 & $3.97 \times 10^{-5}$ & 363.6\\ 
  & 60 7 53 0 - 60 6 55 1 & 261\,942.087 & $5.76 \times 10^{-5}$ & 879.4\\ 
  & 54 3 51 1 - 54 2 52 1 & 261\,971.845 & $7.83 \times 10^{-5}$ & 683.3\\ 
  & 76 9 67 1 - 75 10 65 0 & 261\,983.481 & $3.28 \times 10^{-5}$ & 1412.5\\ 
  & 28 3 25 1 - 27 3 24 1 & 262\,020.774 & $5.56 \times 10^{-4}$ & 198.8\\ 
  & 14 3 11 1 - 14 2 13 0 & 262\,027.209 & $3.59 \times 10^{-5}$ & 64.4\\ 
  & 30 10 21 0 - 31 9 22 0 & 262\,032.199 & $2.19 \times 10^{-5}$ & 344.6\\ 
  & 30 10 20 0 - 31 9 23 0 & 262\,032.199 & $2.19 \times 10^{-5}$ & 344.6\\ 
  & 51 2 50 1 - 50 3 48 0 & 262\,074.734 & $1.89 \times 10^{-5}$ & 587.0\\ 
  & 45 7 39 0 - 45 6 39 1 & 262\,094.475 & $6.80 \times 10^{-5}$ & 526.6\\ 
  & 49 7 42 0 - 49 6 44 1 & 262\,157.573 & $6.62 \times 10^{-5}$ & 610.9\\ 
  & 11 5 6 1 - 12 4 8 0 & 262\,714.706 & $1.60 \times 10^{-5}$ & 69.4\\ 
  & 11 5 7 1 - 12 4 9 0 & 262\,718.054 & $1.60 \times 10^{-5}$ & 69.4\\ 
  & 13 7 7 0 - 13 6 7 1 & 262\,736.376 & $5.96 \times 10^{-5}$ & 108.3\\ 
  & 13 7 6 0 - 13 6 8 1 & 262\,736.376 & $5.96 \times 10^{-5}$ & 108.3\\ 
  & 30 10 20 1 - 31 9 23 1 & 262\,762.028 & $2.21 \times 10^{-5}$ & 350.1\\ 
  & 30 10 21 1 - 31 9 22 1 & 262\,762.028 & $2.21 \times 10^{-5}$ & 350.1\\ 
  & 81 6 75 0 - 81 5 76 0 & 262\,765.813 & $1.42 \times 10^{-4}$ & 1532.1\\ 
  & 30 12 18 0 - 31 11 20 1 & 262\,773.324 & $1.59 \times 10^{-5}$ & 405.5\\ 
  & 30 12 19 0 - 31 11 21 1 & 262\,773.324 & $1.59 \times 10^{-5}$ & 405.5\\ 
  & 28 3 25 0 - 27 3 24 0 & 262\,797.496 & $5.36 \times 10^{-4}$ & 193.8\\ 
  & 14 7 8 0 - 14 6 8 1 & 262\,825.871 & $6.20 \times 10^{-5}$ & 114.5\\ 
  & 14 7 7 0 - 14 6 9 1 & 262\,825.871 & $6.20 \times 10^{-5}$ & 114.5\\ 
  & 37 4 34 0 - 37 2 35 0 & 262\,846.614 & $4.71 \times 10^{-6}$ & 334.1\\ 
  & 46 7 39 0 - 46 6 41 1 & 262\,856.464 & $6.80 \times 10^{-5}$ & 547.0\\ 
  & 43 7 37 0 - 43 6 37 1 & 262\,877.193 & $6.92 \times 10^{-5}$ & 487.2\\ 
  & 15 7 9 0 - 15 6 9 1 & 262\,918.607 & $6.39 \times 10^{-5}$ & 121.2\\ 
  & 15 7 8 0 - 15 6 10 1 & 262\,918.607 & $6.39 \times 10^{-5}$ & 121.2\\ 
  & 45 5 41 1 - 45 4 42 1 & 262\,927.287 & $1.12 \times 10^{-4}$ & 499.1\\ 
  & 27 3 24 0 - 26 2 24 1 & 262\,967.477 & $4.05 \times 10^{-5}$ & 181.1\\ 
  & 45 5 41 0 - 45 4 42 0 & 262\,995.615 & $1.02 \times 10^{-4}$ & 494.0\\ 
  & 16 7 10 0 - 16 6 10 1 & 263\,014.239 & $6.55 \times 10^{-5}$ & 128.3\\ 
  & 16 7 9 0 - 16 6 11 1 & 263\,014.239 & $6.55 \times 10^{-5}$ & 128.3\\ 
  & 57 5 52 0 - 57 4 54 1 & 263\,043.568 & $4.87 \times 10^{-5}$ & 771.6\\ 
  & 45 7 38 0 - 45 6 40 1 & 263\,071.806 & $6.85 \times 10^{-5}$ & 526.6\\ 
  & 17 7 11 0 - 17 6 11 1 & 263\,111.982 & $6.67 \times 10^{-5}$ & 135.8\\ 
  & 17 7 10 0 - 17 6 12 1 & 263\,111.982 & $6.67 \times 10^{-5}$ & 135.8\\ 
  & 42 7 36 0 - 42 6 36 1 & 263\,181.856 & $6.97 \times 10^{-5}$ & 468.1\\ 
  & 15 2 13 1 - 14 2 13 0 & 263\,186.129 & $1.51 \times 10^{-5}$ & 64.5\\ 
  & 18 7 12 0 - 18 6 12 1 & 263\,211.256 & $6.78 \times 10^{-5}$ & 143.8\\ 
  & 18 7 11 0 - 18 6 13 1 & 263\,211.256 & $6.78 \times 10^{-5}$ & 143.8\\ 
  & 62 7 55 0 - 62 6 57 1 & 263\,304.782 & $5.54 \times 10^{-5}$ & 934.0\\ 
  & 19 7 13 0 - 19 6 13 1 & 263\,311.186 & $6.87 \times 10^{-5}$ & 152.2\\ 
  & 19 7 12 0 - 19 6 14 1 & 263\,311.186 & $6.87 \times 10^{-5}$ & 152.2\\ 
  & 55 16 39 0 - 56 15 41 1 & 263\,311.961 & $2.12 \times 10^{-5}$ & 1035.9\\ 
  & 55 16 40 0 - 56 15 42 1 & 263\,311.961 & $2.12 \times 10^{-5}$ & 1035.9\\ 
  & 20 7 14 0 - 20 6 14 1 & 263\,410.999 & $6.95 \times 10^{-5}$ & 161.0\\ 
  & 20 7 13 0 - 20 6 15 1 & 263\,410.999 & $6.95 \times 10^{-5}$ & 161.0\\ 
  & 17 3 14 1 - 17 2 16 0 & 263\,422.063 & $6.50 \times 10^{-5}$ & 85.7\\ 
  & 40 2 39 1 - 40 1 40 1 & 263\,422.917 & $3.38 \times 10^{-5}$ & 369.0\\ 
  & 41 7 35 0 - 41 6 35 1 & 263\,437.433 & $7.02 \times 10^{-5}$ & 449.5\\ 
  & 43 7 36 0 - 43 6 38 1 & 263\,457.612 & $6.95 \times 10^{-5}$ & 487.2\\ 
  & 54 12 42 1 - 55 11 44 0 & 263\,490.25 & $2.54 \times 10^{-5}$ & 862.2\\ 
  & 54 12 43 1 - 55 11 45 0 & 263\,490.252 & $2.54 \times 10^{-5}$ & 862.2\\ 
  & 63 7 57 1 - 62 8 55 0 & 263\,491.36 & $3.37 \times 10^{-5}$ & 966.7\\ 
  & 21 7 15 0 - 21 6 15 1 & 263\,509.996 & $7.01 \times 10^{-5}$ & 170.3\\ 
  & 21 7 14 0 - 21 6 16 1 & 263\,509.996 & $7.01 \times 10^{-5}$ & 170.3\\ 
  & 52 2 51 1 - 51 3 49 0 & 263\,555.984 & $1.88 \times 10^{-5}$ & 609.3\\ 
  & 38 4 35 1 - 38 2 36 1 & 263\,581.59 & $1.52 \times 10^{-6}$ & 356.2\\ 
  & 22 7 16 0 - 22 6 16 1 & 263\,607.225 & $7.06 \times 10^{-5}$ & 180.1\\ 
  & 22 7 15 0 - 22 6 17 1 & 263\,607.225 & $7.06 \times 10^{-5}$ & 180.1\\ 
  & 42 7 35 0 - 42 6 37 1 & 263\,624.45 & $6.99 \times 10^{-5}$ & 468.1\\ 
  & 40 7 34 0 - 40 6 34 1 & 263\,649.421 & $7.06 \times 10^{-5}$ & 431.3\\ 
  & 48 4 44 0 - 47 5 43 0 & 263\,656.627 & $4.52 \times 10^{-5}$ & 547.9\\ 
  & 29 2 28 0 - 28 2 27 0 & 263\,680.233 & $5.59 \times 10^{-4}$ & 196.6\\ 
  & 23 7 17 0 - 23 6 17 1 & 263\,701.584 & $7.10 \times 10^{-5}$ & 190.3\\ 
  & 23 7 16 0 - 23 6 18 1 & 263\,701.948 & $7.10 \times 10^{-5}$ & 190.3
\label{tab:lines_band56}
\end{longtable}

\end{appendix}

\end{document}